\documentclass{llncs}

\usepackage[title]{appendix}
\usepackage{import}
\usepackage{mathpartir,amssymb,mathrsfs,float, lscape, stmaryrd}
\usepackage[fleqn]{mathtools}
\usepackage{xcolor}
\usepackage{bm}
\usepackage{hyperref}
\usepackage{enumitem}
\usepackage{environ}
\usepackage{mwe}
\usepackage{amssymb, stackengine}
\usepackage{ textcomp }
\usepackage{bbm}
\usepackage[mathscr]{euscript}
 \let\mathscr\relax% just so we can load this and rsfs
\usepackage[scr]{rsfso}
\usepackage{centernot}
\usepackage{caption}
\usepackage{algorithm}
\usepackage{algorithmicx}
\usepackage{algpseudocode}
\usepackage{graphicx}
\usepackage{subcaption}
\usepackage{pifont}% http://ctan.org/pkg/pifont
\usepackage{bigints}
\usepackage{xspace}
\usepackage{varwidth}
\usepackage{subcaption}
\usepackage{cleveref}

\usepackage{cite}
\usepackage{amsmath,amsfonts}
\usepackage{textcomp}
\usepackage{wrapfig}

\newcommand{\csre}{\textsf{\textup{CRSE}}\xspace}
\newcommand{\rse}{\textsf{\textup{RSE}}\xspace}
\newcommand{\pfor}{\textsf{\textup{PFOR}}\xspace}

\newcommand{\rpfor}{{\textsf{\textup{RPFOR}}}\xspace}

\newcommand{\distr}[1]{\textbf{\textup{distr}}(#1)}
\newcommand{\sdistr}[1]{\textbf{\textup{sdistr}}(#1)}
\newcommand{\varset}{\mathbb{V}}
\newcommand{\vararrset}{\mathbb{A}}

\newcommand{\len}[1]{{\tt len}(#1)}
\newcommand{\skipp}{{\tt skip}}
\newcommand{\seq}[2]{#1{\tt ;}#2}
\newcommand{\ifte}[3]{{\tt if}~#1~{\tt then}~#2~{\tt else}~#3}
\newcommand{\cfor}[4]{{\tt for}~(#1~{\tt in~}#2{:}#3)~{\tt do}~{#4}~{\tt od}}

\newcommand{\ass}[2]{#1{\tt \leftarrow}#2}
\newcommand{\rass}[2]{#1{\tt \xleftarrow{\$}}#2}
\newcommand{\arracc}[2]{#1[#2]}

\newcommand{\lapp}[2]{lap_{#2}(#1)}

\newcommand{\pexpr}{\mathcal{E}}
\newcommand{\spexpr}{\mathcal{E}_{\textup{s}}}

\newcommand{\pvalset}{\mathcal{V}}

\newcommand{\rpvalset}{\mathcal{V}_{\textup{r}}}
\newcommand{\pcmd}{\mathcal{C}}
\newcommand{\spcmd}{\mathcal{C}_{\textup{s}}}

\newcommand{\memset}{\mathcal{M}}

\newcommand{\cmemset}{\mathcal{M}_{\textup{c}}}

\newcommand{\rulestyle}[1]{{\textbf{\textsc{#1}}}}
\newcommand{\econf}[3]{\langle #1, #2, #3 \rangle}
\newcommand{\reconf}[5]{\langle #1, #2, #3, #4, #5 \rangle}
\newcommand{\estep}{\downarrow_{\textup{c}}}
\newcommand{\restep}{\downarrow_{\textup{rc}}}

\newcommand{\vconf}[2]{\langle #1,#2\rangle}
\newcommand{\rvconf}[3]{\langle #1,#2, #3\rangle}

\newcommand{\cstrset}{\mathcal{S}}

\newcommand{\ints}{\mathbb{Z}}
\newcommand{\symsp}{\mathcal{X}_p}
\newcommand{\syms}{\mathcal{X}}
\newcommand{\conf}[3]{\langle #1,#2, #3 \rangle}

\newcommand{\cstep}{\rightarrow_{\textup{c}}}

\newcommand{\carrayset}{\textbf{Array}}
\newcommand{\sarrayset}{\textbf{Array}_{\textup{s}}}

\newcommand{\setstep}{\Rightarrow_{\textup{c}}}
\newcommand{\setpstep}{\Rightarrow_{\textup{rc}}}
\newcommand{\fresh}[2]{#1\, \textbf{fresh}(#2)}

\newcommand{\pair}[2]{\langle#1|#2\rangle}
\newcommand{\prexpr}{\mathcal{E}_{\textup{r}}}

\newcommand{\prcmd}{\mathcal{C}_{\textup{r}}}
\newcommand{\powerset}{\mathcal{P}}
\newcommand{\proj}[2]{\lfloor #2 \rfloor_{#1} }
\newcommand{\rconf}[5]{\langle #1,#2, #3, #4, #5 \rangle}
\newcommand{\rcstep}{\rightarrow_{\textup{rc}}}
\newcommand{\setsubdistr}{\mathscr{D}}
\newcommand{\setpsubdistr}{\mathscr{R}}
\newcommand{\cdenote}[1]{\llbracket #1 \rrbracket_{\pcmd}}
\newcommand{\final}[1]{\textbf{Final}(#1)}
\newcommand{\denotemp}[2]{\llbracket #1; #2 \rrbracket^{\textbf{mp}}}
\newcommand{\denotep}[1]{\llbracket #1 \rrbracket^{\textbf{p}}}
\newcommand{\denotere}[2]{\llbracket #1 \rrbracket^{\textbf{re}}_{#2}}  
\newcommand{\unit}[1]{\textbf{\textup{unit}}(#1)}
\newcommand{\eunit}{\textbf{\textup{unit}}}
\newcommand{\bind}[2]{\textbf{\textup{bind}}(#1, #2)}
\newcommand{\ebind}{\textbf{\textup{bind}}}
\newcommand{\outputset}{\mathcal{O}}
\newcommand{\pwhile}{\textbf{\textup{pWhile}}}
\newcommand{\match}[2]{#1 \vdash #2}
\newcommand{\select}[2]{\textbf{select}(#1,#2)}
\newcommand{\store}[3]{\textbf{store}(#1,#2, #3)}

\newcommand{\srconf}[6]{\langle #1, #2, #3,  #4, #5, #6 \rangle}

\newcommand{\srsetdistr}{\mathcal{H}_{\textup{sr}}}
\newcommand{\setproof}{\mathscr{G}}
\newcommand{\proofstep}{	\rightsquigarrow}
\newcommand{\ctxt}{\mathcal{CTX}}
\newcommand{\ctxtp}{\mathcal{P}}

\newcommand{\intlog}{\mathcal{I}}
\newcommand{\judg}[5]{#5 \vdash #1: #3 \xrightarrow[]{#2} #4}
\newcommand{\tocstr}[2]{\llbracket #2\rrbracket_{#1}}
\newcommand{\setdistr}{\mathcal{H}}
\newcommand{\setsymsubdistr}{\mathscr{H}}
\newcommand{\psetstep}{\Rightarrow_{\textup{p}}}
\newcommand{\rsetsymsubdistr}{\mathscr{S}\mathscr{R}}
\newcommand{\rsetsubdistr}{\mathscr{R}}
\newcommand{\rsetsteptr}{\rsetstep^{*}}
\newcommand{\rsetstep}{\Rightarrow_{\textup{rp}}}
\newcommand{\proofstephat}{\xRightarrow[\sim]{}}
\newcommand{\proofstepup}{\xRightarrow[]{\sim}}

\newcommand{\srpsetstep}{\Rightarrow_{\textup{srp}}}
\newcommand{\srpsetstepby}[1]{\xRightarrow{#1}_{\textup{srp}}}

\newcommand{\psetsteptr}{\psetstep^{*}}
\newcommand{\sat}[1]{\textup{\textbf{SAT}}(#1)}

\newcommand{\sub}[2]{#2(#1)}
\newcommand{\logvar}{\textsf{\textup{LogVar}}}
\newtheorem{assumption}{Assumption}
\newcommand{\divergence}[3]{\Delta_{#1}(#2,#3)}
\newcommand{\supp}[1]{\textbf{supp}(#1)}

\newcommand{\dbset}{\mathcal{D}}
\newcommand{\reals}{\mathcal{R}}
\newcommand{\crse}{\textsf{\textup{CRSE}}\xspace}

\newcommand{\tbranch}{c_\mathit{tt}}
\newcommand{\fbranch}{c_\mathit{ff}}

\newcommand{\splang}{{\textsf{\textup{SPFOR}}}\xspace}
\newcommand{\srplang}{{\textsf{\textup{SRPFOR}}}\xspace}
\newcommand{\plang}{{\textsf{\textup{PFOR}}}\xspace}
\newcommand{\rplang}{{\textsf{\textup{RPFOR}}}\xspace}

\newcommand{\splangcmd}{\textup{\textbf{SPForCmd}}\xspace}
\newcommand{\splangvalset}{{\ensuremath{\textup{\textbf{V}}}_{\textup{s}}}}
\newcommand{\spmemset}{{\ensuremath{\bm{\mathcal{M}}_{\textup{\tiny{SP}}}}}}
\newcommand{\spintlangvalset}{{\ensuremath{\textup{\textbf{V}}}_{\textup{is}}}}
\newcommand{\plangvalset}{\pvalset}
\newcommand{\cexpr}{\cstrset_{e}}
\newcommand{\langvalset}{\textup{\textbf{V}}}
\newcommand{\spestep}{\downarrow_{\textup{\tiny{SP}}}}
\newcommand{\spcstep}{\rightarrow_{\textup{\tiny{SP}}}}
\newcommand{\speconf}[4]{(#1,#2, #3, #4)}
\newcommand{\spval}[3]{(#1,#2, #3)}
\newcommand{\spcconf}[4]{(#1,#2, #3, #4)}

\newcommand{\spsetstep}{\Rightarrow_{\textup{sp}}}
\newcommand{\spsetstepby}[1]{\xRightarrow{#1}_{\textup{sp}}}

\newcommand{\srplangexpr}{\mathcal{E}_{\textup{rs}}\xspace}
\newcommand{\srplangcmd}{\mathcal{C}_{\textup{rs}}\xspace}
\newcommand{\srplangvalset}{{\ensuremath{\textup{\textbf{V}}}_{\textup{srp}}}}
\newcommand{\srpeconf}[6]{(#1,#2, #3, #4, #5, #6)}
\newcommand{\srpval}[4]{(#1,#2, #3, #4)}
\newcommand{\srpcconf}[6]{(#1,#2, #3, #4, #5, #6)}
\newcommand{\srpestep}{\downarrow_{\textup{\tiny{SRP}}}}
\newcommand{\srpcstep}{\rightarrow_{\textup{\tiny{SRP}}}}
\newcommand{\emptyctxt}[1]{[\ #1\ ]}

\def\full{1}  % 1 means full version

\begin{document}

\title{Coupled Relational Symbolic Execution for Differential Privacy}
\author{Gian Pietro Farina\inst{1} \and
Stephen Chong\inst{2}  \and
Marco Gaboardi\inst{3}}
\institute{
  University at Buffalo, SUNY \and
  Harvard \and
  Boston University
}

\maketitle
\begin{abstract}
Differential privacy is a de facto standard in data
privacy with applications in the private and public
sectors. Most of the techniques that achieve differential
privacy are based on a judicious use of randomness. However, reasoning
about randomized programs is difficult and error prone. For this
reason, several techniques have been recently proposed to support
designer in proving programs differentially private or in finding
violations to it. 

In this work we propose a technique based on  symbolic execution for
reasoning about differential privacy.  Symbolic execution is a classic
technique used for testing, counterexample generation and to prove
absence of bugs. Here we use symbolic execution to support these
tasks specifically for differential privacy. To achieve this goal, we
leverage two ideas that have been already proven useful in formal
reasoning about differential privacy: relational reasoning and
probabilistic coupling. Our technique integrates these two ideas and shows how such a combination can be used to both verify and find violations to differential privacy.

% Recent work
%   have Its relational extension \cite{...} deals with
%   deterministic relational properties but it deos not extend to interesting
%   probabilistic relational properties such as Differential Privacy (DP).
%   This work aims at filling this gap providing some strategies, based on
%   probabilistic coupling theory, useful to extend RSE so it can find violations
%   to Differential Privacy.
  %\keywords{Symbolic execution  \and relational reasoning \and differential privacy \and
  %  probabilistic couplings.}
\end{abstract}

\section{Introduction}
\label{sec:intro}
Differential Privacy \cite{DworkMNS16} has become a de facto gold
standard definition of privacy for statistical analysis.  This success
is mostly due to the generality of the definition, its robustness and
compositionality.
\ifnum\full=1
These valuable properties helped researchers from
many different communities - e.g. machine learning, data analysis, and
security - in coming up with differentially private algorithms for
specific goals.
However, it was quickly understood that getting
differential privacy right in practice is a hard task.
\else
However, getting
differential privacy right in practice is a hard task.
\fi
Even privacy
experts have released fragile code subject to
\ifnum\full=1
attacks~\cite{Haeberlen11,Mironov12,AndryscoKMJLS15,EbadiAS16,Burke17}
\else
attacks~\cite{Haeberlen11,Mironov12}
\fi
and published incorrect algorithms~\cite{Lyu-2017}.  This challenge
has motivated the programming language community to develop techniques to
support programmer to show their algorithms differentially
private. Among the techniques that have been proposed there are
\ifnum\full=1
type systems~\cite{ReedP10,Gaboardi2013,BartheGAHRS15,BartheFGAGHS16,ZhangK17,NearDASGWSZSSS19,Wang19},
methods based on model checking and Markov chains~\cite{TschantzKD11,ChatzikokolakisGPX14,LiuWZ18,ChistikovMP18,ChistikovMP19,Barthe2019AutomatedMF},
and program
logics~\cite{barthe2012probabilistic,BartheGAHKS14,Barthe:2016,BartheFGGHS16,SatoBGHK19}.
\else
type systems~\cite{ReedP10,Gaboardi2013,ZhangK17,NearDASGWSZSSS19,Wang19},
methods based on model checking and Markov chains~\cite{TschantzKD11,LiuWZ18,Barthe2019AutomatedMF},
and program
logics~\cite{barthe2012probabilistic,Barthe:2016,SatoBGHK19}.
\fi
\ifnum\full=1
More recently, the formal methods community have also focused on developing techniques to find violations to differential privacy\cite{DingWWZK18,Bichsel:2018,Barthe2019AutomatedMF}.

\else
Several works have also focused on developing techniques to find violations to differential privacy\cite{DingWWZK18,Bichsel:2018,Barthe2019AutomatedMF}.
\fi
\ifnum\full=1
Most of these works focus on either verifying a program differentially private or finding violations to differential privacy and they do not consider techniques supporting both kind of reasoning. An exception is the recent work by Barthe et al.~\cite{Barthe2019AutomatedMF} which proposes a method based on a decidable logic for a simple while
language over finite input and output domains, that can be used for both verifying
and finding violation to differential privacy.
\else
Most of these works focus on either verifying a program differentially private or finding violationsand they do not address both kinds of reasoning. An exception addressing both is the recent work by Barthe et al.~\cite{Barthe2019AutomatedMF} which proposes a method based on a decidable logic for a simple while
language over finite input and output domains.
\fi

Motivated by this picture, we propose a new technique based on relational symbolic execution, named Coupled Relational Symbolic Execution (\csre),  which supports proving and finding violation to differential privacy for programs.   
Our technique is based on two essential ingredients: the use of a recently introduced notion of relational symbolic execution~\cite{Farina2019} and the use of approximate probabilistic couplings\cite{Barthe:2016} to reason about differential privacy a relational way. This approach allow us also to support reasoning over countable input and output domains.

\indent{\bf Relational Symbolic Execution.}  Symbolic execution
is a classic technique used for bug finding,
testing and proving.  In symbolic execution an evaluator executes the program
which consumes symbolic inputs instead of concrete ones.  The
evaluator follows, potentially, all the execution paths the program
could take and collects constraints over the symbolic values,
corresponding to these paths.
\ifnum\full=1
The evaluator collects in this way a
description of the traces in terms of constraints on symbolic values
or expressions involving them.  Every trace is associated with a set
of constraints and every input satisfiying these constraints
will lead the actual concrete execution along that trace.

Similarly, in relational symbolic execution~\cite{Farina2019} (\rse)
one is concerned with bug finding, testing, or proving for
\emph{relational properties}. These are properties about two
executions of two potentially different programs.  \rse executes two
potentially different programs in a symbolic fashion. RSE exploits
relational assumptions about the two inputs to the two programs in
order to reduce the number of states to analyze. This can be
particularly effective when the codes of the two programs share some
similarities, and when the property under consideration is relational
in nature, as in the case of differential privacy.
\else
Similarly, in relational symbolic execution~\cite{Farina2019} (\rse)
one is concerned with bug finding, testing, or proving for
\emph{relational properties}. These are properties about two
executions of two potentially different programs.  \rse executes two
potentially different programs in a symbolic fashion  and exploits
relational assumptions about the inputs or the programs in
order to reduce the number of states to analyze. This is
effective when the codes of the two programs share some
similarities, and when the property under consideration is relational
in nature, as differential privacy.
\fi
\ifnum\full=1
\indent{\bf Approximate Probabilistic Couplings.}  Probabilistic
coupling \cite{Lindvall1992LecturesOT} is a proof technique useful to
relate two random variables through a common joint probability
distribution. Probabilistic coupling has been used in formal
verification~\cite{Jonsson0L01} to lift a relation over the joint support of two
probability distribution to a relation over the two probability
distributions themselves. This allows one to reason about relations
between probability distributions by reasoning about relations on their support,
which can be usually done in a symbolic way.
In this approach the actual probabilistic reasoning is confined to the soundness of the verification system, rather than being spread everywhere. A relaxation of the notion of
coupling, called \emph{approximate probabilistic coupling}~\cite{barthe2012probabilistic,Barthe:2016},  has been designed to reason about differential privacy. This can be seen as a regular probabilistic coupling with some additional parameter describing how close the two probability distribution are.

\medskip

\else

\indent{\bf Approximate Probabilistic Couplings.} Probabilistic coupling is a proof technique usful to lift a relation over the joint support of two
probability distribution to a relation over the two probability
distributions themselves. This allows one to reason about relations
between probability distributions by reasoning about relations on their support,
which can be usually done in a symbolic way.
In this approach the actual probabilistic reasoning is confined to the soundness of the verification system, rather than being spread everywhere in the verification effort. A relaxation of the notion of
coupling, called \emph{approximate probabilistic coupling}~\cite{barthe2012probabilistic,Barthe:2016},  has been designed to reason about differential privacy. This can be seen as a regular probabilistic coupling with some additional parameter describing how close the two probability distribution are. 
\fi
%
% \indent\emph{Differential Privacy.}
% Intuitively, a differentially private computation should not be too
% sensitive to small input differences. \emph{Small input differences}
% usually denote databases which differ in at most one row which in
% literature is taken to represent the contribute that a single person
% can give to the output.  \emph{Being not too sensitive} is measured
% through the parameter $\epsilon$, requiring that the absolute value of
% the ratio of probabilities of any output event when the algorithm is
% fed with one input and when is fed with another, close input, is no
% more than $e^\epsilon$. Intuitively, the smaller the $\epsilon$
% the more privacy is guaranteed. A qualitatively different notion of
% differential privacy allows for a \emph{slack}, $\delta$, which
% quantifies the probability of a catastrophic event, e.g. the release
% of the secret data. As the reader will have noticed differential privacy
% is an intrinsically probabilistic property. A standard way to
% perform a task in differentially private manner is to
% add judiciously randomness to the computation.
% For instance a standard way to achieve differential privacy
% when releasing a numeric query on a database is to
% add to the query result a value sampled from the
% Laplace distribution with mean 0 and an appropriate scale.

In this work, we combine these two approaches in a framework called
Coupled Relational Symbolic Execution (\csre). In this framework, a
program is executed in a relational and symbolic way. When some
probabilistic primitive is executed, \csre introduces constraints
corresponding to the existence of an approximate probabilistic
coupling on the output. These constraints are combined with the
constraints on the execution traces generated by symbolically and
relationally executing other non-probabilistic commands. These
combined constraints can be exploited to reduce the number of states
to analyze. When the execution is concluded \csre\ checks whether
there is a coupling between the two outputs, or whether there is some
violation to the coupling. We show the soundness of this approach for
both proving and refuting differential privacy. However, for finding violations, one cannot reason only symbolically, and since
checking directly a coupling can be computationally expensive, we devise
several heuristic which can be used to
facilitate this task.  Using these techniques,
\csre allows one to verifying differential privacy for an interesting
class of programs, including programs working on countable input and
output domains, and to find violations to programs that are not
differentially private.
%or to find traces that are potentially leading
%to a failed proof of privacy.

% 
\ifnum\full=1
As we discussed at the begin of this section, other techniques have been devised to achieve similar goals. \csre is not a replacement for them
but it should be seen as an additional method to put in the set of
tools of the privacy developer which provides an high level of
generality. Indeed, by being a totally symbolic technique, it can
leverage on a pletora of current technologies such as SMT solvers, e.g.~\cite{DeMoura:2008}, algebraic solvers, e.g.~\cite{Mathematica}, and
numeric solvers, e.g.~\cite{MatlabOTB}.

Summarizing, the contribution of our work are:
\begin{itemize}
\item We combine relational symbolic execution and approximate probabilistic coupling in a new technique, named Coupled Relational Symbolic Execution (\csre). 
\item We show \csre sound for both proving programs differentially private and for refuting differential privacy. 
\item We devise a set of heuristic that can help a programmer in finding violations to differential privacy. 
\item We show how \csre can help in proving and refuting differential privacy for an interesting class of programs
\end{itemize}
\else
\csre is not a replacement for other techniques that have been proposed for
the same task,  it should be seen as an additional method to put in the set of
tools of the privacy developer which provides an high level of
generality. Indeed, by being a totally symbolic technique, it can
leverage on a pletora of current technologies such as SMT solvers, algebraic solvers,  and numeric solvers.

Summarizing, the contribution of our work are:
\begin{itemize}
\item We combine relational symbolic execution and approximate probabilistic coupling in a new technique: Coupled Relational Symbolic Execution. 
\item We show \csre sound for both proving and refuting  differential privacy. 
\item We devise a set of heuristic for finding violations to differential privacy. 
\item We show how \csre proves and refutes  several examples.
\end{itemize}
\fi
\section{\csre Informally}
\label{sec:highlevel}
In this section, we will motivate in an informal way \csre through
three examples of programs showing potential errors in implementations
of (supposedly) differentially private algorithms. % - this will allow us to discuss how to reason about violations to differential privacy and how to prove programs differentially private.
\ifnum\full=1
In doing this we
will also presenting the notation that will use in the rest of the
paper.

\subsection{Single query with wrong noise parameter.}
\indent\emph{Differential Privacy.}
\else
\paragraph{Single query with wrong noise parameter.}
\fi
Informally, a randomized function $A$
is $\epsilon$-differential privacy if it maps two databases $d_1$ and $d_2$ that differ for the
data of one single individual (denoted $d_1\sim d_2$) to output distributions that are
indistinguishable up to some value
\ifnum\full=1
$\epsilon$ - usually referred to as the privacy budget - this is formalized by requiring that the log-ratio of the two probability distributions is bounded pointwise by $\epsilon$, i.e. for every $u$,
$\Big |\log \frac{\Pr_{x\leftarrow A(d_1)}[x=u]}{\Pr_{x\leftarrow A(d_2)}[x=u]}\Big |\leq\epsilon$  - we will give the
precise definition in Section~\ref{sec:prelim}.  The smaller the
$\epsilon$, the more privacy is guaranteed.
A standard way to achieve differential privacy when we are interested
in a numeric query over a dataset is to add to the query result some
noise sampled from the Laplace distribution with mean 0 and scale
proportional to the \emph{sensitivity} of the function (how far the
function maps two databases differing for the data of one single
individual) over $\epsilon$~\cite{DworkMNS16}.
\else
$\epsilon$ - usually referred to as the privacy budget.  The smaller
the $\epsilon$, the more privacy is guaranteed.  A standard way to
achieve differential privacy for numeric queries is to add to their
result noise sampled from the Laplace distribution with mean 0 and
scale proportional to the \emph{sensitivity} of the function (how far
the function maps two databases $d_1\sim d_2$) over $\epsilon$~\cite{DworkMNS16}.
\fi
\begin{wrapfigure}[15]{L}{0.50\textwidth}
  \vspace{-1.5cm}
\begin{minipage}[t]{0.50\textwidth}
  \begin{algorithm}[H]
  \caption{\\\hspace{\textwidth}A buggy Laplace mechanism}
  \begin{flushleft}
  \hspace*{\algorithmicindent} \textbf{Input:} $q$: $\dbset \rightarrow \mathbb{Z}$,  $d:\dbset, \epsilon:\reals^{+}$\\
  \hspace*{\algorithmicindent} \textbf{Output:}  $o: \mathbb{Z}$\\
  \hspace*{\algorithmicindent} \textbf{Required} $d_1\sim d_2\Rightarrow |q(d_1)-q(d_2)|\leq r$
  \end{flushleft}
  \begin{algorithmic}[1]
  \State $v\gets q(d)$
  \State ${\color{red}\rho\overset{\$}{\gets}\lapp{0}{\epsilon}}$
  \State $o\gets v+\rho$
  \State \Return $o$
\end{algorithmic}
    \label{alg:wrongnoise}
  \end{algorithm}
\end{minipage}
\caption{Example 1. The algorithm is not $\epsilon$-DP.}
\end{wrapfigure}
\ifnum\full=1
Algorithm \ref{alg:wrongnoise} is a wrong implementation of this
principle - more in general it is a simple example of a program that is
implemented with the wrong noise parameters.  Specifically, it takes
in input a numeric query $q$ with type $\dbset \rightarrow \ints$ a
database $d\in\dbset$, and the privacy budget we want to guarantee
$\epsilon\in\reals^{+}$.  It then computes the query on the database,
adds Laplace noise with scale equal to $\frac{1}{\epsilon}$ to the
result of the query\footnote{We actually use the inverse of the scale
  as a parameter. That is the instruction
  $\rass{x}{\lapp{0}{\epsilon}}$ denotes a sample from the Laplace
  distribution with mean 0 and scale $\frac{1}{\epsilon}$. This will
  help in considering $\epsilon$ as a \emph{budget to spend}.}, and
releases the result.

This program is not $\epsilon$-differentially private, because it
doesn't calibrate the Laplace noise to the sensitivity of the
query. In fact, as a precondition we assert that the query $q$ is
$r$-sensitive by the requirement
$d_1\sim d_2\Rightarrow |q(d_1)-q(d_2)|\leq r$, asserting that given
two databases $d_1$ and $d_2$ differing for the data of one individual
the query $q$ returns
two results that are at most at distance $r$. The program implementing
algorithm
\ref{alg:wrongnoise} would be  $\epsilon$-differentially private if we added
noise proportional to $\frac{1}{r\epsilon}$ instead of
$\frac{1}{\epsilon}$, that is using the assignment
$\rho\overset{\$}{\gets}\lapp{0}{r\epsilon}$ instead of
$\rho\overset{\$}{\gets}\lapp{0}{\epsilon}$, in line 2 the algorithm.

To show formally that we have a privacy violation, accordingly to the definition of differential privacy, we need to witness a  query $q$, two databases $d_1$ and $d_2$ in the relation  $d_1\sim d_2$, and a possible output $u$ making the two probability distributions distinguishable for more than $\epsilon$. Approching this task directly is intractable~\cite{GaboardiNP20}. 

Instead, in order to do this, \csre\ will execute the program in a relational
symbolic fashion and it will try to prove that in two runs of the program the output variable has the same value and the privacy budget spent is at
most $\epsilon$. Technically, this is implemented by considering the
postcondition $o_1=o_2 \land \epsilon_c\leq \epsilon$, where
$\epsilon_c$ is a distinguished variable recording the privacy budget
spent. If \csre\ succeed, then the program is
$\epsilon$-differentially private.  If there is an execution that
invalidates this post-condition, then we will have a candidate for a
witness of the violation.

\else
Algorithm \ref{alg:wrongnoise} is a wrong implementation of this
principle - more in general it is a simple example of a program that is
implemented with the wrong noise parameters. It takes
a numeric query $q$ a database $d$, and the privacy budget
$\epsilon$.  It then computes $q(d)$
and adds to it Laplace noise with scale\footnote{The instruction
  $\rass{x}{\lapp{0}{\epsilon}}$ denotes a sample from the Laplace
  distribution with mean 0 and scale $\epsilon$. Using the inverse of the scale makes reasoning about the \emph{budget} more direct.} $\epsilon$, and
releases the result.

This program is not $\epsilon$-differentially private, because it
doesn't calibrate the noise to the sensitivity of the
query. In fact, as a precondition we assert that the query $q$ is
$r$-sensitive by the requirement
$d_1\sim d_2\Rightarrow |q(d_1)-q(d_2)|\leq r$. The program implementing
algorithm
\ref{alg:wrongnoise} would be  $\epsilon$-differentially private if we added
noise proportional to $r\epsilon$ instead of
$\epsilon$%

To show formally that we have a privacy violation, we need to witness
a query $q$, two databases $d_1$ and $d_2$ in the relation
$d_1\sim d_2$, and a possible output $u$ making the two probability
distributions distinguishable for more than $\epsilon$. Approching
this task directly is intractable~\cite{GaboardiNP20}.  Instead, in
order to do this, \csre\ will execute the program in a relational
symbolic fashion and it will try to prove that in two runs of the
program the output variable has the same value and the privacy budget
spent is at most $\epsilon$. Technically, this is implemented by
considering the postcondition $o_1=o_2 \land \epsilon_c\leq \epsilon$,
where $\epsilon_c$ is a distinguished variable recording the privacy
budget spent. If \csre\ succeeds, then the program is
$\epsilon$-differentially private. 
\fi
To avoid resorting to sampling,  when \csre executes the command for Laplace (as in line 2), following the approximate probabilistic coupling idea from~\cite{Barthe:2016},  it couples the
samples ($\rho_1, \rho_2$) in the two runs, and adds the 
constraint $\rho_1+k=\rho_2$, for some $k$.  It also tracks the budget spent with the constraint
$\epsilon_c=\lvert k \rvert \cdot\epsilon$.
The intuition behind this
constraint is that we can ensure the two samples to be at some
distance if we \emph{pay} enough budget.  From this we can see that if
$o_1$ is to be equal to $o_2$ then $k$ needs to be necessarily equal
to $v_1-v_2$. Since, $q(d_1)=v_1, q(d_2)=v_2$, the difference
$v_1-v_2$ is bounded above by $r$, and we get that, in the worst case
$\epsilon_c=r\epsilon$.  This means that in order to achieve equality
of the output variables and hence, $\epsilon$ differential privacy, we
need to spend at least $r$ times the budget $\epsilon$. So, if we are trying
to use less budget, the constraints will give us a candidate for a
witness of the violation.

\ifnum\full=1
\subsection{Two buggy Sparse Vector implementations.}
\else
\paragraph{Two buggy Sparse Vector implementations.}
\fi
\begin{figure*}
\vspace{-1.5cm}
  \begin{minipage}[t]{0.46\textwidth}
  \begin{algorithm}[H]
  \caption{A buggy Above Threshold}
  \begin{flushleft}
  \hspace*{\algorithmicindent} \textbf{Input:}   $t,\epsilon \in\mathbb{R},d\in\dbset, q[i]:\dbset\rightarrow \mathbb{N}$ \\
  \hspace*{\algorithmicindent} \textbf{Output:}  $o: [\bot^{i}, z,\bot^{n-i-1}]$\\
  \hspace*{\algorithmicindent} \textbf{Required} $d_1\sim d_2\Rightarrow |q[i](d_1)-q[i](d_2)|\leq 1$
  \end{flushleft}
  \begin{algorithmic}[1]
    \State $o \gets \bot^{n}; r \gets n+1$
    \State $\hat{t} \gets \lapp{t}{\frac{\epsilon}{2}}$\\
    {\tt for}~($i$~{\tt in~}1{:}$n$)~{\tt do}
    \State\ \ \ \ $\hat{s}\gets \lapp{q[i](d)}{\frac{\epsilon}{4}}$\\
    \ \ \ \ {\tt if}~$\hat{s}>\hat{t}\wedge r=n+1$~{\tt then}
    \State \ \ \ \ \ \ \ \ ${\color{red}o[i]\gets \hat{s}}; r\gets i$
    \State \Return \emph{o}
  \end{algorithmic}
\label{alg:wrongsvt-1}
\end{algorithm}
\end{minipage}
\begin{minipage}[t]{0.48\textwidth}
  \begin{algorithm}[H]
  \caption{Another buggy Above Threshold}
  \begin{flushleft}
  \hspace*{\algorithmicindent} \textbf{Input:}   $t,\epsilon \in\mathbb{R},d\in\dbset, q[i]:\dbset\rightarrow \mathbb{N}$\\
  \hspace*{\algorithmicindent} \textbf{Output:}  $o\in \{\bot,\top\}^{n}$  \\
  \hspace*{\algorithmicindent} \textbf{Required} $d_1\sim d_2\Rightarrow |q[i](d_1)-q[i](d_2)|\leq 1$
  \end{flushleft}
  \begin{algorithmic}[1]
    \State $\hat{t} \gets \lapp{t}{\frac{\epsilon}{2}}$\\
    {\tt for}~($i$~{\tt in~}1{:}$n$)~{\tt do}\\
    \ \ \ \ {\tt if}~${\color{red}q[i](d)}\geq \hat{t}$~{\tt then}
    \State \ \ \ \ \ \ \ \ $o[i]\gets \top$\\
    \ \ \ \ {\tt else}
    \State \ \ \ \ \ \ \ \ $o[i]\gets \bot$
    \State \Return \emph{o}
\end{algorithmic}
  \vspace{.6mm}
\label{alg:wrongsvt-2}
  \end{algorithm}
\end{minipage}
\end{figure*}
The next two examples are variations of the same algorithm: \emph{above threshold}, a component of the
\emph{sparse vector} technique~\cite{Lyu-2017}.
Given a numeric
threshold, an array of numeric queries of length $n$, and a dataset, this algorithm returns the
index of the first query whose result exceeds the
threshold - and potentially it should also return the value of that query.
This should be done in a way that preserves differential privacy. To do this
in the right way, a program should add noise to the threshold (even if it is not a sensitive data), add noise to each query, compare the values, and return the index of the first query for which this comparison succeed. The analysis of this algorithm is rather complex: it uses the noise on the threshold as a way to pay only once for all the queries that are below the threshold, and the noise on the queries to pay for the first and only query that is above the threshold, if any. Due to this complex analysis, this algorithm has been a benchmark for tools for reasoning about differential privacy~\cite{Barthe:2016,ZhangK17,Barthe2019AutomatedMF}.

Algorithm \ref{alg:wrongsvt-1} has a bug making the whole algorithm not differentially
private, for values of  $n$ greater than 4.
\ifnum\full=1
The program takes in input an array of queries of
type $\dbset \rightarrow \ints$, a privacy budget $\epsilon$ and a thresold $t$.
\else
\fi
The program  initializes an array of outputs $o$ to all bottoms values, and a variable $r$ to $n+1$ which will be used
as guard in the main loop. It then adds noise to the threshold, and iterates over all the queries adding 
noise to their results. If one of the noised-results is above the noisy threshold it saves
the value in the array of outputs and updates the value of the guard variable,
causing it to exit the main loop. Otherwise it keeps iterating.
The bug is returning the value of the noisy query that is above the threshold and not only its index, as done by the instruction in red in line 6  - this is indeed not enough for guaranteeing differential privacy.  
%
% \newpage
% \begin{minipage}[t]{0.48\textwidth}
%   \begin{algorithm}[H]
%   \caption{Another buggy sparse vector}
%   \begin{flushleft}
%   \hspace*{\algorithmicindent} \textbf{Input:}   $t,\epsilon \in\mathbb{R},d\in\dbset, q[i]:\dbset\rightarrow \mathbb{N}$\\
%   \hspace*{\algorithmicindent} \textbf{Output:}  $o\in \{\bot,\top\}^{n}$  \\
%   \hspace*{\algorithmicindent} \textbf{Required} $d_1\sim d_2\Rightarrow |q[i](d_1)-q[i](d_2)|\leq 1$
%   \end{flushleft}
%   \begin{algorithmic}[1]
%     \State $\hat{t} \gets \lapp{t}{\frac{\epsilon}{2}}$\\
%     {\tt for}~($i$~{\tt in~}1{:}$n$)~{\tt do}\\
%     \ \ \ \ {\tt if}~${\color{red}q[i](d)}\geq \hat{t}$~{\tt then}
%     \State \ \ \ \ \ \ \ \ $o[i]\gets \top$\\
%     \ \ \ \ {\tt else}
%     \State \ \ \ \ \ \ \ \ $o[i]\gets \bot$
%     \State \Return \emph{o}
% \end{algorithmic}
%     \label{alg:wrongsvt-2}
%   \end{algorithm}
%   \strut{ }
% \end{minipage}
For $n<5$ this program can be shown $\epsilon$-differentially private by using
the composition property of differential privacy that says that the k-fold
composition of $\epsilon$-DP programs is $k\epsilon$-differentially private(Section \ref{sec:prelim}). However, for $n\geq 5$ the more sophisticated analysis
we described above fails. 
%
% The algorithm initializes the output array to all bottom values, after
% that it sets $r$ to a value that can never be set to inside the loop
% so that it performs as a flag.  After, it runs through all the queries
% noising the results and then performing a test on the noisy
% values. The first time the noisy value is greater or equal than the
% noisy threshold it unsets the flag, and it sets the right index of the
% array to the noisy value. It then keeps iterating without ever entering the
% true branch again, and finally it releases the array.
The proof
principle \csre will use to try to show this program $\epsilon$-differentially private is
to prove the assertion $o_1=\iota \implies o_2=\iota
\land\epsilon_c\leq \epsilon$, for every $\iota \leq n$ - the soundness of this principle has been proved in~\cite{Barthe:2016}. That is,
\csre will try to prove the following assertions (which would prove the program without bug $\epsilon$-differentially private):
\begin{itemize}
\item[$\bullet$] $o_1=[\hat{s}_1,\bot,\dots,\bot] \implies o_2=[\hat{s}_1,\bot,\dots,\bot] \land\epsilon_c\leq \epsilon$
\item[$\bullet$] $o_1=[\bot,\hat{s}_1,\dots,\bot] \implies o_2=[\bot,\hat{s}_1,\dots,\bot] \land\epsilon_c\leq \epsilon$
\item[] $\dots$
  \item[$\bullet$] $o_1=[\bot,\dots,\hat{s}_1] \implies o_2=[\bot,\dots,\hat{s}_1] \land\epsilon_c\leq \epsilon$
  \end{itemize}
While proving the first assertion,  \csre will first couple at line 3 the threshold  as $\hat{t}_1+k_0=\hat{t}_2$, for $k_0>1$ where $1$ is the sensitivity of the queries, which is needed to guarantee that all the query results below the threshold in one run stay below the threshold in the other run, then, it will  increase appropriately the privacy budget by $k_0\frac{\epsilon}{2}$. As a second step it will couple $\hat{s}_1+k_1=\hat{s}_2$ in line 4.
Now, the only way for the assertion $o_1=[\hat{s}_1,\bot,\bot]\implies o_2=[\hat{s}_1,\bot,\bot]$ to hold,
is guaranting that both $\hat{s_1}=\hat{s}_2$ and $\hat{s}_1\geq t_1 \implies \hat{s_2}\geq t_2$ hold.
But these two assertions are not consistent with each other because $k_0\geq 1$.
That is, the only way, using these coupling rules, to guarantee that the run
on the right follows the same branches of the run on the left (this being necessary for proving the postcondion)
is to couple the samples $\hat{s}_1$ and $\hat{s}_2$ so that they are different,
this necessarily implying the negation of the postcondition. This would not the the case, if we were returning only the index of the query, since we can have that both the queries are above the threshold but return different values. Indeed,
by substituting line 7 with $\rass{o[i]}{\top}$ the program can be proven $\epsilon$-differentially private.
So the \emph{refuting} principle \csre will use here is the one that finds a trace on the left run
such that the only way the right run can be forced to follow it is by making
the output variables different.

A second example with bug of the above threshold algorithm is shown in
Figure \ref{alg:wrongsvt-2}.  In this example, in the body of the
loop, the test is performed between the noisy threshold and the actual
value of the query on the database - that is, we don't add noise to
the query. \csre will use for this example another \emph{refuting}
principle based on reachability. In particular, it will vacuously
couple the two thresholds at line 1. That is it will not introduce any
relation between $\hat{t}_1$, and $\hat{t}_2$. \csre will then search
for a trace which is satisfiable in the first run but not in the
second one. This translates in an output event which has positive
probability on the first run but 0 probability in the second one
leading to an unbounded privacy loss, and making the algorithm not
$\epsilon$-differentially private for all finite
$\epsilon$. Interestingly this unbounded privacy loss can be achieved
with just 2 iterations.
\section{Preliminaries}
\label{sec:prelim}
\ifnum\full=1
\subsubsection*{Discrete Probability Distributions}
Let $A$ be a
denumerable set, a \emph{subdistribution} over A is a function
$\mu:A\to [0,1]$ with weight $\sum_{a\in A}\mu(a)$ less or equal than
1.  We can think abour subdistributions as functions assigning to each
subset of A a probability mass. We denote the set of subdistributions
over $A$ as $\sdistr{A}$.  When a subdistribution has weight equal to
1, then we call it a \emph{distribution}. We denote the set of
distributions over $A$ by $\distr{A}$.
An example of a subdistribution that we will use in the sequel is the \emph{null} subdistribution $\mu_0:A\to[0,1]$, assigning to  every element of $A$ mass 0. Another example is the Dirac's distribution $\eunit(a):A\to[0,1]$, defined for $a\in A$ as
\[
  \eunit(a)(x)\equiv         \left \{
        \begin{array}{rcl}
          1 && \text{if}\ x=a \cr
           0 && \text{otherwise}
        \end{array}
      \right .\ 
\]
This is a distribution assigning  all the mass to the element $a\in A$. The set of subprobability distributions can be given the structure of a \emph{monad}, with unit the function $\eunit$ associating with each element its Dirac's distribution - this is why we chose this notation.
We have also a function $\ebind:\sdistr{A}\rightarrow
(A\rightarrow \sdistr{B})\rightarrow\sdistr{B}$ allowing us to compose subdistributions (as we compose monads). This is defined as
$\ebind\equiv \lambda \mu.\lambda f.\lambda
a.\displaystyle\sum_{b\in\outputset'}\mu(b)\cdot f(b)(a)$. We will use these constructions to give a semantics to our language in Section~\ref{sec:conf_to_distr}.

We will also use the following notion of $\epsilon$-divergence to define a notion of approximate coupling at the end of this section.
\begin{definition}
  Let $\epsilon\geq 0$. The \emph{$\epsilon$-divergence} between two
  subdistributions $\mu_1, \mu_2\in\sdistr{A}$, denoted by
  $\divergence{\epsilon}{\mu_1}{\mu_2}$,
  is defined as:
  \[
    \divergence{\epsilon}{\mu_1}{\mu_2}\equiv
    \sup_{E\subseteq O}\bigg (\mu_1(E) -\exp(\epsilon)\cdot \mu_2(E)\bigg) 
  \]
\end{definition}

\else
\paragraph{Discrete Probability Distributions}
Let $A$ be a denumerable set, a \emph{subdistribution} over A is a
function $\mu:A\to [0,1]$ with weight $\sum_{a\in A}\mu(a)$ less or
equal than 1. We denote the set of
subdistributions over $A$ as $\sdistr{A}$.  When a subdistribution has
weight equal to 1, then we call it a \emph{distribution}. We denote
the set of distributions over $A$ by $\distr{A}$.
The \emph{null} subdistribution $\mu_0:A\to[0,1]$  assigns to  every element of $A$ mass 0. The Dirac's distribution $\eunit(a):A\to[0,1]$, defined for $a\in A$ as $\eunit(a)(x)\equiv 1$ if $x=a$, and  $\eunit(a)(x)\equiv 0$, otherwise. The set of subprobability distributions can be given the structure of a \emph{monad}, with unit the function $\eunit$.
We have also a function $\ebind\equiv \lambda \mu.\lambda f.\lambda
a.\displaystyle\sum_{b\in\outputset'}\mu(b)\cdot f(b)(a)$ allowing us to compose subdistributions (as we compose monads).
We will use the notion of $\epsilon$-divergence  $\divergence{\epsilon}{\mu_1}{\mu_2}$ between two
  subdistributions $\mu_1, \mu_2\in\sdistr{A}$ to define approximate coupling, this is defined  as:$
    \divergence{\epsilon}{\mu_1}{\mu_2}\equiv
    \sup_{E\subseteq O}\big (\mu_1(E) -\exp(\epsilon)\cdot \mu_2(E)\big) 
$.
\fi
\ifnum\full=1
\subsubsection*{Differential Privacy}
\label{sec:dp}
Differential Privacy intuitively guarantees that  computation over any two inputs differing for the data of one individual result in close distributions over outputs. Formally, it is defined as follows.
\begin{definition}[Differential Privacy\cite{DworkMNS16}]
  Let $\epsilon\geq 0$ and $0\leq \delta\leq 1$. Let $\sim\subseteq \dbset\times\dbset$.
  An algorithm $\mathcal{A}:\dbset\rightarrow \distr{\mathcal{O}}$ is $(\epsilon,\delta)$-differentially private w.r.t $\sim$ iff
  $\forall D\sim D'.\forall o\subseteq\mathcal{O}. \Pr[\mathcal{A}(D)\in o] \leq e^{\epsilon} \Pr[\mathcal{A}(D')\in o] + \delta$.
\end{definition}The
relation $\sim$ models which pairs of input databases should be
considered sensitive, i.e., what data should be nearly
indistinguishable for an adversary.  In this work we will mostly
consider the vanilla definition of differential privacy where
$\delta=0$.  Differential privacy implies a number of interesting
properties. Here we will describe the most interesting ones for this work.

\begin{lemma}[Sequential Composition\cite{DworkMNS16}]
  \label{lem:sequential}
  Given an $A_1$ and $A_2$, respectively $(\epsilon_1,\delta_1)$-dp and  $(\epsilon_2,\delta_2)$-dp,
  their sequential composition $A(d)\equiv A_2(\langle A_1(d),d\rangle)$ is $(\epsilon_1+\epsilon_2, \delta_1+\delta_2)$-dp.
\end{lemma}
In the specific case of $A_2$ being 0-d.p, for instance when $A_2$
ignores or does not depend on $d$, the property of sequential
composition is called \emph{post-processing}. It intuitively means
that an $(\epsilon,\delta)$ differentially private answer remains such
when arbitrarly post processed, as long as the post processing does
not depend on the data. Any differentially private version of a
numeric query has necessarily to hide the difference in output of
two adjacent inputs\cite{Vadhan827361}. This difference in output is
captured by the following notion of sensitivity of a function.
\begin{definition}
  Let $\sim\subseteq \dbset\times\dbset$, and $f:\dbset\rightarrow\mathbb{Z}$. Then $f$ is $k$ sensitive if
  $\lvert f(x)-f(y)\rvert\leq k$, for all $x\sim y$.
\end{definition}
The following lemma provides the first differentially private primitive.
\begin{lemma}[Laplace Mechanism\cite{DworkMNS16}]
  \label{lem:laplace}
  Let $\epsilon>0$, and assume that $f:\dbset\mapsto \ints$ is a $k$ sensitive function
  with repsect to $\sim\subseteq \dbset\times\dbset$. Then the randomized algorithm
  mapping $D$ to $f(D)+\nu$, where $\nu$ is sampled from the Laplace distribution with
  scale $\frac{1}{\epsilon}$, is $k\epsilon$-differentially private w.r.t to $\sim$.
\end{lemma}
\else
\paragraph{Differential Privacy}
\label{sec:dp}
Formally, differential privacy~\cite{DworkMNS16} is a property of a program defined as:
\begin{definition}
  Let $\epsilon\geq 0$ and  $\sim\subseteq \dbset\times\dbset$.
  A program $\mathcal{A}:\dbset\rightarrow \distr{\mathcal{O}}$ is $\epsilon$-differentially private with respect to $\sim$ if and only if
  $\forall D\sim D'.\forall o\in\mathcal{O}. \Pr[\mathcal{A}(D)=o] \leq e^{\epsilon} \Pr[\mathcal{A}(D')=o]$\footnote{We use the vanilla definition of differential privacy for simplicity of explanation, but allowing a $\delta>0$ would not change
how \csre works.}.
\end{definition}
The adjacency
relation $\sim$ models which pairs of input databases should be 
indistinguishable to an adversary. Differentially private program can be composed~\cite{DworkMNS16}:
  given programs $A_1$ and $A_2$, respectively $\epsilon_1$ and  $\epsilon_2$ differentially private,
  their sequential composition $A(d)\equiv A_2(\langle A_1(d),d\rangle)$ is $\epsilon_1+\epsilon_2$-differentially private.
The following lemma is a classical result in differential privacy~\cite{DworkMNS16}.
\begin{lemma}
  \label{lem:laplace}
  Let $\epsilon>0$, and assume that $f:\dbset\mapsto \ints$ is a $k$ sensitive function
  with repsect to $\sim\subseteq \dbset\times\dbset$. Then the randomized algorithm
  mapping $D$ to $f(D)+\nu$, where $\nu$ is sampled from a discrete version of the Laplace distribution with
  scale $\frac{1}{\epsilon}$, is $k\epsilon$-differentially private w.r.t to $\sim$.
\end{lemma}
where a program $f:\dbset\rightarrow\mathbb{Z}$ is $k$ sensitive if
  $\lvert f(x)-f(y)\rvert\leq k$, for all $x\sim y$.
\fi
\ifnum\full=1
\subsubsection*{Approximate Probabilistic Liftings }
In this section we will give the formal and precise defintion of
probabilistic approximate liftings and
we will make explicit their connection with differential privacy.
\begin{definition}
  Given two sub-distributions $\mu_1\in\sdistr{A}, \mu_2\in\sdistr{B}$, a relation
  $\Psi\subseteq A\times B$, and $\epsilon\in\mathbb{R},\delta\in[0,1]$,
  we say that $\mu_1,\mu_2$ are related by the $(\epsilon,\delta)$ approximate lifting
  of $\Psi$ iff there exists $\mu_L,\mu_R\in\distr{A\times B}$ such that:
  \begin{itemize}
  \item $\pi_1(\mu_L)=\mu_1$ and $\pi_2(\mu_R)=\mu_2$
  \item $\supp{\mu_L}\cup\supp{\mu_R}\subseteq \Psi$
  \item $\divergence{\epsilon}{\mu_L}{\mu_R}\leq \delta$
  \end{itemize}
\end{definition}

\begin{lemma}[Foundamental Property of Liftings\cite{Barthe:2016}]
  \label{lem:foundamental}
  Let $\mu_1,\mu_2\in\distr{A}, \epsilon,\delta\geq 0$. Then  $\divergence{\epsilon}{\mu_1}{\mu_2}\leq \delta$ iff $\mu_1 (=)^{\epsilon,\delta} \mu_2$.
\end{lemma}
From Lemma \ref{lem:foundamental} we can derive that an algorithm $A$
is $(\epsilon,\delta)$-dp w.r.t to an dajcency relation $\sim$ iff
$A(d_1) (=)^{\epsilon,\delta} A(d_2)$ for all $d_1\sim d_2$.
The following lemma states another useful proof principle.
\begin{lemma}[Pointwise Differential Privacy\cite{Barthe:2016}]
  \label{lem:pointwise}
  An algorithm $A:\dbset\rightarrow \distr{B}$ is $(\epsilon,\delta)$-dp w.r.t $\sim$ iff
  there exists $\{\delta_b\mid \delta_b\geq 0\}_{b\in B}$ such that $\displaystyle\sum \delta_b\leq \delta$
  and $A(d_1) (\Psi_{b})^{\epsilon,\delta_b}A(d_2)$ for every $d_1\sim d_2$.
  Where $\Psi_b\equiv\{(x_1,x_2)\mid x_1=b \implies x_2=b\}\subseteq B\times B$.
\end{lemma}
The next lemma, finally, casts the Laplace mechanisms in terms of couplings.
\begin{lemma}
  \label{em:laplace_gen}
  Let $L_{v_1,b}, L_{v_2,b}$ two random variables with law Laplace distribution with mean $v_1$, and $v_2$ respectively, and $b$ as scale.
  Then $L_{v_1, b}  \{(z_1,z_2)\mid z_1+k=z_2\in \ints\times\ints\}^{\mid k + v_1 - v_2\mid\epsilon} L_{v_2,b}$, for all $k\in\ints, \epsilon\geq 0$.
\end{lemma}
\else
\paragraph*{Probabilistic Liftings and Coupling}
The notion of approximate probabilistic coupling is internalized by the notion of approximate lifting. We will focus here on distributions and we will use a simplified version of the notion presented in~\cite{Barthe:2016} since we are focusing on pure differential privacy. 
\begin{definition}
  Given two sub-distributions $\mu_1\in\distr{A}, \mu_2\in\distr{B}$, a relation
  $\Psi\subseteq A\times B$, and $\epsilon\in\mathbb{R}$,
  we say that $\mu_1,\mu_2$ are related by the $\epsilon$ approximate lifting
  of $\Psi$, denoted $\mu_1 (\Psi)^{\epsilon} \mu_2$, iff there exists $\mu_L,\mu_R\in\distr{A\times B}$ such that:
 1) $\lambda a.\sum_b\mu_L(a,b)=\mu_1$ and $\lambda b.\sum_a\mu_R(a,b)=\mu_2$,
 2) $\{(a,b)| \mu_L(a,b)>0 \lor \mu_R(a,b)>0\}\subseteq \Psi$,
 3) $\divergence{\epsilon}{\mu_L}{\mu_R}\leq 0$.
\end{definition}
Approximate lifting satisfies the following fundamental property.
\begin{lemma}
  \label{lem:foundamental}
  Let $\mu_1,\mu_2\in\distr{A}, \epsilon \geq 0$. Then  $\divergence{\epsilon}{\mu_1}{\mu_2}\leq 0$ iff $\mu_1 (=)^{\epsilon} \mu_2$.
\end{lemma}
From Lemma \ref{lem:foundamental} we have that an algorithm $A$
is $\epsilon$-differentially private w.r.t to $\sim$ iff
$A(d_1) (=)^{\epsilon} A(d_2)$ for all $d_1\sim d_2$.
% The following is another useful proof principle.
% \begin{lemma}[Pointwise Differential Privacy\cite{Barthe:2016}]
%   \label{lem:pointwise}
%   An program $A:\dbset\rightarrow \distr{B}$ is $\epsilon$-differentially private  w.r.t $\sim$ iff $A(d_1) (\{(x_1,x_2)\mid x_1=b \implies x_2=b\})^{\epsilon}A(d_2)$ for every $d_1\sim d_2$ and $b\in B$.
% \end{lemma}
The next lemma, finally, casts the Laplace mechanisms in terms of couplings.
\begin{lemma}
  \label{em:laplace_gen}
  Let $L_{v_1,b}, L_{v_2,b}$ two Laplace random variables with mean $v_1$, and $v_2$ respectively, and scale $b$.
  Then $L_{v_1, b}\,  \{(z_1,z_2)\mid z_1+k=z_2\in \ints\times\ints\}^{\mid k + v_1 - v_2\mid\epsilon}\, L_{v_2,b}$, for all $k\in\ints, \epsilon\geq 0$.
\end{lemma}
\fi

\section{Concrete languages}
\label{sec:conc_lang}
\ifnum\full=1
In this section we will describe the syntax and the semantics
of the concrete languages \pfor and \rpfor. We call them
\emph{concrete} languages, as opposed to \emph{symbolic}, as it is
standard in the symbolic execution literature(\cite{King:1976,Farina2019}).  The first
language, \pfor, is the language in which our programs will be written
in and is a simple imperative language with for loops and random assignment.
In order to prove relational properties about them we will define
\rpfor that, with a relational semantics, will capture pairs of \pfor
programs and their paired semantics. We start off with \pfor.
\else
In this section we  sketch the two \csre  concrete languages, the unary one \pfor and the relational one \rpfor. More details are in the Appendix.
\fi
\ifnum\full=1
\begin{wrapfigure}{L}{0.6\textwidth}
  \vspace{-0.8cm}
  \fbox{
\begin{minipage}[t]{0.55\textwidth}
    \begin{align*}
    \pexpr \ni e::=  &v\mid x\mid \arracc{a}{e}\mid \len{a}\mid e \oplus e\\
    \pcmd\ni c ::=  &\skipp\mid \seq{c}{c}\mid \ass{x}{e}\mid \ass{\arracc{a}{e}}{e}\mid\\
    &\rass{x}{\lapp{e}{e}}\mid \ifte{e}{c}{c}\mid\\
    &\cfor{x}{e}{e}{c}
  \end{align*}
  \end{minipage}
}
  \caption{Syntax of \pfor}
  \label{fig:pfor-syntax}
\end{wrapfigure}
\subsection{\pfor syntax}
\label{sec:pfor}
\pfor is a basic FOR-like language with array and probabilistic sampling from the Laplace distribution. The syntax ispretty standard  and we  present it in Figure
\ref{fig:pfor-syntax}.  We let $n,n_1,n_2$ range
over $\ints$. We let $e_1,e_2, e$ range over the set of
arithmetic expressions $\pexpr$ which is inductively defined.  Expressions are
basic values $v\in\pvalset\equiv\ints\cup\symsp$, where the set $\symsp$
contains values denoting random expressions and will be explained more at the semantic level in the next
section. Program variables are also expressions $x\in\varset$ as well as
arithmetic operations $e_1\oplus e_2$ where $\oplus\in\{+,-,*,/ \}$. Finally,
array accesses $\arracc{a}{e}$, and $\len{a}$ when $a$ is an array name in $\vararrset$.
We assume $\varset,\vararrset$, and $\symsp$ to be pairwise disjoint.  The set
of commands $\pcmd$ includes assignments, array assignments, the
$\skipp$ command, sequencing, branching, and a looping construct.  Finally, we
also include a primitive instruction $\rass{x}{\lapp{e_1}{e_2}}$ to
model random sampling from the laplace distribution.
\subsection{\pfor Semantics}
As mentioned above, the set $\symsp$ contains values denoting
random expressions. We call values in $\symsp$ distribution values.
We will use capital letters such as $X,Y,\dots$ to
denote arbitrary elements in $\symsp$
In Figure \ref{fig:pconstraints-syntax}, we introduce a grammar of
random expressions,  where $X$ ranges over $\symsp$ and
$n,n_1,n_2\in\ints$. The simple constraints in the syntactic
categories $ra$ and $re$ record that a random
value is either associated with a specific distribution, or that the
computation is conditioned on some random expression being greater than
0 or less than or equal than 0. The former constraints, as we will
see, come from branching instructions. We treat constraint lists $p,
p'$, in Figure \ref{fig:pconstraints-syntax} as lists of simple
constraints and hence, from now on, we will use the infix operators
$::$ and $@$, respectively, for appending a simple constraint to a
constraint and for concatenating two constraints. The symbol $[]$
denotes the empty list of probabilistic constraints.  Environments in
the set $\memset$, or probabilistic memories, map program variables to values in
$\pvalset$, and array names to elements in
$\carrayset\equiv\bigcup_{i}\pvalset^i$, so the type of a
memory $m\in\memset$ is $\varset\rightarrow\pvalset
\cup\vararrset\rightarrow \carrayset$.
We will distinguish between probabilistic concrete memories in $\memset$ and concrete
memories in the set $\cmemset\equiv \varset\rightarrow \ints \cup
\vararrset\rightarrow\bigcup_{i}\ints^i$.
Probabilistic concrete memories are meant to denote subdistributions
over the set of concrete memories $\memset_c$, more about this connection
in Section \ref{sec:conf_to_distr}.
\begin{wrapfigure}{L}{0.5\textwidth}
  \vspace{-0.8cm}
  \fbox{
    \begin{minipage}{0.4\textwidth}
      \begin{flalign*}
        ra&::=\rass{X}{\lapp{n_1}{n_2}}&\\
        re&::=n\mid X\mid re\oplus re&\\
        P \ni p&::=X=re\mid re>0\mid re\leq 0\mid &\\
        &\ \ \ ra \mid p::P\mid []&
      \end{flalign*}
\end{minipage}
}
\caption{Concrete probabilistic constraints}
\label{fig:pconstraints-syntax}
\end{wrapfigure}
Expressions in \pfor
are given meaning through a big-step evaluation semantics specified by
a judgment of the form: $\econf{m}{e}{p}\estep \vconf{v}{p'}$, where
$m\in\memset, e\in\pexpr, p,p'\in P, v\in\pvalset$.  The judgments
reads as: expression $e$ reduces to the value $v$ and probabilistic
constraints $p'$ in an enviroment $m$ with probabilistic concrete constraints
$p$.  Commands are given meaning through a small-step evaluation
semantics specified by a judgment of the form: $\conf{m}{c}{p}\cstep
\conf{m'}{c'}{p'}$, where $m,m'\in\memset, c,c'\in\pcmd,p,p'\in
P$. The judgment reads as: the probabilistic concrete configuration $\conf{m}{c}{p}$
steps in to the probabilistic concrete configuration $\conf{m'}{c'}{p'}$.  We call a
probabilistic concrete configuration of the form $\conf{m}{\skipp}{p}$ final. A
set of concrete configurations $\setsubdistr$ is called final and we denote it by $\final{\setsubdistr}$ if
all its concrete configurations are final. We will use this predicate even for sets of sets of concrete configurations
with the obvious lifted meaning. Figure
(\ref{fig:pfor-selrules}) shows a selection of the rules defining
these judgments. 
\begin{figure}
  \fbox{
  \begin{mathpar}
    \inferrule[\rulestyle{if-false}]
    {\econf{m}{e}{p}\estep \vconf{v}{p'}\and
      v\in\ints\and v\leq 0
    }
    {\conf{m}{\ifte{e}{c_1}{c_2}}{p} \cstep \conf{m}{c_2}{p'}}
       
    \inferrule[\rulestyle{if-true-prob}]
    {\econf{m}{e}{p}\estep \vconf{v}{p'} \and v\in\symsp
      \and p''\equiv p'@v>0
    }
    {\conf{m}{\ifte{e}{c_1}{c_2}}{p} \cstep \conf{m}{c_1}{p''}}

    \inferrule[\rulestyle{lap-ass}]
    {
      \econf{m}{e_1}{p} \estep\vconf{n_1}{p_1} \and
      \econf{m}{e_2}{p_1} \estep\vconf{n_2}{p_2} \\\\
      n_2>0 \and
      \fresh{X}{\symsp} \\\\  p' \equiv p_1@X=\lapp{n_1}{n_2}
      }
    {
      \conf{m}{\rass{x}{\lapp{e_1}{e_2}}}{p} \cstep \conf{m[x\mapsto X]}{\skipp}{p'}
    }
  \end{mathpar}
  }
  \caption{\pfor selected rules}
      \label{fig:pfor-selrules}
    \end{figure}

    Most of the rules are self-explanatory so we only describe the ones
which are non standard. Rule $\rulestyle{lap-ass}$
handles the random assignment. It evaluates the mean $e_1$ and the
scale $e_2$ of the distribution and checks that $e_2$ actually denotes
a positive number. The semantic predicate $\fresh{\cdot}{\cdot}$
asserts that the first argument is drawn non deterministically from
the second argument and that it was never used before in the
computation. Notice that if one of these two expressions reduces to a
probabilistic symbolic value the computation halts.  Rule
$\rulestyle{if-true-prob}$ (and $\rulestyle{if-false-prob}$) reduces
the guard of a branching instruction to a value. If the value is a
probabilistic symbolic constraint then it will nondeterministically
choose one of the two branches recording the choice made in the list
of probabilistic constraints.  If instead the value of the guard is
a numerical constant it will choose the right branch
deterministically using the rules $\rulestyle{if-false}$ and
$\rulestyle{if-true}$ (not showed).
As clear from the rules a run of a \pfor program can generate many
different final concrete configurations.
A different judgment of the form $\setsubdistr \setstep
\setsubdistr'$, where
$\setsubdistr,\setsubdistr'\in\powerset(\memset\times\pcmd\times P)$,
and in particular its transitive and reflexive closure ( $\setstep^{*}$), will help us in collecting
all the possible final configurations stemmign from a computation.
\begin{wrapfigure}{L}{0.6\textwidth}
  \vspace{-0.5cm}
  \fbox{
    \begin{minipage}{0.55\textwidth}
      \begin{mathpar}
\inferrule[\rulestyle{Sub-distr-step}]
        {
          \setsubdistr_t\equiv\{\conf{m'}{c'}{p'}\mid
          \conf{m}{c}{p} \cstep \conf{m'}{c'}{p'} \}\\\\
          \conf{m}{c}{p} \in \setsubdistr\\\\
          \setsubdistr'\equiv\bigg(\setsubdistr\setminus\{\conf{m}{c}{p}\}\bigg ) \cup \setsubdistr_t
        }
        {
          \setsubdistr \setstep \setsubdistr'
        }
      \end{mathpar}
    \end{minipage}
  }
  \caption{\rulestyle{Sub-distr-rule}}
  \label{fig:rule_confs}
\end{wrapfigure}
The only rule that defines the judgment, $\rulestyle{Sub-distr-step}$,
is presented in Figure \ref{fig:rule_confs}.
Rule $\rulestyle{Sub-distr-step}$ selects non
deterministically one configuration $s=\conf{m}{c}{p}$ from
$\setsubdistr$, removes $s$ from it, and adds to $\setsubdistr'$ all the
configurations $s'$ that are reachable from $s$.
\subsection{From configurations to subdistribution}
  \label{sec:conf_to_distr}
  In section \ref{sec:prelim} we defined the notions of lifting,
coupling and differential privacy using subdistributions in the form
of functions from a set of atomic events to the interval $[0,1]$.  The
semantics of the languages proposed so far though only deal with
subidstributions represented as set of concrete probabilistic
configurations. In this section we will map the latter to the former.
We start by giving two operators used to compose and define new
subdistridbutions in the functional form, that is: $\unit{\cdot}$, and $\bind{\cdot}{\cdot}$.
In the following, we use lambda notation and the denumerable sets $\outputset, \outputset'$
are universally quantified. The first one is defined as:
$\textbf{\textup{unit}}: \outputset \rightarrow \sdistr{\outputset} \equiv \lambda a. \lambda x. \left \{
        \begin{array}{rcl}
          1 && \text{if}\ x=a \cr
           0 && \text{otherwise}
        \end{array}
      \right .\ 
      $.

The second one, is defined as
\[
  \textbf{\textup{bind}}: \sdistr{\outputset'}\rightarrow
(\outputset'\rightarrow \sdistr{\outputset})\rightarrow\sdistr{\outputset}\equiv \lambda \mu.\lambda f.\lambda
a.\displaystyle\sum_{b\in\outputset'}\mu(b)\cdot f(b)(a)
\]
In particular $\unit{\cdot}$ takes an arbitrary element $a$ in a set $\outputset$
and returns a delta distribution centered in $a$.
$\bind{\cdot}{\cdot}$
builds a new subdistribution starting from a family an initial
subdistribution and a family of conditional distributions.  Using
$\unit{\cdot}, \bind{\cdot}{\cdot}$ it is possible to give a monadic
structure to the semantics of the language as it is done in (\cite{barthe2012probabilistic})
for the language \pwhile.
In Figure \ref{fig:translation} we define a translation function
($\denotemp{\cdot}{\cdot}$) and, auxiliary functions as well, between
a single probabilistic concrete configuration and a subdistribution
defined using the $\unit{\cdot}/\bind{\cdot}{\cdot}$ constructs. We make use of the
constant subdistribution $\mu_0$ which maps every element to mass 0,
and is usually referred to as the \emph{null} subdistribution, also by $\lapp{n_1}{n_2}(z)$
we denote the mass of (discrete version of) the Laplace distribution centered in $n_1$ with scale $n_2$ at the point $z$.
\begin{figure*}
  \fbox{
\[
  \begin{array}{lll}
    \denotemp{m_s}{p}&=&\bind {\denotep{p}} {(\lambda s_o. \unit{s_o(m_s)})}\\
    \denotep{[]}&=&\unit{[]}\\
    \denotep{X=re::p'}&=& \bind{\denotep{p'}} {\lambda s_o. \bind{\denotere{re}{s_o}}{\lambda z_o. \unit{X=z_o::s_o}}}\\
    \denotep{re>0::p'}&=& \bind{\denotep{p'}}{\lambda s_o. \bind{\denotere{re}{s_o}}{\lambda z_o. \text{if}\ (z_o>0)\ \text{then}\ \unit{z_o}\ \text{else}\ \mu_0}}\\
    \denotep{re\leq 0::p'}&=& \bind{\denotep{p'}}{\lambda s_o. \bind{\denotere{re}{s_o}}{\lambda z_o. \text{if}\ (z_o\leq 0)\ \text{then}\ \unit{z_o}\ \text{else}\ \mu_0}}\\
    \denotere{\lapp{n_1}{n_2}}{s}&=&\lambda z.\lapp{n_1}{n_2}(z)\\
    \denotere{n}{s}&=&\unit{n}\\
    \denotere{X}{s}&=&\unit{s(X)}\\
    \denotere{re_1\oplus re_2}{s}&=&\bind{\denotere{re_1}{s}}{\lambda v_1.\bind{\denotere{re_2}{s}}{\lambda v_2.\unit{v_1\oplus v_2}}}
  \end{array}
\]
}
\caption{Translation from configuration to $\unit{\cdot}/\bind{\cdot}{\cdot}$ representation of subdistribution}
\label{fig:translation}
\end{figure*}
The idea of the translation is that we can transform a probabilistic
concrete memory  $m_s\in\memset$ into a distribution over fully concrete memories in
$\memset_c$ by sampling from the distributions of the
probabilistic variables defined in $m_s$ in the order they were
decleared which is specified by the probabilistic path constraints.
To do this we first build a substitution for the probabilistic variable
which maps them into integers and then we perform the substitution
on $m_s$.
Given a set of probabilistic concrete memories we can then turn them
in a subdistribution by summing up all the translations of the
single probabilistic configurations. Indeed, given two
subdistributions $\mu_1,\mu_2$ defined over the same set we can always
define the subdistribution $\mu_1+\mu_2$ by the mapping
$(\mu_1+\mu_2)(a)=\mu_1(a)+\mu_2(a)$.

The following Lemma states an equivalence between these two
representations of probability subdistributions.  The hypothesis of
the theorem involve a judgment, $\match{m}{p}$, which has not been specified for lack
of space but can be found in the appendix, it deals with
well-formedness of the probabilistic path constraint $p$ with respect
to the concrete probabilistic memory $m$. 
\begin{lemma}\label{thm:op_to_denot} If $\match{m}{p}$ and
  $\{\conf{m}{c}{p}\}\setstep^{*}\{\conf{m_1}{\skipp}{p_1},\dots,\conf{m_n}{\skipp}{p_n}\}$
  then

  $\bind{\denotemp{m}{p}}{\cdenote{c}}=\displaystyle\sum_{i=1}^{n}\denotemp{m_i}{p_i}$
\end{lemma}
We can now hence take, in this work, the following as definition of
full denotational semantics of a program executed in a memory:
    \begin{definition}
      \label{def:full_semantics} The semantics of a program $c$
executed on memory $m$ and probability path constraint $p_0$ is
$\cdenote{c}(m_0,p_0)\equiv
\displaystyle\sum_{(m,\skipp,p)\in\setsubdistr}\denotemp{m}{p}$,
when $\{\conf{m}{c}{p}\}\setstep^{*}\setsubdistr$, $\final{\setsubdistr}$, and $\match{m_0}{p_0}$.
If $p_0=[]$ we write $\cdenote{c}(m_0)$.
\end{definition}

\else
\pfor is a basic for language with array and probabilistic sampling from the Laplace distribution $\rass{x}{\lapp{e}{e}}$. 
For simplicity, we assume that the parameters of the Laplace distribution are expressions which do not depend on other probability distributions - this is sufficient for the most common examples in differential privacy.  The rest of the syntax is pretty standard and we omit it here.
For executing programs, we use a distribution transformer semantics based on constraints  $p$ over a set $\symsp$ whose elements denote
random values - we call them distribution values and use capital letters such as $X,Y,\dots$ for them.
% In Figure \ref{fig:pfor-syntax}, we introduce also a grammar of
% constraints $p$. The simple constraints in the syntactic
% categories $ra$ and $re$ record that a random
% value is either associated with a specific distribution or an expression over random variables. % Combined constraints $p$ (or probabilistic path constraints)  assert
% that a computation is conditioned on some random expression - as when executing a branching instructions - or that we have a list of constraints (we will use $@$ for the concatenation of constraints).
% We require in general that probabilistic path constraints are well-formed with respect to a probabilistic memory (notation $\match{m}{p}$), details are in the appendix. 
% %
% \begin{figure}
%   \vspace{-0.5cm}
%   \centering
%   \fbox{
% \begin{minipage}[t]{.9\textwidth}
%     \begin{align*}
%     \pcmd\ni c ::=  &\skipp\mid \seq{c}{c}\mid \ass{x}{e}\mid \ass{\arracc{a}{e}}{e}\mid \rass{x}{\lapp{e}{e}}\mid \ifte{e}{c}{c}\mid \cfor{x}{e}{e}{c}\\
%       \pexpr \ni e::=  &v\mid x\mid \arracc{a}{e}\mid \len{a}\mid e \oplus e
%                          \qquad         P \ni p::=X=re\mid re>0\mid re\leq 0\mid  ra \mid p::P\mid []\\
%       ra::=&\rass{X}{\lapp{n_1}{n_2}}\qquad
%         re::=n\mid X\mid re\oplus re\\[-4mm]
%     \end{align*}
%   \end{minipage}
% }
%  \caption{Syntax and Semantics structures for \pfor}
%  \label{fig:pfor-syntax}
% \end{figure}
Expressions in \pfor
are given meaning through a big-step evaluation semantics specified by
a judgment of the form: $\econf{m}{e}{p}\estep \vconf{v}{p'}$, where
$m\in\memset$ is a memory containing potentially distribution values.
Commands are given meaning through a small-step evaluation
semantics specified by a judgment of the form: $\conf{m}{c}{p}\cstep
\conf{m'}{c'}{p'}$.  We call a
probabilistic concrete configuration of the form $\conf{m}{\skipp}{p}$ final. A
set of concrete configurations $\setsubdistr$ is called final and we denote it by $\final{\setsubdistr}$ if
all its concrete configurations are final.
We also have a collecting semantics
proving judgment of the form $\setsubdistr \setstep
\setsubdistr'$ which can be  mapped to the corresponding semantics over probability distributions.
\fi
\ifnum\full=1
\subsection{\rpfor syntax}
\label{sec:rpfor}
\pfor's semantics is unary, syntactically meaning that the configurations it deals with
are characterized by memories mapping variables and arrays names to single objects.
Semantically, it means that it captures only the computation of a program over a single memory.
In order to be able to reason about a relational property, such as differential
privacy, we will build on top of it a relational language called \rpfor with
a relational semantics dealing with pair of traces. Intuitively,
an execution of a single \rpfor program represents the execution of two
\pfor programs. Inspired by the approach of
\cite{pottier2002information}, we extend the grammar of \pfor with a
pair constructor $\pair{\cdot}{\cdot}$ which can be used at the
level of values $\pair{v_1}{v_2}$,
expressions~$\pair{e_1}{e_2}$, or commands $\pair{c_1}{c_2}$.
Notice that $c_i, e_i, v_i$ for $i\in\{1,2\}$ are commands,
expressions, and values in \pfor, hence nested pairing is not
allowed. This syntactic invariant is preserved by the rules handling
the branching instruction.  Pair constructs are used to indicate where
commands, values, or expressions might be different in the two unary
executions represented by a single \rpfor execution.  To define the
semantics for \rpfor, we first extend memories to allow program
variables to map to pairs of integers, and array variables to map to
pairs of arrays.
The set of expressions and commands in \rpfor, $\prexpr, \prcmd$
are generated by the grammars: 
\begin{align*}
  \prexpr\ni e_r&::=v\mid e \mid \pair{e_1}{e_2} &\prcmd\ni c_r &::=\ass{x}{e_r}\mid \rass{x}{\lapp{e_{r}}{e_{r}}} \mid c \mid \pair{c_1}{c_2}
\end{align*}
where $v\in\rpvalset , e, e_1,e_2 \in \pexpr, c,c_1,c_2\in\pcmd$.
Values can now be also pairs of unary values, that is $\rpvalset\equiv\pvalset \cup \pvalset^2$.
\subsection{\rpfor semantics}
In the following we will use the following projection functions
$\proj{i}{\cdot}$ for $i\in\{1,2\}$, which project, respectively, the
first (left) and second (right) elements of a pair construct (i.e.,
$\proj{i}{\pair{c_1}{c_2}}=c_i$, $\proj{i}{\pair{e_1}{e_2}}=e_i$ with
$\proj{i}{v}=v$ when $v\in\pvalset$), and are homomorphic for other
constructs.  The semantics of expressions in \rpfor is specified
through the following judgment $\reconf{m_1}{m_2}{e}{p_1}{p_2}\restep
\rvconf{v}{p'_1}{p'_2}$, where $m_1,m_2\in\memset, p_1,p_2,
p'_1,p'_2\in P, e\in\prexpr, v\in\rpvalset$.
Similarly, for commands, we have the following judgment
$\rconf{m_1}{m_2}{c}{p_1}{p_2}\rcstep\rconf{m'_1}{m'_2}{c'}{p'_1}{p'_2}$.
Again, we use the predicate $\final{\cdot}$ for configurations
$\rconf{m_1}{m_2}{c}{p_1}{p_2}$ such that $c=\skipp$, and lift the
predicate to sets of configurations as well.
Intuitively a relational probabilistic concrete configuration
$\rconf{m_1}{m_2}{c}{p_1}{p_2}$ denotes a pair of probabilistic
concrete states, that is a pair of subdistributions over the space of
concrete memories. In Figure \ref{fig:rpfor-selrules} a selection of the rules defining
the the judgements is presented. Most of the rules are quite natural.
Notice how branching instructions combine both probabilistic and relational
nondeterminism.
\begin{figure}
\fbox{
  \begin{mathpar}
    \inferrule[\rulestyle{r-expr-1}]
    {
      \econf{m_1}{\proj{1}{e}}{p_1}\estep \vconf{v_1}{p'_1} \\\\
      \econf{m_2}{\proj{2}{e}}{p_2}\estep \vconf{v_2}{p'_2} \\\\
      v_1,v_2 \in \ints \and v_1=v_2
    }
    {\reconf{m_1}{m_2}{e}{p_1}{p_2}\restep\rvconf{v_1}{p'_1}{p'_2}}
    
    \inferrule[\rulestyle{r-expr-2}]
    {
      \econf{m_1}{\proj{1}{e}}{p_1}\estep \vconf{v_1}{p'_1} \\\\
      \econf{m_2}{\proj{2}{e}}{p_2}\estep \vconf{v_2}{p'_2} \\\\ (\exists i\in\{1,2\}. v_i\not\in\ints \vee v_1\neq v_2)
    }
    {\reconf{m_1}{m_2}{e}{p_1}{p_2}\restep\rvconf{\pair{v_1}{v_2}}{p'_1}{p'_2}}
    % 
    % \inferrule[\rulestyle{r-ass}]
    % {
    %   \reconf{m_1}{m_2}{e}{p_1}{p_2} \restep\rvconf{v}{p'_1}{p'_2}
    % }
    % {
    %   \rconf{m_1}{m_2}{\ass{x}{e}}{p_1}{p_2} \rcstep
    %   \rconf{m_1[x\mapsto \proj{1}{v}]}{m_2[x\mapsto \proj{2}{v}]}{\skipp}{p'_1}{p'_2}
    % }
    %
    % \inferrule[\rulestyle{r-if-conc-conc-true-true}]
    % {
    %   \reconf{m_1}{m_2}{e}{p_1}{p_2} \restep\rvconf{v}{p'_1}{p'_2}\\\\
    %   \proj{1}{v},\proj{2}{v}\in\ints \and \proj{1}{v}>0 \and \proj{2}{v}>0
    % }
    % {
    %   \rconf{m_1}{m_2}{\ifte{e}{c_1}{c_2}}{p_1}{p_2}\rcstep  \rconf{m_1}{m_2}{c_1}{p'_1}{p'_2}
    % }

    \inferrule[\rulestyle{r-if-conc-conc-true-false}]
    {
      \reconf{m_1}{m_2}{e}{p_1}{p_2} \restep\rvconf{v}{p'_1}{p'_2}\\\\
      \proj{1}{v},\proj{2}{v}\in\ints \and \proj{1}{v}>0 \and \proj{2}{v}\leq 0
    }
    {
      \rconf{m_1}{m_2}{\ifte{e}{c_1}{c_2}}{p_1}{p_2}\rcstep \\\\ \rconf{m_1}{m_2}{\pair{\proj{1}{c_1}}{\proj{2}{c_2}}}{p'_1}{p'_2}
    }

    \inferrule[\rulestyle{r-if-prob-prob-true-false}]
    {
      \reconf{m_1}{m_2}{e}{p_1}{p_2} \restep\rvconf{v}{p'_1}{p'_2}\\\\ 
      \proj{1}{v},\proj{2}{v}\in\symsp
    }
    {
      \rconf{m_1}{m_2}{\ifte{e}{c_1}{c_2}}{p_1}{p_2}\rcstep\\\\
      \rconf{m_1}{m_2}{\pair{\proj{1}{c_1}}{\proj{2}{c_2}}}{\proj{1}{v}>0@p'_1}{\proj{2}{v}\leq 0@p'_2}
    }

    \inferrule[\rulestyle{r-pair-step}]
    {
      \{i,j\}=\{1,2\}
      \and 
      \conf{\proj{i}{m}}{c_{i}}{p_i}\cstep\conf{m'_{i}}{c'_{i}}{p'_i}
      \\\\
      c'_{j}=c_{j}\and p'_{j}=p_{j} \and
      m'_{j}= \proj{j}{m}
    }
    {
      \rconf{m_1}{m_2}{\pair{c_{1}}{c_{2}}}{p_1}{p_2}\rcstep   \\\\    \rconf{m'_1}{m'_2}{\pair{c'_{1}}{c'_{2}}}{p'_1}{p'_2}
    }
  \end{mathpar}
  }
 \caption{\rpfor selected rules}
 \label{fig:rpfor-selrules}
\end{figure}
So, as in the case of \pfor, we
collect sets of relational configurations using the judgment
$\setpsubdistr \setpstep \setpsubdistr'$
with $\setpsubdistr,\setpsubdistr'\in\powerset(\memset\times\memset\times \prcmd\times P\ \times P)$,
defined by only one rule presented in Figure \ref{fig:rule_pconfs}: $\tiny{\rulestyle{SUB-PDISTR-STEP}}$.
\begin{figure}
    \begin{mathpar}
       \inferrule[\tiny{\rulestyle{SUB-PDISTR-STEP}}]
      {
        \rconf{m_1}{m_2}{c}{p_1}{p_2} \in \setpsubdistr \\\\
        \setpsubdistr_t\equiv\{\rconf{m'_1}{m'_2}{c'}{p'_1}{p'_2}\\\\
        \rconf{m_1}{m_2}{c}{p_1}{p_2} \rcstep \rconf{m'_1}{m'_2}{c'}{p'_1}{p'_2} \}\\\\
        \setpsubdistr'\equiv\bigg(\setpsubdistr\setminus\{\rconf{m_1}{m_2}{c}{p_1}{p_2}\}\bigg ) \cup \setpsubdistr_t
      }
      {
        \setpsubdistr \setpstep \setpsubdistr'
      }  
    \end{mathpar}
    \caption{The only rule defining the $\setpstep$ relation.}
    \label{fig:rule_pconfs}
  \end{figure}
  The  rule picks and remove non
deterministically one relational configuration from a set and adds to
it all those configurations that are reachable from it.
As mentioned before a run of a program in \rpfor corresponds to the
execution of two runs the program in \pfor.  Before making this
precise we extend projection functions to relational configurations in
the following way:
$\proj{i}{\rconf{m_1}{m_2}{c}{p_1}{p_2}}=\conf{m_i}{c}{p_i}$, for
$i\in\{1,2\}$. Projection functions extend in the obvious way also to
sets of relational configurations.  We are now ready to state the
following lemma:
\begin{lemma}
  Let $i\in\{1,2\}$ then $\setpsubdistr\setpstep^{*}\setpsubdistr'$  iff
  $\proj{i}\setpsubdistr\setstep^{*}\proj{i}{\setpsubdistr'}$.
\end{lemma}

\else
\rpfor
 extends the grammar of \pfor with a
pair constructor $\pair{\cdot}{\cdot}$, which can be used at the
level of values, expressions, or commands but which cannot be nested. Intuitively,
an execution of a single \rpfor program represents the execution of two
\pfor programs and the pair construct is used to indicate where
programs might be different in two unary
executions represented by a single \rpfor execution.
Values in \rpfor can now be pairs of \pfor values.
% We will use projection functions
% $\proj{i}{\cdot}$ for $i\in\{1,2\}$, which project the
% first (left) and second (right) elements of a pair, respectively (with
% $\proj{i}{v}=v$ when $v$ is a value in \pfor).
%
The semantics of expressions in \rpfor extends naturally the one of \pfor where  judgments
$\reconf{m_1}{m_2}{e}{p_1}{p_2} \restep \rvconf{v}{p'_1}{p'_2}$ reads as: starting in the enviroments $m_1$ and $m_2$ with
probabilistic constraint $p_1$ and $p_2$, the expression e reduces to
the value $v$ (which can be a pair of values) and probabilistic
constraints $p_1'$ and $p_2'$, respectively.  Similarly, for commands,
we have the following judgment
$\rconf{m_1}{m_2}{c}{p_1}{p_2}\rcstep\rconf{m'_1}{m'_2}{c'}{p'_1}{p'_2}$.
Again, we use the predicate $\final{\cdot}$ for configurations
$\rconf{m_1}{m_2}{c}{p_1}{p_2}$ such that $c=\skipp$, and lift the
predicate to sets of configurations as well.  Intuitively a relational
probabilistic concrete configuration $\rconf{m_1}{m_2}{c}{p_1}{p_2}$
denotes a pair of probabilistic concrete states, that is a pair of
subdistributions over the space of concrete memories.
We also have the corresponding collecting semantics  $\setpsubdistr \setpstep \setpsubdistr'$.
\fi
\ifnum\full=1
\section{Symbolic languages}
In this section we proceed to lift the concrete languages, metioned in Section
(\ref{sec:conc_lang}), to their
symbolic versions (respectively, \splang and \srplang).
As it is standard in symbolic execution literature,
the first step is to extend, with symbolic values $X\in\syms$, the set
of values of the concrete languages, after that rules of semantics
execution will be defined.  We start off by doing this for the
language \plang.
\subsection{\splang: Syntax}
\begin{wrapfigure}{L}{0.6\textwidth}
  \begin{minipage}{0.55\textwidth}
    \vspace{-0.8cm}
  \fbox{
  \[
  \begin{array}{rcl}  
    \spexpr \ni e::=  &v\mid x\mid X\mid \arracc{a}{e}\mid \len{a}\mid e \oplus e\\
    \spcmd\ni c ::=  &\skipp\mid \seq{c}{c}\mid \ass{x}{e}\mid \ass{\arracc{a}{e}}{e}\mid\\
    &\rass{x}{\lapp{e}{e}}\mid \ifte{e}{c}{c}\mid\\
                     &\cfor{x}{e}{e}{c}
  \end{array}
\]
}
\end{minipage}
\caption[\splang syntax]{\splang syntax. $X\in\syms$}
\label{fig:splang-syntax}
\end{wrapfigure}
We now extend \plang expressions with symbolic values
$X\in\syms$. Syntax of \splang is presented in Figure
\ref{fig:splang-syntax}.
As we can see the syntax is similar to that of \plang except that the
set of expressions has been increased with symbolic values from
$\syms$ denoting integers.  We assume $\symsp\cap
\syms=\emptyset$. This assumption reflects the idea that symbolic
values in $\syms$ do not denote unknown or sets of probability
distributions but only unknown sets of integers.  The set of
values is now $\splangvalset\equiv\plangvalset\cup\syms$, we will also
need the set $\spintlangvalset\equiv\ints\cup\syms$.  Notice how
symbolic values can very well appear in probabilistic expressions.
\subsection{\splang: Semantics of expressions}
In order to collect constraints on symbolic values we extend
configurations with set of constraints over integer values, drawn from
the set $\cstrset$, not to be confused with probabilistic path
constraints.  The former express constraints over integer values, for
instance parameters of the distributions. The grammar of constraints
over integers and array values is presented in Figure \ref{fig:splang-constraints-grammar}.
\begin{figure}
  \begin{subfigure}{0.5\textwidth}
  \[
    \begin{array}{rcl}
      \cexpr\ni e &::=& n \ \mid X \mid i\mid  e \oplus e  \mid\store{e}{e}{e}\mid \\ &&  \select{e}{e} \mid | e |\\
      \cstrset\ni s  &::=& \top\mid e\circ e\ \mid s \wedge s\mid \neg s\mid\forall i.s
    \end{array}
  \]
  \caption[Grammar of constraints, again]{Grammar of constraints. $X\in\syms, n\in\langvalset$.}
  \label{fig:splang-constraints-grammar}
\end{subfigure}
\begin{subfigure}{0.5\textwidth}
\[
  \begin{array}{rcl}
    sra&::=&\rass{Y}{\lapp{c_e}{c_e}}\\
    sre&::=&n\mid X\mid Y\mid re\oplus re\\
    SP&::=&Y=re\mid re>0\mid re\leq 0\mid ra
  \end{array}
\]
\caption[Grammar of symbolic probabilistic constraints]{Grammar of symbolic probabilistic constraints. $c_e\in\cstrset, X\in\syms, Y\in\symsp$}
\label{fig:probcstr_syntax}
\end{subfigure}
\end{figure}
In particular constraint expressions include standard arithmetic
expressions with values being symbolic or integer constants, and array
selection. Actual constraints include first order logic formulas over
arithmetic expressions.  Finally, probabilistic path constraints now
can also contain symbolic integer values and hence the grammar gets
updated to what is shown in Figure \ref{fig:probcstr_syntax}.
What changes with respect to Figure \ref{fig:pconstraints-syntax} is
that now arithmetic expressions can also include symbolic integer
values.  Indeed, also probabilistic path constraints now can be
symbolic.  Since values have been extended, also memories, now
properly symbolic, change type in:
$\spmemset\equiv\varset\rightarrow\splangvalset \cup\vararrset\rightarrow \sarrayset$,
where $\sarrayset\equiv\{(X,v)\mid X\in \syms, v\in\spintlangvalset\}$.  In particular, we
represent arrays in memory as pairs $(X,v)$, where $v$ is a (concrete
or symbolic) integer value representing the length of the array, and
$X$ is a symbolic value representing the array contents. The content
of the arrays is kept and refined in the set of constraints by means
of $\select{\cdot}{\cdot}$ and $\store{\cdot}{\cdot}{\cdot}$ relation
symbols.  Evaluation judgments, both for expressions and commands,
will also include a set of constraints over integers. This is because
constraints can also be generated during evaluation of expressions.
We can no proceed with the semantics of expressions.
\begin{figure}
  \vspace{-0.7cm}
    \centering
    \fbox{
      $\speconf{m}{e}{p}{s}\spestep \spval{v}{p'}{s'}$
    }
  \begin{minipage}{0.7\textwidth}
  \fbox{
      \begin{mathpar}
        \inferrule[\rulestyle{S-P-Op-2}]
        {
          \speconf{m}{e_1}{p}{s} \spestep \spval{v_1}{p'}{s'} \\\\
          \speconf{m}{e_2}{p'}{s'} \spestep \spval{v_2}{p''}{s''}\\\\
          v_1,v_2\in \symsp \and \fresh{X}{\symsp}
        }
        {\speconf{m}{e_1\oplus e_2}{p}{s}\spestep \spval{X}{p''@[X=v_1\oplus v_2]}{s''}}

        \inferrule[\rulestyle{S-P-Op-5}]
        {
          \speconf{m}{e_1}{p}{s} \spestep \spval{v_1}{p'}{s'} \\\\
          \speconf{m}{e_2}{p'}{s'} \spestep \spval{v_2}{p''}{s''} \\\\
          \{i,j\}=\{1,2\}\and v_i\in\ints \and v_j\in\syms \and \fresh{X}{\syms}
        }
        {\speconf{m}{e_1\oplus e_2}{p}{s}\spestep \spval{X}{p''}{s''\cup\{X=v_1\oplus v_2\}}}

        \inferrule[\rulestyle{S-P-Op-6}]
        {
          \speconf{m}{e_1}{p}{s} \spestep \spval{v_1}{p'}{s'} \\\\
          \speconf{m}{e_2}{p'}{s'} \spestep \spval{v_2}{p''}{s''} \\\\
          \{i,j\}=\{1,2\}\and v_i\in\syms \and v_j\in\symsp \and \fresh{X}{\symsp}
        }
        {\speconf{m}{e_1\oplus e_2}{p}{s}\spestep \spval{X}{p''@[X=v_1\oplus v_2]}{s''}}
      \end{mathpar}
    }
    \end{minipage}
        \caption[\splang: Semantics of expressions]{\splang: Semantics of expressions, selected rules.}
    \label{fig:sem-splang-expr}
 \end{figure}  

Figure \ref{fig:sem-splang-expr} shows the judgment form.
The judgment is then inductively defined.
We only show few rules for this judgment. We briefly describe the rules
presented. Rule \rulestyle{S-P-Op-2}  applies
when an arithmetic operation has both of its operands that reduce
respectively to elements $\symsp$. Appropriately it updates
the set of probabilistic constraints.
Rules \rulestyle{S-P-Op-5}  instead fires when one of them is an integer
and the other is a symbolic value. In this case only the
list of symbolic constraints needs to be updated.  
Finally, in rule $\rulestyle{S-P-Op-6}$  one of the operands
reduces to an element in $\symsp$ and the other to an element in
$\syms$. We only update the list of probabilistic constraints
appropriately, as integer constraints cannot contain symbols in
$\symsp$.

\else
\section{Symbolic languages}
In this section we lift the concrete languages, mentioned in Section
\label{sec:conc_lang}, to their
symbolic versions (respectively, \splang and \srplang) by extending them with symbolic values $X\in\syms$.
\paragraph{\splang}
% \begin{wrapfigure}{L}{0.46\textwidth}
%   \begin{minipage}{0.44\textwidth}
%     \vspace{-0.8cm}
%   \fbox{
%   \[
%   \begin{array}{rcl}  
%     \spexpr \ni e::=  &v\mid x\mid X\mid \arracc{a}{e}\mid \len{a}\mid e \oplus e\\
%     \spcmd\ni c ::=  &\skipp\mid \seq{c}{c}\mid \ass{x}{e}\mid \ass{\arracc{a}{e}}{e}\mid\\
%     &\rass{x}{\lapp{e}{e}}\mid \ifte{e}{c}{c}\mid\\
%                      &\cfor{x}{e}{e}{c}
%   \end{array}
% \]
% }
% \end{minipage}
% \caption[\splang syntax]{\splang syntax. $X\in\syms$}
% \label{fig:splang-syntax}
% \end{wrapfigure}
 expressions extend \pfor expressions with symbolic values
$X\in\syms$:
$\spexpr \ni e::=  v\mid x\mid X\mid \arracc{a}{e}\mid \len{a}\mid e \oplus e$.
We assume $\symsp\cap
\syms=\emptyset$ - this because we want symbolic
values in $\syms$ to denote only unknown sets of integers, rather than sets of probability
distributions.  % The set of
% values is now $\splangvalset\equiv\plangvalset\cup\syms$, we will also
% need the set $\spintlangvalset\equiv\ints\cup\syms$.
Commands in \splang are the same as in \plang but now
symbolic values can appear in probabilistic expressions.

In order to collect constraints on symbolic values we extend
configurations with set of constraints over integer values, drawn from
the set $\cstrset$ (Figure~\ref{fig:splang-constraints-grammar}), not to be confused with probabilistic path
constraints (Figure~\ref{fig:probcstr_syntax}).  The former express constraints over integer values, for
instance parameters of the distributions. 
\begin{figure*}
  \vspace{-0.5cm}
    \centering
  \fbox{
\begin{minipage}{0.95\textwidth}  
  \begin{subfigure}{0.5\textwidth}
  \[
    \begin{array}{rcl}
      \cexpr\ni e &::=& n \ \mid X \mid i\mid  e \oplus e  \mid\store{e}{e}{e}\mid \\ &&  \select{e}{e} \mid | e |\\
      \cstrset\ni s  &::=& \top\mid e\circ e\ \mid s \wedge s\mid \neg s\mid\forall i.s\\[-3mm]
    \end{array}
  \]
  \caption[Grammar of constraints, again]{Symbolic constraints. $X\in\syms, n\in\langvalset$.}
  \label{fig:splang-constraints-grammar}
\end{subfigure}
\begin{subfigure}{0.5\textwidth}
\[
  \begin{array}{rcl}
    sra&::=&\rass{Y}{\lapp{c_e}{c_e}}\\
    sre&::=&n\mid X\mid Y\mid re\oplus re\\
    SP&::=&Y=re\mid re>0\mid re\leq 0\mid ra\\[-3mm]
  \end{array}
\]
\caption[Grammar of symbolic probabilistic constraints]{Prob. constraints. $c_e\in\cstrset, X\in\syms, Y\in\symsp$}
\label{fig:probcstr_syntax}
\end{subfigure}
\end{minipage}
}
\caption{Grammar of constraints}
\end{figure*}
In particular constraint expressions include standard arithmetic
expressions with values being symbolic or integer constants, and array
selection. Probabilistic path constraints now
can also contain symbolic integer values. Hence, also probabilistic path constraints now can be
symbolic.
Memories can now contain symbolic values and  we
represent arrays in memory as pairs $(X,v)$, where $v$ is a (concrete
or symbolic) integer value representing the length of the array, and
$X$ is a symbolic value representing the array content. The content
of the arrays is kept and refined in the set of constraints by means
of the $\select{\cdot}{\cdot}$ and $\store{\cdot}{\cdot}{\cdot}$ operations.
The semantics of expressions is captured by the judgment
$\speconf{m}{e}{p}{s}\spestep \spval{v}{p'}{s'}$
including now  a set of constraints over integers.
\begin{figure*}
  \vspace{-0.5cm}
    \centering
  \fbox{
      \begin{mathpar}
        \inferrule[\rulestyle{S-P-Op-2}]
        {
          \speconf{m}{e_1}{p}{s} \spestep \spval{v_1}{p'}{s'} \and  v_1,v_2\in \symsp\\\\
          \speconf{m}{e_2}{p'}{s'} \spestep \spval{v_2}{p''}{s''}
          \and \fresh{X}{\symsp}
        }
        {\speconf{m}{e_1\oplus e_2}{p}{s}\spestep \spval{X}{p''@[X=v_1\oplus v_2]}{s''}}
        \and
        \inferrule[\rulestyle{S-P-Op-5}]
        {
          \speconf{m}{e_1}{p}{s} \spestep \spval{v_1}{p'}{s'}\quad  \{i,j\}=\{1,2\},\,v_i\in\ints \\\\
          \speconf{m}{e_2}{p'}{s'} \spestep \spval{v_2}{p''}{s''} \quad
       v_j\in\syms,\, \fresh{X}{\syms}
        }
        {\speconf{m}{e_1\oplus e_2}{p}{s}\spestep \spval{X}{p''}{s''\cup\{X=v_1\oplus v_2\}}}
        \and
        \inferrule[\rulestyle{S-P-Op-6}]
        {
          \speconf{m}{e_1}{p}{s} \spestep \spval{v_1}{p'}{s'} \\\\
          \speconf{m}{e_2}{p'}{s'} \spestep \spval{v_2}{p''}{s''} \\\\
          \{i,j\}=\{1,2\}\and v_i\in\syms \and v_j\in\symsp \and \fresh{X}{\symsp}
        }
        {\speconf{m}{e_1\oplus e_2}{p}{s}\spestep \spval{X}{p''@[X=v_1\oplus v_2]}{s''}}
      \end{mathpar}
    }
        \caption[\splang: Semantics of expressions]{\splang: Semantics of expressions, selected rules.}
    \label{fig:sem-splang-expr}
 \end{figure*}  
Figure \ref{fig:sem-splang-expr} gives few rules for this judgment. We briefly describe the rules
presented. Rule \rulestyle{S-P-Op-2}  applies
when an arithmetic operation has both of its operands that reduce
respectively to elements in $\symsp$. Appropriately it updates
the set of probabilistic constraints.
Rules \rulestyle{S-P-Op-5}  instead fires when one of them is an integer
and the other is a symbolic value. In this case only the
list of symbolic constraints needs to be updated.  
% Finally, in rule $\rulestyle{S-P-Op-6}$  one of the operands
% reduces to an element in $\symsp$ and the other to an element in
% $\syms$. We only update the list of probabilistic constraints
% appropriately, as integer constraints cannot contain symbols in
% $\symsp$.
% 
\fi
\ifnum\full=1
\subsection{\splang: Semantics of commands}
We can now formalize the semantics of commands of \splang.
Again, we provide a selection of the rules of the small step semantics in Figure \ref{fig:sem-splang-cmd}.

Rule \rulestyle{S-P-If-sym-true} fires when a branching instruction is
to be executed and the guard is reduced to either an integer or a
value in $\syms$.  In this case we can very well proceed with the true
branch recording in the set of integer constraints the fact that the
guard is greater than 0. Notice that if the guard is an integer
actually less than or equal than 0 then there will never be a ground
substitution for that set of constraints and hence this is not
unsound.

\begin{figure}
  \centering
  \fbox{
    $\spcconf{m}{c}{p}{s}\spcstep \spcconf{m'}{c'}{p'}{s'}$
  }
  \begin{mathpar}
    \inferrule[\rulestyle{S-P-If-sym-true}]
    {
      \speconf{m}{e}{p}{s} \spestep \spval{v}{p'}{s'} \\\\ v\in\spintlangvalset 
    }
    { \spcconf{m}{\ifte{e}{\tbranch}{\fbranch}}{p}{s} \spcstep \\\\\spcconf{m}{\tbranch}{p'}{s'\cup\{v>0\}} }
    
    \inferrule[\rulestyle{S-P-If-prob-false}]
    {
      \speconf{m}{e}{p}{s} \spestep \spval{v}{p'}{s'} \\\\ v\in\symsp 
    }
    { \spcconf{m}{\ifte{e}{\tbranch}{\fbranch}}{p}{s} \\\\ \spcconf{m}{\fbranch}{p'@[v\leq 0]}{s'} }
    
    \inferrule[\rulestyle{S-P-Lap-Ass}]
    {
      \speconf{m}{e_a}{p}{s} \spestep \spval{v_a}{p'}{s'} \\\\
      \speconf{m}{e_b}{p'}{s'} \spestep \spval{v_b}{p''}{s''} \\\\
      \fresh{X}{\symsp} \and v_a,v_b\in\spintlangvalset \\\\
      s'''=s''\cup\{v_b>0\} \and p'''=p''@[\rass{X}{\lapp{v_a}{v_b}}] \\\\
      m'\equiv m[x\mapsto X]
    }
    { \spcconf{m}{\rass{x}{\lapp{e_a}{e_b}}}{p}{s} \\\\ \spcconf{m'}{\skipp}{p'''}{s'''} }
  \end{mathpar}
  \caption{Semantics of \splang (selected rules)}
  \label{fig:sem-splang-cmd}
\end{figure}

In that case the rule, showed in the appendix,
\rulestyle{S-P-If-sym-false} would instead lead to a satisfiable
constraint. Rule \rulestyle{S-P-If-prob-false} handles a branching
instruction which has a guard reducing to a value in $\symsp$. In this
case we can proceed in both branches, even though here we only show
one of the two rules, by recording the conditioning fact on the list
of probabilistic constraints. Finally, rule \rulestyle{S-P-Lap-Ass} handles
probabilistic assignment. After having reduced both the expression for the mean
and the expression for the scale to values we check that those are both
either integers or symbolic integers, if that's the case we make sure
that the scale is greater than 0 and we add a probabilistic constraints
recording the fact that the modified variable now points to a probabilistic
symbolic value related to a Laplace distribution.
Notice that again we don't handle situations where the expression for
the mean or the expression for the scale reduces to a probabilistic
symbolic value.
\else
The semantics of commands of \splang is described by small step semantics
judgments of the form: $\spcconf{m}{c}{p}{s}\spcstep \spcconf{m'}{c'}{p'}{s'}$,
including a set of constraints over integers.
We provide a selection of the rules in Figure \ref{fig:sem-splang-cmd}.
Rule \rulestyle{S-P-If-sym-true} fires when a branching instruction is
to be executed and the guard is reduced to either an integer or a
value in $\syms$.  In this case we can proceed with the true
branch recording in the set of integer constraints the fact that the
guard is greater than $0$.
\begin{figure*}
 \begin{minipage}{0.95\textwidth}
  \fbox{
  \begin{mathpar}
    \inferrule[\rulestyle{S-P-If-sym-true}]
    {
      \speconf{m}{e}{p}{s} \spestep \spval{v}{p'}{s'} \and v\in\spintlangvalset 
    }
    { \spcconf{m}{\ifte{e}{\tbranch}{\fbranch}}{p}{s} \spcstep \\\\\spcconf{m}{\tbranch}{p'}{s'\cup\{v>0\}} }
    
    \inferrule[\rulestyle{S-P-If-prob-false}]
    {
      \speconf{m}{e}{p}{s} \spestep \spval{v}{p'}{s'} \and v\in\symsp 
    }
    { \spcconf{m}{\ifte{e}{\tbranch}{\fbranch}}{p}{s} \spcstep\\\\ \spcconf{m}{\fbranch}{p'@[v\leq 0]}{s'} }
    
    \inferrule[\rulestyle{S-P-Lap-Ass}]
    {
      \speconf{m}{e_a}{p}{s} \spestep \spval{v_a}{p'}{s'} \and
      \speconf{m}{e_b}{p'}{s'} \spestep \spval{v_b}{p''}{s''} \and
      \fresh{X}{\symsp} \\\\ v_a,v_b\in\spintlangvalset \and
      s'''=s''\cup\{v_b>0\} \and p'''=p''@[\rass{X}{\lapp{v_a}{v_b}}]     }
    { \spcconf{m}{\rass{x}{\lapp{e_a}{e_b}}}{p}{s} \spcstep \spcconf{ m[x\mapsto X]}{\skipp}{p'''}{s'''} }
  \end{mathpar}
}
\end{minipage}
  \caption{\splang: Semantics of commands (selected rules)}
  \label{fig:sem-splang-cmd}
\end{figure*}
Rule \rulestyle{S-P-If-prob-false} handles a branching
instruction which has a guard reducing to a value in $\symsp$. In this
case we can proceed in both branches, even though here we only show
one of the two rules, by recording the conditioning fact on the list
of probabilistic constraints. Finally, rule \rulestyle{S-P-Lap-Ass} handles
probabilistic assignment. After having reduced both the expression for the mean
and the expression for the scale to values we check that those are both
either integers or symbolic integers, if that's the case we make sure
that the scale is greater than 0 and we add a probabilistic constraints
recording the fact that the modified variable now points to a probabilistic
symbolic value related to a Laplace distribution.
% Notice that again we don't handle situations where the expression for
% the mean or the expression for the scale reduces to a probabilistic
% symbolic value.
% 
\fi
\ifnum\full=1
\subsection{\splang: Collecting semantics}
Semantics of \splang introduces two levels of nondeterminism, the
first one is given by branching instructions whose guard reduces to a
symbolic value, the second one comes from branching instructions whose
guard reduces to a probabilistic symbolic value. The collecting
semantics of \splang, specified by the judgment with form $
\setsymsubdistr \spsetstep \setsymsubdistr' $ (where $
\setsymsubdistr,
\setsymsubdistr'\in\powerset{\spmemset\times\splangcmd\times SP\times
\cstrset}$) and whose only rule is specified in Figure
\ref{fig:sem-splang-collecting}, takes care of both of them. Unlike in
the deterministic case of the rule $\rulestyle{Set-Step}$, where only
one configuration was chosen nondeterministically from the initial
set, here we select nondeterministically a (maximal) set of
configurations all sharing the same symbolic constraints.
\begin{figure}
    \begin{mathpar}
    \inferrule[\rulestyle{s-p-collect}]
    {
      \setsubdistr_{[s]}\subseteq \setsymsubdistr \\\\
      \setsymsubdistr'\equiv\{\spcconf{m'}{c'}{p'}{s'} \mid \exists \spcconf{m}{c}{p}{s}\in\setsubdistr_{[s]} \text{ s.t. }
      \spcconf{m}{c}{p}{s} \spcstep \spcconf{m'}{c'}{p'}{s'} \land \sat{s'} \}}
    {\setsymsubdistr\spsetstep \bigg(\setsymsubdistr\setminus  \setsubdistr_{[s]} \bigg)\cup \setsymsubdistr' }
  \end{mathpar}
  \caption{Rule for $\spsetstep$ }
  \label{fig:sem-splang-collecting}
\end{figure}
The notation $\setsubdistr_{[s]}\subseteq \setsymsubdistr$ means that
$\setsubdistr$ is the maximal subset of configuration in
$\setsymsubdistr$ which have $s$ as set of constraints. That is
$\setsubdistr_{[s]}\equiv \{\spcconf{m}{c}{p}{s}\mid
\spcconf{m}{c}{p}{s}\in\setsymsubdistr\}$.
Again, we extend the notation and use $\setsymsubdistr\spsetstepby{\setsubdistr_{[s]}}\setsymsubdistr'$
when we want to make explicit the set of symbolic configurations, $\setsubdistr_{[s]}$, 
that we are using to make the step.
So \rulestyle{s-p-collect}, starts from a set of configurations
and reaches all of those that are reachable from it.  By reachable we
mean that they have a satisfiable set of configurations and are
reachable from one of the original configurations with only one step of the
symbolic semantics. 
Similarly to deterministic case, the following lemma of coverage connects the \plang with \splang.
The difference is in the use of sets of configurations instead of single configurations.
Notice that in a set of constraints can appear constraints involving probabilistic symbols.
For instance if the i-th element of an array is associated with a random expression.
The predicate $\sat{\cdot}$ does not take in consideration relations involving probabilistic symbolic
constraints but only relations involving symbolic values denoting integers.
\begin{lemma}[Probabilistic Unary Coverage]
  \label{lem:plang-coverage}
  If $\setsymsubdistr\spsetstepby{\setsubdistr_{[s]}} \setsymsubdistr'$ and $\sigma\models_{\intlog} \setsubdistr_{[s]}$ then
  $\exists \sigma', \setsubdistr_{[s']}\subseteq \setsymsubdistr'$ such that $\sigma'\models_{\intlog}  \setsubdistr_{[s']}$,  and
  $\sub{\setsubdistr_{[s]}}{\sigma} \psetsteptr\sub{\setsubdistr_{[s']}}{\sigma'}$.
\end{lemma}

Intuitively, Lemma \ref{lem:plang-coverage} ensures us that
a concrete execution is covered by a symbolic one.
\else

The semantics of \splang has two sources of nondeterminism, from guards which reduce to 
symbolic values, and from guards which reduce to a probabilistic symbolic value. The collecting
semantics of \splang, specified by judgments as $
\setsymsubdistr \spsetstep \setsymsubdistr'$ (for sets of configurations $\setsymsubdistr$ and $\setsymsubdistr'$)  takes care of both of them.
The rule for this judgment form is: 
    \begin{mathpar}
    \inferrule[\rulestyle{s-p-collect}]
    {
      \setsubdistr_{[s]}\subseteq \setsymsubdistr \\\\
      \setsymsubdistr'\equiv\{\spcconf{m'}{c'}{p'}{s'} \mid \exists \spcconf{m}{c}{p}{s}\in\setsubdistr_{[s]} \text{ s.t. }
      \spcconf{m}{c}{p}{s} \spcstep \spcconf{m'}{c'}{p'}{s'} \\\\\land \sat{s'} \}}
    {\setsymsubdistr\spsetstep \big(\setsymsubdistr\setminus  \setsubdistr_{[s]} \big)\cup \setsymsubdistr' }
  \end{mathpar}
Unlike in
the deterministic case of the rule $\rulestyle{Set-Step}$, where only
one configuration was chosen nondeterministically from the initial
set, here we select nondeterministically a (maximal) set of
configurations all sharing the same symbolic constraints.
The notation $\setsubdistr_{[s]}\subseteq \setsymsubdistr$ means that
$\setsubdistr$ is the maximal subset of configuration in
$\setsymsubdistr$ which have $s$ as set of constraints. 
We use $\setsymsubdistr\spsetstepby{\setsubdistr_{[s]}}\setsymsubdistr'$
when we want to make explicit the set of symbolic configurations, $\setsubdistr_{[s]}$, 
that we are using to make the step.
Intuitively, \rulestyle{s-p-collect} starts from a set of configurations
and reaches all of those that are reachable from it - all the configurations that have a satisfiable set of constraints and are
reachable from one of the original configurations with only one step of the
symbolic semantics.
Notice that in a set of constraints we can have constraints involving probabilistic symbols, e.g. if the i-th element of an array is associated with a random expression. Nevertheless, the predicate $\sat{\cdot}$ does not need to take in consideration relations involving probabilistic symbolic
constraints but only relations involving symbolic values denoting integers.
The following lemma of coverage connects \plang with \splang ensuring  that
a concrete execution is covered by a symbolic one.
\begin{lemma}[Probabilistic Unary Coverage]
  \label{lem:plang-coverage}
  If $\setsymsubdistr\spsetstepby{\setsubdistr_{[s]}} \setsymsubdistr'$ and $\sigma\models_{\intlog} \setsubdistr_{[s]}$ then
  $\exists \sigma', \setsubdistr_{[s']}\subseteq \setsymsubdistr'$ such that $\sigma'\models_{\intlog}  \setsubdistr_{[s']}$,  and
  $\sub{\setsubdistr_{[s]}}{\sigma} \psetsteptr\sub{\setsubdistr_{[s']}}{\sigma'}$.
\end{lemma}

\fi
\ifnum\full=1
\subsection{\srplang}
\label{sec:rsplang}
We finally arrived to the last rung on this ladder of languages.  The
language presented in this section is the the symbolic extension of
the concrete language \rplang. It can also be seen as the relational
extension of \splang.  The key part of this language's semantics will
be the handling of the probabilistic assignment. For that construct we
will provide 2 rules instead of one. The first one is the obvious one
which carries on a standard symbolic probabilistic assignment. The
second one will implement a coupling semantics on the basis of Section
\ref{sec:prelim}.
We start off by providing the syntax of the language.
\subsection{\srplang: Syntax}
\begin{wrapfigure}[9]{L}{0.55\textwidth}
  \begin{minipage}{0.53\textwidth}
    \vspace{-0.8cm}
    \fbox{
  \[
  \begin{array}{rcl}  
    \srplangexpr \ni e_{sr} ::=  &e_{s} \mid \pair{e_{s}}{e_{s}} \mid e_{sr} \oplus e_{sr} \mid \arracc{a}{e_{sr}}\\
    \srplangcmd \ni c_{sr}  ::=  &c_{s} \mid  \pair{c_{s}}{c_{s}} \mid \seq{c_{sr}}{c_{sr}} \mid \ass{x}{e_{sr}} \mid \\
                                 &\ass{\arracc{a}{e_{sr}}}{e_{sr}}\mid \rass{x}{\lapp{e_{sr}}{e_{s}}}\mid \\
                                 &\ifte{e_{sr}}{c_{sr}}{c_{sr}} \mid\\
                                 & \cfor{x}{e_{sr}}{e_{sr}}{c_{sr}}
  \end{array}
\]
}
\caption{\srplang syntax. $e_{s}\in\spexpr, c_{s}\in\spcmd$.}
\label{fig:srplang-syntax}
\end{minipage}
\end{wrapfigure}
Figure \ref{fig:srplang-syntax} shows the semantics of the language
\srplang.  We extended the language $\splang$ with the pairing
construct, both at the level of expressions and commands. Importantly,
only unary symbolic expressions and commands are admitted in the
pairing construct.  This invariant is maintained during branching by
projection functions. In fact, projection function $\proj{i}{\cdot}$ for $i\in\{1,2\}$,
extend to also relational symbolic expressions and commands in the following way:
$\proj{i}{\pair{e_{1}}{e_{2}}}=e_{i}, \proj{i}{\pair{c_{1}}{c_{2}}}=c_{i}$.
With $\proj{i}{v}=v$, for $v\in\splangvalset$. Also, the projection
functions behave homomorphically on the other constructs.

\subsection{\srplang: Semantics of expressions}
\begin{wrapfigure}[11]{L}{0.65\textwidth}
\centering
\fbox{
  $\srpeconf{m_1}{m_2}{e}{p_1}{p_2}{s}\srpestep \srpval{v}{p'_1}{p'_2}{s'}$
}
\begin{minipage}{0.6\textwidth}
\fbox{
\begin{mathpar}
    \inferrule[\rulestyle{S-R-P-Lift}]
    {\speconf{m_1}{\proj{1}{e}}{p_1}{s}\spestep \spval{v_1}{p'_1}{s'} \\\\
      \speconf{m_2}{\proj{2}{e}}{p_2}{s'}\spestep \spval{v_2}{p'_2}{s''} \\\\
      v=\left \{
        \begin{array}{rcl}
          (v_1,v_2) && \text{if}\ (v_i\not\in\ints, i\in\{1,2\})\vee v_1\neq v_2 \cr
                       v_1 && \text{otherwise}
        \end{array}
      \right .\
    }
    {\srpeconf{m_1}{m_2}{e}{p_1}{p_2}{s} \srpestep  \srpval{v}{p'_1}{p'_2}{s''}}
  \end{mathpar}
}
\end{minipage}
\caption{\srplang: Semantics of expressions.}
\label{fig:srplang-sem-expr}
\end{wrapfigure}
As usual we provide a big-step evaluation semantics for expressions.
The judgment form and a selection of the rules defining the judgment
are provided in Figure \ref{fig:srplang-sem-expr}. The set of values now is
$\srplangvalset\equiv \splangvalset\cup\splangvalset^2$.
The only rule defining the judgment $\srpestep$ is
\rulestyle{S-R-P-Lift}.  It project the symbolic relational expression
first on the left and evaluates it to a unary symbolic value,
potentially updating the probabilistic symbolic constraints and the
symbolic constraints. It then does the same projecting the expression
on the right but starting from the potentially previously updated
constraints.  Now, the only case when the value returned is unary is
when both the previous evaluation returned equal integers, in all the
other cases an element in $\splangvalset^2$ is returned. So, the
relational symbolic semantics leverages on the unary semantics.
\else
\paragraph{\srplang}
\label{sec:rsplang}
The
language presented in this section is the the symbolic extension of
the concrete language \rplang. It can also be seen as the relational
extension of \splang.  The key part of this language's semantics will
be the handling of the probabilistic assignment. For that construct we
will provide 2 rules instead of one. The first one is the obvious one
which carries on a standard symbolic probabilistic assignment. The
second one will implement a coupling semantics.
The syntax of the \srplang extends the syntax of \rplang by adding symbolic values, since it is almost identical to the one of \rplang we omit it here.
%
% \begin{wrapfigure}[9]{L}{0.47\textwidth}
%   \begin{minipage}{0.45\textwidth}
%     \vspace{-0.8cm}
%     \fbox{
%   \[
%   \begin{array}{rcl}  
%     \srplangexpr \ni e_{sr} ::=  &e_{s} \mid \pair{e_{s}}{e_{s}} \mid e_{sr} \oplus e_{sr} \mid \arracc{a}{e_{sr}}\\
%     \srplangcmd \ni c_{sr}  ::=  &c_{s} \mid  \pair{c_{s}}{c_{s}} \mid \seq{c_{sr}}{c_{sr}} \mid \ass{x}{e_{sr}} \mid \\
%                                  &\ass{\arracc{a}{e_{sr}}}{e_{sr}}\mid \rass{x}{\lapp{e_{sr}}{e_{s}}}\mid \\
%                                  &\ifte{e_{sr}}{c_{sr}}{c_{sr}} \mid\\
%                                  & \cfor{x}{e_{sr}}{e_{sr}}{c_{sr}}
%   \end{array}
% \]
% }
% \caption{\srplang syntax. $e_{s}\in\spexpr, c_{s}\in\spcmd$.}
% \label{fig:srplang-syntax}
% \end{minipage}
% \end{wrapfigure}
As in the case of \rplang, only unary symbolic expressions and
commands are admitted in the pairing construct.  This invariant is
maintained by the semantics rules.
As for the other languages, we provide a big-step evaluation semantics for expressions proving judgments of the form $\srpeconf{m_1}{m_2}{e}{p_1}{p_2}{s}\srpestep \srpval{v}{p'_1}{p'_2}{s'}$.
The only rule defining the judgment $\srpestep$ is the following:
\begin{mathpar}
    \inferrule[\rulestyle{S-R-P-Lift}]
    { \begin{array}{c}
       \speconf{m_1}{\proj{1}{e}}{p_1}{s}\spestep \spval{v_1}{p'_1}{s'}\cr
      \speconf{m_2}{\proj{2}{e}}{p_2}{s'}\spestep \spval{v_2}{p'_2}{s''} \end{array}\and
      \hspace{-0.4cm}v=\left \{
        \begin{array}{rcl}
          (v_1,v_2) && \text{if}\ (v_i\not\in\ints, i\in\{1,2\})\vee v_1\neq v_2 \cr
                       v_1 && \text{otherwise}
        \end{array}
      \right .\
    }
    {\srpeconf{m_1}{m_2}{e}{p_1}{p_2}{s} \srpestep  \srpval{v}{p'_1}{p'_2}{s''}}
  \end{mathpar}

It project the symbolic relational expression
first on the left and evaluates it to a unary symbolic value,
potentially updating the probabilistic symbolic constraints and the
symbolic constraints. It then does the same projecting the expression
on the right but starting from the potentially previously updated
constraints.  Now, the only case when the value returned is unary is
when both the previous evaluation returned equal integers, in all the
other cases a pair of values is returned. So, the
relational symbolic semantics leverages on the unary semantics.
\fi
\ifnum\full=1
\subsection{\srplang: Semantics of commands}
\begin{wrapfigure}[8]{L}{0.35\textwidth}
  \vspace{-0.8cm}
  \begin{minipage}{0.30\textwidth}
    \fbox{
      \[
        \begin{array}{rcl}
          \ctxt&::=& \emptyctxt{\cdot} \mid \seq{\ctxt}{c}\\
          \ctxtp&::=&\pair{\seq{\cdot}{c}}{\cdot} \mid \pair{\cdot}{\seq{\cdot}{c}} \\
               &&\pair{\cdot}{\cdot} \mid \pair{\seq{\cdot}{c}}{\seq{\cdot}{c}}
        \end{array}
      \]
    }
  \end{minipage}
\caption{Grammars of evaluation contexts}
\label{fig:contexts}
\end{wrapfigure}
Before proceeding with the semantics of commands, we need to introduce the following
grammars of contexts. We use evaluation contexts to simplify the exposition of the rules since
for specific constructs such as relational probabilistic assignment we will
have more than one rule we could fire.
The two grammars for the evaluation contexts are shown in Figure \ref{fig:contexts}.
Notice how $\ctxtp$ gets saturated by pairs of commands.
Before proceeding we want to make a syntactical distinction between
commands. In particular, we call \emph{synchronizing}  all the commands in $\srplangcmd$
with the following shapes $\rass{x}{\lapp{e_1}{e_2}}$, $\pair{\rass{x}{\lapp{e_1}{e_2}}}{\rass{x'}{\lapp{e'_1}{e'_2}}}$.
We call commands with this structure: synchronizing because
they allow synchronization of two runs as we will see later on.
In particular, synchronizing commands are the ones that allow
the use of coupling semantics and coupling rules.
We call non synchronizing all the other commands in $\srplangcmd$
The semantics of commands is again provided 
\subsubsection{Non synchronizing commands}
In this section we provide the semantics for non synchronizing
commands.  In particular we are defining a judgment for a small-step
semantics with the form in Figure \ref{fig:judg-form-srplang-sem-cmds}.
A selection of the rules inductively defining the judgment is specified in Figure
\ref{fig:srplang-sem-cmds-non-synch}.
\begin{figure}
\centering
    \fbox{
      $\srpcconf{m_1}{m_2}{c}{p_1}{p_2}{s}\srpcstep \srpcconf{m'_1}{m'_2}{c'}{p'_1}{p'_2}{s'}$
    }
    \caption[\srplang: Judgment form for semantics of commands.]{\srplang: Judgment form for semantics of non synchronizing commands.
      $m_1,m_2,m'_1,m'_2\in\spmemset, c,c'\in\srplangcmd, p,p'\in SP, s,s'\in\cstrset$.}
    \label{fig:judg-form-srplang-sem-cmds}
  \end{figure}

  \begin{figure}
    \begin{mathpar}
    \inferrule[\rulestyle{s-r-if-prob-prob-true-false}]
    {
      \srpeconf{m_1}{m_2}{e}{p_1}{p_2}{s}\srpestep \srpval{v}{p'_1}{p'_2}{s'} \and \proj{1}{v},\proj{2}{v}\in\symsp\\\\
      p''_1\equiv p'_1@[\proj{1}{v}>0] \and p''_2\equiv p'_2@[\proj{2}{v}\leq 0]
    }
    {\srpcconf{m_1}{m_2}{\ifte{e}{\tbranch}{\fbranch}}{p_1}{p_2}{s}\srpcstep \srpcconf{m_1}{m_2}{\pair{\proj{1}{\tbranch}}{\proj{2}{\fbranch}}}{p''_1}{p''_2}{s'}}

    \inferrule[\rulestyle{s-r-if-prob-sym-true-false}]
    {
      \srpeconf{m_1}{m_2}{e}{p_1}{p_2}{s}\srpestep \srpval{v}{p'_1}{p'_2}{s'} \and \proj{1}{v}\in\symsp \and \proj{2}{v}\in\syms\\\\
      p''_1\equiv p'_1@[\proj{1}{v}>0] \and s'''\equiv s''\cup\{\proj{2}{v}\leq 0\}
    }
    {\srpcconf{m_1}{m_2}{\ifte{e}{\tbranch}{\fbranch}}{p_1}{p_2}{s}\srpcstep \srpcconf{m_1}{m_2}{\tbranch}{p''_1}{p'_2}{s'''}}

    \inferrule[\rulestyle{s-r-pair-lap-skip}]
    {
      \spcconf{m_1}{\rass{x}{\lapp{e_a}{e_b}}}{p_1}{s} \spcstep \spcconf{m'_1}{\skipp}{p'_1}{s'}
    }
    {\srpcconf{m_1}{m_2}{\pair{\rass{x}{\lapp{e_a}{e_b}}}{\skipp}}{p_1}{p_2}{s} \srpcstep
      \srpcconf{m'_1}{m_2}{\pair{\skipp}{\skipp}}{p'_1}{p_2}{s'}}

    \inferrule[\rulestyle{s-r-pair-lapleft-sync}]
    {
      c\not\equiv\rass{x}{\lapp{e'_a}{e'_b}} \and
      \spcconf{m_2}{c}{p_2}{s} \spcstep \spcconf{m'_2}{c'}{p'_2}{s'}  \and \ctxtp\equiv \pair{\cdot}{\cdot}
      %\\\\
      %m''=\pair{\proj{1}{m}}{m'}%[\epsilon_c\mapsto 0]
    }
    {\srpcconf{m_1}{m_2}{\ctxtp(\rass{x}{\lapp{e_a}{e_b}}, c)}{p_1}{p_2}{s} \srpcstep
      \srpcconf{m_1}{m'_2}{\pair{\rass{x}{\lapp{e_a}{e_b}}}{c'}}{p_1}{p'_2}{s'}}

    \inferrule[\rulestyle{s-r-pair-ctxt-1}]
        {
           \rass{x}{\lapp{e_a}{e_b}}\notin\{c_1,c_2\} \and |\{c_1,c_2\}|=2 \\\\
           \{1,2\}=\{i,j\}   \and
           m'_{i}\equiv m_i \and       \spcconf{m_j}{c_j}{p_j}{s} \spcstep \spcconf{m'_j}{c'_j}{p'_j}{s'} \\\\
           c'_i\equiv c_i \and p'_i\equiv p_i \and
           %m_1''=m'_1[\epsilon_c\mapsto 0] \and m_2''=m'_2[\epsilon_c\mapsto 0]
        }
        {\srpcconf{m_1}{m_2}{\ctxtp(c_1,c_2)}{p_1}{p_2}{s} \srpcstep
          \srpcconf{m'_1}{m'_2}{\ctxtp(c'_1,c'_2)}{p'_1}{p'_2}{s'}}

        \inferrule[\rulestyle{s-r-pair-ctxt-2}]
        {
          \ctxtp\not\equiv \pair{\cdot}{\cdot}\\\\
          \srpcconf{m_1}{m_2}{\pair{c_1}{c_2}}{p_1}{p_2}{s} \srpcstep \srpcconf{m'_1}{m'_2}{\pair{c'_1}{c'_2}}{p'_1}{p'_2}{s'}
        }
        {
          \srpcconf{m_1}{m_2}{\ctxtp(c_1,c_2)}{p_1}{p_2}{s} \srpcstep
          \srpcconf{m'_1}{m'_2}{\ctxtp(c'_1,c'_2)}{p'_1}{p'_2}{s'}
        }
  \end{mathpar}
  \caption{\srplang: Semantics of non synchronizing commands. Selected rules.}
  \label{fig:srplang-sem-cmds-non-synch}
\end{figure}
An explanation of the rules follows.
Rule \rulestyle{s-r-if-prob-prob-true-false} fires when evaluating
a branching instruction. In particular, it fires when the guard
evaluates on both side to a probabilistic symbolic value.
In this case the semantics can continue with the true branch
on the left run and with the false branch on the right one.
Notice that commands are projected to avoid pairing commands
appearing in a nested form.
In the case where the guard of a branching instruction evaluates to
a probabilistic symbolic value on the left run and a symbolic integer
value on the right one, rule  \rulestyle{s-r-if-prob-sym-true-false}
can apply. The rule allows to continue on the true branch on the left
run and on the false branch on the right one. Notice that in one case
the probabilistic list of constraints is updated, while on the
other the symbolic set of constraints. Rule \rulestyle{s-r-pair-lap-skip}
handles the pairing command where on the left hand side we have
a probabilistic assignment and on the right a skip instruction.
In this case, there is no \emph{hope for synchronization}  between the two runs
and hence we can just unarily perform the left probabilistic assignment relying
on the unary symbolic semantics.  Rule \rulestyle{s-r-pair-lapleft-sync} instead
applies when on the left we have a probabilistic assignment and on the right we have
another arbitrary command. In this case we can hope to reach a situation where
on the right run another probabilistic assignment appears. Hence,
it makes sense to continue the computation unarily on the right side.
Rule \rulestyle{s-r-pair-ctxt-1} applies when a pairing command is built out
of two different unary commands neither of which is a probabilistic assignment, or a sequence of commands.
In this case we can just rely on the unary semantics and execute one step
on side.
Rule \rulestyle{s-r-pair-ctxt-2} instead applies in all the other cases by recursively relying
on the $\srpcstep$ semantics.
\else
For the semantics of commands we use the following evaluation contexts to simplify the exposition: $\ctxt::= \emptyctxt{\cdot} \mid \seq{\ctxt}{c}$ and 
$\ctxtp::= \pair{\seq{\cdot}{c}}{\cdot} \mid \pair{\cdot}{\seq{\cdot}{c}}\mid  \pair{\cdot}{\cdot} \mid \pair{\seq{\cdot}{c}}{\seq{\cdot}{c}}$.
Notice how $\ctxtp$ gets saturated by pairs of commands.
Before proceeding we want to make a
Moreover, we separate commands in two classes. We call \emph{synchronizing}  all the commands in $\srplangcmd$
with the following shapes $\rass{x}{\lapp{e_1}{e_2}}$, $\pair{\rass{x}{\lapp{e_1}{e_2}}}{\rass{x'}{\lapp{e'_1}{e'_2}}}$, since
they allow synchronization of two runs using coupling rules.
We call non synchronizing all the other commands. We relegate the explanation of their semantics to the appendix.
\fi
\ifnum\full=1
\subsubsection{\srplang: Collecting semantics for non synchronizing commands}
Again $\srpcstep$ is a non deterministic semantics.  The
non determinism comes from the use of probabilistic symbols as guards
in branching instructions, as well as from symbolic values used as
guards.  Finally the layer of non determinism given by the relational
approach which allows runs to take different branches in a branching
instruction.
So, in order to collect all the possible traces stemming from such non determinism
we define a collecting semantics relating set of configurations to set of configurations.
The semantics is specified through a judgment with the form: $ \rsetsymsubdistr \srpsetstep  \rsetsymsubdistr' $,
with  $ \rsetsymsubdistr,  \rsetsymsubdistr'\in\powerset(\spmemset\times\spmemset\times\srplangcmd\times SP\times SP\times\cstrset)$.
The only rule defining the judgment is given in Figure \ref{fig:sem-srplang-collecting}.
\begin{figure}
    \begin{mathpar}
    \inferrule[\rulestyle{s-r-p-collect}]
    {
      \rsetsubdistr_{[s]}\subseteq \rsetsymsubdistr \\\\
      \rsetsymsubdistr'\equiv\{\srpcconf{m'_1}{m'_2}{c'}{p'_1}{p'_2}{s'} \mid
      \\\\\exists \srpcconf{m_1}{m_2}{c}{p_1}{p_2}{s}\in\rsetsubdistr_{[s]} \text{ s.t. }
      \srpcconf{m_1}{m_2}{c}{p_1}{p_2}{s} \srpcstep \srpcconf{m'_1}{m'_2}{c'}{p'_1}{p'_2}{s'} \land \sat{s'} \}}
    {\rsetsymsubdistr\srpsetstep \bigg(\rsetsymsubdistr\setminus  \rsetsubdistr_{[s]} \bigg)\cup \rsetsymsubdistr' }
  \end{mathpar}
  \caption{Rule for $\srpsetstep$ }
  \label{fig:sem-srplang-collecting}
\end{figure}
The rule, and the auxiliary notation $\rsetsubdistr_{[s]}$,
is pretty similar to the one in Figure \ref{fig:sem-splang-collecting}.
The only difference is that here sets of symbolic relational probabilistic configurations
are considered instead of symbolic (unary) probabilistic configurations.

\else
Again $\srpcstep$ is a non deterministic semantics.  The
non determinism comes from the use of probabilistic symbols and symbolic values as guards, and by the relational
approach. So, in order to collect all the possible traces stemming from such non determinism
we define a collecting semantics relating set of configurations to set of configurations.
The semantics is specified through a judgment of the form: $ \rsetsymsubdistr \srpsetstep  \rsetsymsubdistr' $,
with  $ \rsetsymsubdistr,  \rsetsymsubdistr'\in\powerset(\spmemset\times\spmemset\times\srplangcmd\times SP\times SP\times\cstrset)$.
The only rule defining it is a natural lifting of the one for the unary semantics.
% The only rule defining the judgment is:
%     \begin{mathpar}
%     \inferrule[\rulestyle{s-r-p-collect}]
%     {
%       \rsetsubdistr_{[s]}\subseteq \rsetsymsubdistr \and
%       \rsetsymsubdistr'\equiv\{\srpcconf{m'_1}{m'_2}{c'}{p'_1}{p'_2}{s'} \mid
%       \\\\ \exists \srpcconf{m_1}{m_2}{c}{p_1}{p_2}{s}\in\rsetsubdistr_{[s]} \text{ s.t. }
%       \srpcconf{m_1}{m_2}{c}{p_1}{p_2}{s} \srpcstep \srpcconf{m'_1}{m'_2}{c'}{p'_1}{p'_2}{s'} \\\\\land \sat{s'} \}}
%     {\rsetsymsubdistr\srpsetstep \big(\rsetsymsubdistr\setminus  \rsetsubdistr_{[s]} \big)\cup \rsetsymsubdistr' }
%   \end{mathpar}
% The rule, and the auxiliary notation $\rsetsubdistr_{[s]}$,
% is pretty similar to the one for \splang, the only difference is that here sets of symbolic relational probabilistic configurations
% are considered instead of symbolic (unary) probabilistic configurations.
We can extend the coverage lemma  to the relational setting:
\begin{lemma}[Probabilistic Relational Coverage]
\label{lem:rplang-coverage}
  If $\rsetsymsubdistr \srpsetstepby{\rsetsubdistr_{[s]}} \rsetsymsubdistr'$ and $\sigma\models_{\intlog} \rsetsubdistr_{[s]}$ then
  $\exists \sigma', \rsetsubdistr_{[s']}\in\rsetsymsubdistr'$ such that $\rsetsubdistr_{[s']}\subseteq\rsetsymsubdistr',
  \sigma'\models_{\intlog}  \rsetsubdistr_{[s']}$,  and
  $\sub{\rsetsubdistr_{[s]}}{\sigma} \rsetsteptr\sub{\rsetsubdistr_{[s']}}{\sigma'}$.
\end{lemma}
\fi
\ifnum\full=1
\subsubsection{Synchronizing commands}
So far all the semantics rules for $\srplang$ we presented are uniquely determined
by the syntactic construct they are associated to. Also, the rules of
the $\srpcstep$ semantics and, hence, also of the $\srpsetstep$
semantics, don't deal with synchronizing commands.  For those we want
to be able to apply different rules.  In order to consider all the
possibilities we define a new judgment with form   $ \setproof \proofstep  \setproof' $, with
$\setproof,\setproof'\in\powerset(\powerset(\spmemset\times\spmemset\times\srplangcmd\times SP\times SP\times\cstrset))$
  \begin{figure}
    \begin{mathpar}
      \inferrule[\rulestyle{Proof-Step-No-Sync}]
        {
          \rsetsymsubdistr\in \setproof \and
          \rsetsymsubdistr \srpsetstep \rsetsymsubdistr'
          \and \setproof'\equiv \bigg (\setproof\setminus\{\rsetsymsubdistr\}\bigg )\cup\{\rsetsymsubdistr'\}
        }
        {\setproof \proofstep \setproof'}
    
        \inferrule[\rulestyle{Proof-Step-No-Coup}]
      {
        \srpcconf{m_1}{m_2}{\ctxt[\rass{x}{\lapp{e_a}{e_b}}]}{p_1}{p_2}{s} \in \rsetsymsubdistr\in \setproof \\\\
        \srpeconf{m_1}{m_2}{e_a}{p_1}{p_2}{s}      \srpestep\srpval{v_a}{p'_1}{p'_2}{s_a} \and
        \srpeconf{m_1}{m_2}{e_b}{p'_1}{p'_2}{s_a} \srpestep \srpval{v_b}{p''_1}{p''_2}{s_b} \\\\
        \fresh{X_1,X_2}{\symsp} \and  m'_1\equiv m_1[x\mapsto X_1] \and m'_2\equiv m_2[x\mapsto X_2] \\\\
        p'''_1\equiv p''_1@[\rass{X_1}{\lapp{\proj{1}{v_a}}{\proj{1}{v_b}}}]\and p'''_2\equiv p''_2@[\rass{X_2}{\lapp{\proj{2}{v_a}}{\proj{2}{v_b}}}] \\\\
        \rsetsymsubdistr'\equiv \bigg (\rsetsymsubdistr\setminus \{\srpcconf{m_1}{m_2}{\ctxt[\rass{x}{\lapp{e_a}{e_b}}]}{p_1}{p_2}{s}\}\bigg)\cup
        \{\srpcconf{m'_1}{m'_2}{\ctxt[\skipp]}{p'''_1}{p'''_2}{s''}\}\\\\
        \setproof'\equiv \bigg( \setproof\setminus\{\rsetsymsubdistr\}\bigg ) \cup \{\rsetsymsubdistr'\}
      }
        {
        \setproof \proofstep       \setproof' 
      }

      \inferrule[\rulestyle{Proof-Step-Avoc}]
      {
        \srpcconf{m_1}{m_2}{\ctxt[\rass{x}{\lapp{e_a}{e_b}}]}{p_1}{p_2}{s} \in \rsetsymsubdistr\in \setproof \\\\
        \srpeconf{m_1}{m_2}{e_a}{p_1}{p_2}{s}      \srpestep\srpval{v_a}{p'_1}{p'_2}{s_a} \and
        \srpeconf{m_1}{m_2}{e_b}{p'_1}{p'_2}{s_a} \srpestep \srpval{v_b}{p''_1}{p''_2}{s_b} \\\\
        \fresh{X_1,X_2}{\syms} \and  m'_1\equiv m_1[x\mapsto X_1] \and m'_2\equiv m_2[x\mapsto X_2] \\\\
        \rsetsymsubdistr'\equiv \bigg (\rsetsymsubdistr\setminus \{\srpcconf{m_1}{m_2}{\ctxt[\rass{x}{\lapp{e_a}{e_b}}]}{p_1}{p_2}{s}\}\bigg)\cup
        \{\srpcconf{m'_1}{m'_2}{\ctxt[\skipp]}{p''_1}{p''_2}{s''}\}\\\\
        \setproof'\equiv \bigg( \setproof\setminus\{\rsetsymsubdistr\}\bigg ) \cup \{\rsetsymsubdistr'\}
      }
      {
        \setproof \proofstep       \setproof' 
      }
    \end{mathpar}
    \caption[\srplang: Proof collecting semantics]{\srplang: Proof collecting semantics, selected rules.}
    \label{fig:proof-semantics-1}
  \end{figure}

   \begin{figure}[h]
     \begin{mathpar}
       \inferrule[\rulestyle{Proof-Step-Lap-Gen}]
      {
        \srpcconf{m_1}{m_2}{\ctxt[\rass{x}{\lapp{e_a}{e_b}}]}{p_1}{p_2}{s} \in \rsetsymsubdistr\in \setproof \\\\
        \srpeconf{m_1}{m_2}{e_a}{p_1}{p_2}{s} \srpestep\srpval{v_a}{p'_1}{p'_2}{s_a} \and
        \srpeconf{m_1}{m_2}{e_b}{p'_1}{p'_2}{s_a} \srpestep \srpval{v_b}{p''_1}{p''_2}{s_b} \\\\
        s'\equiv s_b\cup \{\proj{1}{v_b}=\proj{2}{v_b}, \proj{1}{v_b}>0\} \and m_1(\epsilon_c)=E'=m'_2(\epsilon_c) \\\\
        \fresh{E'',X_1, X_2, K, K'}{\syms} \and
        m'_1\equiv m_1[x\mapsto X_1][\epsilon_c\mapsto E''],\and m'_2=m_2[x\mapsto X_2][\epsilon_c\mapsto E'']\\\\
         m(\epsilon)=E \and
        s''\equiv s'\cup \{X_1+K=X_2, K\leq K', K'\cdot E=\proj{1}{v_b}, E''=E'+|\proj{1}{v_a}-\proj{2}{v_a}|\cdot K'\}\\\\
        p'''_1\equiv p''_1@[\rass{X_1}{\lapp{\proj{1}{v_a}}{\proj{1}{v_b}}}]\and p'''_2\equiv p''_2@[\rass{X_2}{\lapp{\proj{2}{v_a}}{\proj{2}{v_b}}}] \\\\
        \rsetsymsubdistr'\equiv \bigg (\rsetsymsubdistr\setminus \{\srpcconf{m_1}{m_2}{\ctxt[\rass{x}{\lapp{e_a}{e_b}}]}{p_1}{p_2}{s}\}\bigg)\cup
        \{\srpcconf{m'_1}{m'_2}{\ctxt[\skipp]}{p'''_1}{p'''_2}{s''}\}\\\\
        \setproof'\equiv \bigg( \setproof\setminus\{\rsetsymsubdistr\}\bigg ) \cup \{\rsetsymsubdistr'\}
      }
        {
        \setproof \proofstep       \setproof' 
      }

      \inferrule[\rulestyle{Proof-Step-If}]
      {
        \rho=\srpcconf{m_1}{m_2}{\ctxt[\ifte{e}{c_1}{c_2}]}{p_1}{p_2}{s} \in \rsetsymsubdistr\in \setproof\\\\
        \srpeconf{m_1}{m_2}{e}{p_1}{p_2}{s} \srpestep \srpval{v}{p'_1}{p'_2}{s'}\and \{\oplus_1,\oplus_2\}=\{>,\leq\}\\\\
      \models s'\implies \proj{1}{v}\oplus_{i} 0 \iff \proj{2}{v}\oplus_{i} 0\\\\
        \rsetsymsubdistr \srpsetstepby{\{\rho\}} \rsetsymsubdistr' \and 
        \setproof'\equiv \bigg( \setproof\setminus\{\rsetsymsubdistr\}\bigg ) \cup \{\rsetsymsubdistr'\}
      }
      {
        \setproof \proofstep       \setproof' 
      }

      \inferrule[\rulestyle{Proof-Step-Other-Cmds}]
      {
       \rho= \srpcconf{m_1}{m_2}{\ctxt[c]}{p_1}{p_2}{s} \in \rsetsymsubdistr\in \setproof\\\\
       c\neq \ifte{\cdot}{\cdot}{\cdot}\and \rsetsymsubdistr \srpsetstepby{\{\rho\}} \rsetsymsubdistr'\\\\
        \setproof'\equiv \bigg( \setproof\setminus\{\rsetsymsubdistr\}\bigg ) \cup \{\rsetsymsubdistr'\}
      }
      {
        \setproof \proofstep       \setproof' 
      }
    \end{mathpar}
    \caption[\srplang: Proof collecting semantics]{\srplang: Proof collecting semantics, other rules}
    \label{fig:proof-semantics}
  \end{figure}
  In Figures \ref{fig:proof-semantics-1} and \ref{fig:proof-semantics},
  a selected collection of the
rules defining the judgment is presented.  One of the rules defining
this last judgment relies on the previously defined collecting
semantics.  Indeed, rule \rulestyle{Proof-Step-No-Sync} applies when
no synchronizing commands are involved, and hence there is no
possible coupling rule to be applied.  Before proceeding with the
explanation of the other rules there is the need to explain the
variable $\epsilon_c$ which is used in the latter rules.  The variable
$\epsilon_c$ is a variable that symbolically counts the current level
of privacy in the current relational execution.  The variable gets
increased when the rule $\rulestyle{Proof-Step-Lap-Gen}$
fires.  We have chosen to omit a similar ghost counter variable for
$\delta$ so rules would be more readable.
The variable gets increased also when the rules handling the pairing command
$\pair{\rass{x}{\lapp{e_a}{e_b}}}{\rass{x}{e'_a}{e'_b}}$.  This latter
rule is shown in Appendix.  This symbolic counter
variable is useful when trying to prove equality of certain variables
without spending more than a specific budget.
In the set of sets of configurations $\setproof$, a set of configurations, $\rsetsymsubdistr$,
is nondeterministically chosen. Among elements in $\rsetsymsubdistr$ a configuration
is also nondeterministically chosen.
Using contexts we check that in the selected configuration the
next command to execute is the probabilistic assignment.
After reducing to values both the mean and scale expression,
and verified (that is, assumed in the set of constraints)
that in the two runs the scales have the same value, the rule adds to the set of
constraints a new element, that is,
$E''=E'+|\proj{1}{v_a}-\proj{2}{v_a}|\cdot K'$, where $K, K', E''$ are
fresh symbols denoting integers and $E'$ is the symbolic integer to which the budget
variable $\epsilon_c$ maps to.  Notice that $\epsilon_c$ needs to
point to the same symbol in both memories.  This because it's a shared
variable tracking the privacy budget spent so far in both runs.  This
new constraint increases the budget spent. The other constraint added
is the real coupling relation, that is $X_1+K=X_2$. Where $X_1, X_2$ are fresh in $\syms$.
Later, $K$ will be existentially quantified in order to search for a proof of
$\epsilon$-indistinguishability.  Rule $\rulestyle{Proof-Step-Avoc}$
does not use any coupling rule but treats the samples in a purely in
a symbolic manner. It intuitively asserts that the two samples are
drawn from the distributions and assigns to them arbitrary integers
free to vary on the all domain of Thai Laplace distribution.  Finally,
$\rulestyle{Proof-Step-No-Coup}$ applies to synchronizing commands as
well.  It does not add any relational constraints to the samples.
This rules intuitively means that we are not correlating in any way
the two samples. Notice that since we are not using any coupling rule
we don't need to check that the scale value is the same in the two
runs as it is requested in the previous rule.  We could think of this
as a way to encode the relational semantics of the program in an
expression which later can be fed in input to other tools.  The main
difference with the previous rule is that here we treat the sampling
instruction symbolically and that's why the fresh symbols are in
$\symsp$ denoting full distributions, while in the previous rule the
fresh symbols are in $\syms$ denoting sampled integers, even though in
no particular relation.  When the program involves a synchronizing
command we basically fork the execution when it's time to execute it.
In particular the set of configurations gets to continue the
computation in different ways, one for every rule applicable.
\subsection{Coverage}
The coverage lemma can be extended also to the relational setting.  To do
that though we need to consider only the fragment of the $\proofstep$
semantics that only uses the rules $\rulestyle{Proof-Step-No-Sync}$, and
$\rulestyle{Proof-Step-No-Coupl}$.  We denote this fragment with
the notation $\proofstephat$.
While the semantics that uses
only the following rules will be denoted by $\proofstepup$: $\rulestyle{Proof-Step-Lap-Gen}$,
$\rulestyle{Proof-Step-Other-Cmds}$, $\rulestyle{Proof-Step-If}$.
A similar relational coverage Lemma holds for the $\srpsetstep$ semantics
and hence it trivially extends to $\proofstephat$. 
\begin{lemma}[Probabilistic Relational Coverage]
\label{lem:rplang-coverage}
  If $\rsetsymsubdistr \srpsetstepby{\rsetsubdistr_{[s]}} \rsetsymsubdistr'$ and $\sigma\models_{\intlog} \rsetsubdistr_{[s]}$ then
  $\exists \sigma', \rsetsubdistr_{[s']}\in\rsetsymsubdistr'$ such that $\rsetsubdistr_{[s']}\subseteq\rsetsymsubdistr',
  \sigma'\models_{\intlog}  \rsetsubdistr_{[s']}$,  and
  $\sub{\rsetsubdistr_{[s]}}{\sigma} \rsetsteptr\sub{\rsetsubdistr_{[s']}}{\sigma'}$.
\end{lemma}
\else
\paragraph{Semantics of synchronizing commands}
We define a new judgment with form   $ \setproof \proofstep  \setproof' $, with
$\setproof,\setproof'\in\powerset(\powerset(\spmemset\times\spmemset\times\srplangcmd\times SP\times SP\times\cstrset))$.
  \begin{figure*}
 \begin{minipage}{0.98\textwidth}
  \fbox{
    \begin{mathpar}
      \inferrule[\rulestyle{Proof-Step-No-Sync}]
        {
          \rsetsymsubdistr\in \setproof \and
          \rsetsymsubdistr \srpsetstep \rsetsymsubdistr'
          \and \setproof'\equiv \big (\setproof\setminus\{\rsetsymsubdistr\}\big )\cup\{\rsetsymsubdistr'\}
        }
        {\setproof \proofstep \setproof'}
    
      %   \inferrule[\rulestyle{Proof-Step-No-Coup}]
      % {
      %   \srpcconf{m_1}{m_2}{\ctxt[\rass{x}{\lapp{e_a}{e_b}}]}{p_1}{p_2}{s} \in \rsetsymsubdistr\in \setproof \\\\
      %   \srpeconf{m_1}{m_2}{e_a}{p_1}{p_2}{s}      \srpestep\srpval{v_a}{p'_1}{p'_2}{s_a} \and
      %   \srpeconf{m_1}{m_2}{e_b}{p'_1}{p'_2}{s_a} \srpestep \srpval{v_b}{p''_1}{p''_2}{s_b} \\\\
      %   \fresh{X_1,X_2}{\symsp} \and  m'_1\equiv m_1[x\mapsto X_1] \and m'_2\equiv m_2[x\mapsto X_2] \\\\
      %   p'''_1\equiv p''_1@[\rass{X_1}{\lapp{\proj{1}{v_a}}{\proj{1}{v_b}}}]\and p'''_2\equiv p''_2@[\rass{X_2}{\lapp{\proj{2}{v_a}}{\proj{2}{v_b}}}] \\\\
      %   \rsetsymsubdistr'\equiv \bigg (\rsetsymsubdistr\setminus \{\srpcconf{m_1}{m_2}{\ctxt[\rass{x}{\lapp{e_a}{e_b}}]}{p_1}{p_2}{s}\}\bigg)\cup
      %   \{\srpcconf{m'_1}{m'_2}{\ctxt[\skipp]}{p'''_1}{p'''_2}{s''}\}\\\\
      %   \setproof'\equiv \bigg( \setproof\setminus\{\rsetsymsubdistr\}\bigg ) \cup \{\rsetsymsubdistr'\}
      % }
      %   {
      %   \setproof \proofstep       \setproof' 
      % }
%
      \inferrule[\rulestyle{Proof-Step-Avoc}]
      {
        \srpcconf{m_1}{m_2}{\ctxt[\rass{x}{\lapp{e_a}{e_b}}]}{p_1}{p_2}{s} \in \rsetsymsubdistr\in \setproof \\\\
        \srpeconf{m_1}{m_2}{e_a}{p_1}{p_2}{s}      \srpestep\srpval{v_a}{p'_1}{p'_2}{s_a} \\\\
        \srpeconf{m_1}{m_2}{e_b}{p'_1}{p'_2}{s_a} \srpestep \srpval{v_b}{p''_1}{p''_2}{s_b} \\\\
        \fresh{X_1,X_2}{\syms} \and  m'_1\equiv m_1[x\mapsto X_1] \and m'_2\equiv m_2[x\mapsto X_2]\\\\
        \setproof'\equiv \big( \setproof\setminus\{\rsetsymsubdistr\}\big ) \cup \{\rsetsymsubdistr'\}\\\\
        \rsetsymsubdistr'\equiv \big (\rsetsymsubdistr\setminus \{\srpcconf{m_1}{m_2}{\ctxt[\rass{x}{\lapp{e_a}{e_b}}]}{p_1}{p_2}{s}\}\big)\\\\\cup
        \{\srpcconf{m'_1}{m'_2}{\ctxt[\skipp]}{p''_1}{p''_2}{s''}\}
      }
      {
        \setproof \proofstep       \setproof' 
      }
  %   \end{mathpar}
  %   \caption[\srplang: Proof collecting semantics]{\srplang: Proof collecting semantics, selected rules.}
  %   \label{fig:proof-semantics-1}
  % \end{figure}
  %
  %  \begin{figure}[h]
  %    \begin{mathpar}
\and
      \inferrule[\rulestyle{Proof-Step-Lap-Gen}]
      {
        \srpcconf{m_1}{m_2}{\ctxt[\rass{x}{\lapp{e_a}{e_b}}]}{p_1}{p_2}{s} \in \rsetsymsubdistr\in \setproof \\\\
        \srpeconf{m_1}{m_2}{e_a}{p_1}{p_2}{s} \srpestep\srpval{v_a}{p'_1}{p'_2}{s_a} \\\\
        \srpeconf{m_1}{m_2}{e_b}{p'_1}{p'_2}{s_a} \srpestep \srpval{v_b}{p''_1}{p''_2}{s_b} \\\\
        s'\equiv s_b\cup \{\proj{1}{v_b}=\proj{2}{v_b}, \proj{1}{v_b}>0\} \and m_1(\epsilon_c)=E'=m'_2(\epsilon_c) \\\\
        \fresh{E'',X_1, X_2, K, K'}{\syms} \and
        m'_1\equiv m_1[x\mapsto X_1][\epsilon_c\mapsto E''] \\\\ m'_2=m_2[x\mapsto X_2][\epsilon_c\mapsto E'']\and
         m(\epsilon)=E \\\\
        s''\equiv s'\cup \{X_1+K=X_2, K\leq K', K'\cdot E=\proj{1}{v_b}, E''=E'+|\proj{1}{v_a}-\proj{2}{v_a}|\cdot K'\}\\\\
        p'''_1\equiv p''_1@[\rass{X_1}{\lapp{\proj{1}{v_a}}{\proj{1}{v_b}}}]\and p'''_2\equiv p''_2@[\rass{X_2}{\lapp{\proj{2}{v_a}}{\proj{2}{v_b}}}]
        \\\\
        \setproof'\equiv \big( \setproof\setminus\{\rsetsymsubdistr\}\big ) \cup \{\rsetsymsubdistr'\}
        \\\\
        \rsetsymsubdistr'\equiv \big (\rsetsymsubdistr\setminus \{\srpcconf{m_1}{m_2}{\ctxt[\rass{x}{\lapp{e_a}{e_b}}]}{p_1}{p_2}{s}\}\big)\\\\\cup
        \{\srpcconf{m'_1}{m'_2}{\ctxt[\skipp]}{p'''_1}{p'''_2}{s''}\}
      }
        {
        \setproof \proofstep       \setproof' 
      }
      %
      % \inferrule[\rulestyle{Proof-Step-If}]
      % {
      %   \rho=\srpcconf{m_1}{m_2}{\ctxt[\ifte{e}{c_1}{c_2}]}{p_1}{p_2}{s} \in \rsetsymsubdistr\in \setproof\\\\
      %   \srpeconf{m_1}{m_2}{e}{p_1}{p_2}{s} \srpestep \srpval{v}{p'_1}{p'_2}{s'}\and \{\oplus_1,\oplus_2\}=\{>,\leq\}\\\\
      % \models s'\implies \proj{1}{v}\oplus_{i} 0 \iff \proj{2}{v}\oplus_{i} 0\\\\
      %   \rsetsymsubdistr \srpsetstepby{\{\rho\}} \rsetsymsubdistr' \and 
      %   \setproof'\equiv \bigg( \setproof\setminus\{\rsetsymsubdistr\}\bigg ) \cup \{\rsetsymsubdistr'\}
      % }
      % {
      %   \setproof \proofstep       \setproof' 
      % }
      %
      % \inferrule[\rulestyle{Proof-Step-Other-Cmds}]
      % {
      %  \rho= \srpcconf{m_1}{m_2}{\ctxt[c]}{p_1}{p_2}{s} \in \rsetsymsubdistr\in \setproof\\\\
      %  c\neq \ifte{\cdot}{\cdot}{\cdot}\and \rsetsymsubdistr \srpsetstepby{\{\rho\}} \rsetsymsubdistr'\\\\
      %   \setproof'\equiv \bigg( \setproof\setminus\{\rsetsymsubdistr\}\bigg ) \cup \{\rsetsymsubdistr'\}
      % }
      % {
      %   \setproof \proofstep       \setproof' 
      % }
    \end{mathpar}
  }
  \end{minipage}
    \caption[\srplang: Proof collecting semantics]{\srplang: Proof collecting semantics, selected rules}
    \label{fig:proof-semantics}
  \end{figure*}
  In Figure \ref{fig:proof-semantics},
  we give a selection  of the
rules. %  One of the rules defining
% this last judgment relies on the previously defined collecting
% semantics. 
Rule \rulestyle{Proof-Step-No-Sync} applies when no synchronizing
commands are involved, and hence there is no possible coupling rule to
be applied. In the other rules, we use the variable $\epsilon_c$ to
symbolically count the privacy budget in the current relational
execution.  The variable gets increased when the rule
$\rulestyle{Proof-Step-Lap-Gen}$ fires. %  The variable gets increased also when the rules handling
% the pairing command
% $\pair{\rass{x}{\lapp{e_a}{e_b}}}{\rass{x}{e'_a}{e'_b}}$.  This latter
% rule is shown in Appendix \ref{app:appendixA}. 
This symbolic counter
variable is useful when trying to prove equality of certain variables
without spending more than a specific budget.  In the set of sets of
configurations $\setproof$, a set of configurations,
$\rsetsymsubdistr$, is nondeterministically chosen. Among elements in
$\rsetsymsubdistr$ a configuration is also nondeterministically
chosen.  Using contexts we check that in the selected configuration
the next command to execute is the probabilistic assignment.  After
reducing to values both the mean and scale expression, and verified
(that is, assumed in the set of constraints) that in the two runs the
scales have the same value, the rule adds to the set of constraints a
new element, that is, $E''=E'+|\proj{1}{v_a}-\proj{2}{v_a}|\cdot K'$,
where $K, K', E''$ are fresh symbols denoting integers and $E'$ is the
symbolic integer to which the budget variable $\epsilon_c$ maps to.
Notice that $\epsilon_c$ needs to point to the same symbol in both
memories.  This because it's a shared variable tracking the privacy
budget spent so far in both runs.  This new constraint increases the
budget spent. The other constraint added is the real coupling
relation, that is $X_1+K=X_2$. Where $X_1, X_2$ are fresh in $\syms$.
Later, $K$ will be existentially quantified in order to search for a
proof of $\epsilon$-indistinguishability.  Rule
$\rulestyle{Proof-Step-Avoc}$ does not use any coupling rule but
treats the samples in a purely in a symbolic manner. It intuitively
asserts that the two samples are drawn from the distributions and
assigns to them arbitrary integers free to vary on the all domain of
the Laplace distribution.
%Finally,
% Rule $\rulestyle{Proof-Step-No-Coup}$
% applies to synchronizing commands as well.  It does not add any
% relational constraints to the samples.  This rules intuitively means
% that we are not correlating in any way the two samples. Notice that
% since we are not using any coupling rule we don't need to check that
% the scale value is the same in the two runs as it is requested in the
% previous rule.  We could think of this as a way to encode the
% relational semantics of the program in an expression which later can
% be fed in input to other tools.  The main difference with the previous
% rule is that here we treat the sampling instruction symbolically and
% that's why the fresh symbols are in $\symsp$ denoting full
% distributions, while in the previous rule the fresh symbols are in
% $\syms$ denoting sampled integers, even though in no particular
% relation.
When the program involves a synchronizing command we
basically fork the execution when it's time to execute it.  In
particular the set of configurations gets to continue the computation
in different ways, one for every rule applicable.

\fi
\ifnum\full=1
\section{Derivations}
\label{sec:derivations}
The language of relational assertions is defined using  first order predicate logic formulas
involving relational program expressions and logical variables in an unspecified set $\logvar$. The interpretation of
a relational assertions is naturally defined as a subset of $\cmemset\times\cmemset$,
that is the set of pairs of memories modelling the assertion. We will let capital greek letters
such as $\Phi, \Psi\dots$ range of the set of relational assertions. We will also need an additional substituting
function $\tocstr{\cdot}{\cdot}$ taking an assertion and a memory in input and returning the assertion where all the
program variables have been substituted with the values to which the memory maps variables to.
That is, given a memory, relational or unary, $m$ and an assertion, relational or unary, $\Phi$,
$\tocstr{m}{\Phi}$ is a constraint. More details can be found in \cite{Farina2019}.
\begin{definition}
  Let $\Phi, \Psi$ be relational assertions, $c\in\prcmd$, $\intlog:\logvar\rightarrow\mathbb{R}$ defined on $\epsilon,\delta$.
  We say that, $\Phi$ yields $\Psi$ through $c$  within $\epsilon$ and $\delta$ under $\intlog$
  (and we write $\judg{c}{(\epsilon,\delta)}{\Phi}{\Psi}{\intlog}$) iff
  \begin{itemize}
  \item $\{\{ \{\srconf{m_{I_{1}}}{m_{I_{2}}}{c}{[]}{[]}{\tocstr{m_{I}}{\Phi}}\}\}  \} \proofstep^{*}\setproof$
  \item $\exists  \srsetdistr=\{\setsymsubdistr{s_1}, \dots, \setsymsubdistr{s_t}\}\in\setproof$ such that
    \begin{itemize}
  \item $\final{\srsetdistr}$
  \item $\forall  \srconf{m_1}{m_2}{\skipp}{p_1}{p_2}{s}\in\bigcup_{\setsubdistr\in\srsetdistr}\setsubdistr.\exists \vec{k}$.
    \[s\implies \tocstr{\pair{m_1}{m_2}}{\Psi\land \epsilon_c\leq \epsilon \land \delta_c\leq \delta}\]
  \end{itemize}
  where $m_{I}\equiv\pair{m_{I_1}}{m_{I_2}}=
  \pair{m'_{I_1}[\epsilon_c\mapsto 0][\delta_c\mapsto 0]}{m'_{I_2}[\epsilon_c\mapsto 0][\delta_c\mapsto 0]}$,
  $m'_{I_1}$, and $m'_{I_2}$ are fully symbolic memories,   and $\vec{k}=k_1, k_2,\dots$ are the symbols generated by the rules for synchronizing commands.
\end{itemize}

\end{definition}
The idea of this definition is to automatize the proof search.
When proving differential privacy we will
usually consider $\Psi$ as being equality of the output variables in
the two runs and $\Phi$ as being our preconditions.
\section{Soundness of proofs and refutations}
In this section we will connect the material presented in Section
\ref{sec:derivations} with the one presented in Section \ref{sec:prelim}. 

\subsection{Soundness}
\begin{lemma}
  \label{lemma:sound}
  Let $c\in\prcmd$ with $o$ its output variable taking values over the set $O$. Let $c\in\prcmd$.
  If $\judg{c}{(\epsilon,\delta)}{d_1\sim d_2}{o_1=o_2}{\intlog}$ then $c$ is ($\epsilon$,$\delta$)-differentially private.
  Also, if $\judg{c}{(\epsilon,\delta)}{d_1\sim d_2}{o_1=\iota \implies o_2=\iota}{\intlog}$ forall $\iota\in O$
  then $c$ is ($\epsilon$,$\delta$)-differentially private.
  \end{lemma}
\subsection{Refutations through pure semantics}
\begin{lemma}
  \label{lem:refutations}
  $\{\{\{\srconf{m_1}{m_2}{c}{[]}{[]}{\tocstr{\pair{m_1}{m_2}}{\Phi}}\}\}\}\hat{\proofstep}\setproof$,
  and $\setsymsubdistr{s}\in\setdistr\in\setproof$ and, $\exists \sigma\models_{\ints} s$ such that
  $\Delta_{\epsilon}(\cdenote{\proj{1}{c}}(\sub{m_1}{\sigma}), \cdenote{\proj{2}{c}}(\sub{m_2}{\sigma}))>\delta$
  then $c$ is not differentially private.
  \begin{proof}
    It suffices to notice that the two distributions $\cdenote{\proj{1}{c}}(\sub{m_1}{\sigma})$, and $\cdenote{\proj{2}{c}}(\sub{m_2}{\sigma})$
    violate the $\delta$ bound of the $\Delta_{\epsilon}$ distance. Then $\sub{m_1}{\sigma}$ and $\sub{m_2}{\sigma}$ are counterexamples.
  \end{proof}
\end{lemma}

\else
\paragraph{\bf Metatheory}
\label{sec:derivations}
The language of relational assertions  $\Phi, \Psi\dots$ is defined using  first order predicate logic formulas
involving relational program expressions and logical variables in $\logvar$. The interpretation of
a relational assertions is naturally defined as a subset of $\cmemset\times\cmemset$,
that is the set of pairs of memories modelling the assertion.  We will denote by
$\tocstr{\cdot}{\cdot}$ the substitution 
function  mapping the variables in an assertion to the values they have in a memory (unary or relational). More details are in \cite{Farina2019}.
\begin{definition}
  Let $\Phi, \Psi$ be relational assertions, $c\in\prcmd$, $\intlog:\logvar\rightarrow\mathbb{R}$ be an interpretation defined on $\epsilon$.
  We say that, $\Phi$ yields $\Psi$ through $c$  within $\epsilon$  under $\intlog$
  (and we write $\judg{c}{\epsilon}{\Phi}{\Psi}{\intlog}$) iff
  1) $\{\{ \{\srconf{m_{I_{1}}}{m_{I_{2}}}{c}{[]}{[]}{\tocstr{m_{I}}{\Phi}}\}\}  \} \proofstep^{*}\setproof$
  2) $\exists  \srsetdistr=\{\setsymsubdistr{s_1}, \dots, \setsymsubdistr{s_t}\}\in\setproof$ such that $\final{\srsetdistr}$ and
  $\forall  \srconf{m_1}{m_2}{\skipp}{p_1}{p_2}{s}$ $\in\bigcup_{\setsubdistr\in\srsetdistr}\setsubdistr.$ $\exists\vec{k}$.
    $s\implies \tocstr{\pair{m_1}{m_2}}{\Psi\land \epsilon_c\leq \epsilon }$
  where $m_{I}\equiv\pair{m_{I_1}}{m_{I_2}}=
  \pair{m'_{I_1}[\epsilon_c\mapsto 0]}{m'_{I_2}[\epsilon_c\mapsto 0]}$,
  $m'_{I_1}$, and $m'_{I_2}$ are fully symbolic memories,   and $\vec{k}=k_1, k_2,\dots$ are the symbols generated by the rules for synchronizing commands.

\end{definition}
The idea of this definition is to automatize the proof search.
When proving differential privacy we will
usually consider $\Psi$ as being equality of the output variables in
the two runs and $\Phi$ as being our preconditions.
We can now prove the soundness of our approach. 
\begin{lemma}
  \label{lemma:sound}
  Let $c\in\prcmd$. If $\judg{c}{\epsilon}{d_1\sim d_2}{o_1=o_2}{\intlog}$ then $c$ is $\epsilon$-differentially private.
\end{lemma}
% \begin{lemma}
%    Let $c\in\prcmd$ with $o$ its output variable taking values over the set $O$.
%    If $\judg{c}{(\epsilon,\delta)}{d_1\sim d_2}{o_1=\iota \implies o_2=\iota}{\intlog}$ forall $\iota\in O$
%   then $c$ is ($\epsilon$,$\delta$)-differentially private.
%   \begin{proof}
%     The proof is by structural induction on $c$, using Lemma \ref{lem:pointwise}.
%   \end{proof}
%   \end{lemma}
We can also prove the soundness of refutations obtained by the semantics.
\begin{lemma}
  \label{lem:refutations}
  $\{\{\{\srconf{m_1}{m_2}{c}{[]}{[]}{\tocstr{\pair{m_1}{m_2}}{\Phi}}\}\}\}{\proofstep}\setproof$,
  and $\setsymsubdistr{s}\in\setdistr\in\setproof$ and, $\exists \sigma\models_{\ints} s$ such that
  $\Delta_{\epsilon}(\cdenote{\proj{1}{c}}(\sub{m_1}{\sigma}), \cdenote{\proj{2}{c}}(\sub{m_2}{\sigma}))>0$
  then $c$ is not differentially private.
\end{lemma}

\fi
\ifnum\full=1
\section{Strategies for counterexample finding}
Lemma \ref{lem:refutations} is hard to use to find counterexamples
because given two arbitrary probability distributions computing their
$\epsilon$-divergence is hard in general. For this reasons we will now
describe three strategies that might help in reducing the effort in
counterexample finding. This strategies help in isolating traces that could
potentially lead to violations. For this we need first some notation.
Given a set of constraints $s$ we define the triple
$\Omega=\langle\Omega_1, \Omega_2, C(\vec{k})\rangle\equiv\langle
\proj{1}{s}, \proj{2}{s},s\setminus(\proj{1}{s}\cup \proj{2}{s})
\rangle$.
Given a relational symbolic configuration we can always split its
symbolic set of constraints along the $\Omega$ pattern.  Given a set
of constraint $s$ of a relational trace its $\Omega_1$ projection is
the set of constraints generated by the branching instruction
performed by the left run, simlarly for $\Omega_2$. Finally,
$C(\vec{k})$ is the set of relational constraints coming from either
preconditions or invariants or, from the rule
$\tiny{\rulestyle{PROOF-STEP-LAP-GEN}}$. The, potentially empty, vector $\vec{k}=K_1,\dots K_n$
is the set of fresh symbols $K$ generated by that rule. We sometimes abuse notation
and consider $\Omega$ also as a set of constraints given by the union
of its first, second and third projection. Given a set of constraints we will also
consider it as a single proposition given by the conjunction of its elements.
\subsection{A simplifying assumption on traces and events}
In this section we will formalize the main assumption that we will use
in order to apply the strategies for counterexample finding presented
in the following two subsections. 
\begin{assumption}
\label{ass:assumption}
  Consider $c\in\prcmd$ with output variable $o$, then $c$ is such that $ \{\{\{\srconf{m_1}{m_2}{c}{[]}{[]}{s}\}\}\}\proofstep^{*}\setproof$ and 
  \[
    \forall \setsymsubdistr{\langle \Omega_1, C(\vec{k}), \Omega_2\rangle}\in\setdistr\in\setproof.\final{\setdistr}\wedge o_1=o_2 \implies \Omega_1\Leftrightarrow \Omega_2
  \]
\end{assumption}
The idea of this assumption is to consider only programs for which
it is necessary, for the output variable on both runs to assume the same values,
that the two runs follow the same branches. That is,  if the two output variables
are different then the two executions must have, at some point, taken different branches.
\paragraph{} The following definition will be used to distinguish
relational traces which are reachable on one run but not on the
other. We call this traces \emph{orthogonal}.
\begin{definition} A final relational symbolic trace is orthogonal
  when  its set of constraints is such that $\exists\sigma.\sigma \not\models \Omega_2$
  and $\sigma \models \Omega_1\wedge C(\vec{k})$. That is a trace for which the
  following formula is satisiable: $\neg(\Omega_1\wedge C(\vec{k})\implies \Omega_2)$
\end{definition}
\paragraph{} The next definition, instead, will be used to isolate relational
traces for which it's not possible that the left one is executed but the right one
is not. We call this traces \emph{specular}.
\begin{definition} A final relational symbolic trace
  is specular when its set of constraints is such that
  $\exists \vec{k}.\Omega_1\wedge C(\vec{k}) \implies \Omega_2$.
\end{definition}
The constraint $\Omega_1\wedge C(\vec{k})$ includes all the
constraints coming from the left projection's branching of the
symbolic execution and all the relational assumptions such as the
adjacency condition, and all constraints added by the potentially
fired $\tiny{\rulestyle{PROOF-STEP-LAP-GEN}}$ rule.  A specular trace
is such that its left projection constraints plus the relational
assumptions imply the right projection constraints. 
We will now describe three strategies that will be used to isolate
relational symbolic traces potentially leading to counterexamples.
\subsection{Strategy A}
In this strategy \csre uses only the rule
$\tiny{\rulestyle{PROOF-STEP-AVOC}}$ for sampling instructions, also
this strategy searches for orthogonal relational traces.  Under
assumption \ref{ass:assumption}, if this happens for a program then it
must be the case that the progam can output one value on one run with
some probability but the same value has 0 probability of being output
on the second run.  This very fact implies that for some input the
program has an unbounded privacy loss. To implement this strategy
\csre looks for orthogonal relational traces
$\srconf{m_1}{m_2}{\skipp}{p_1}{p_2}{\Omega}$ such that: $ \exists
\sigma. \sigma \models \Omega_1\wedge C(\vec{k})$ but $\sigma
\not\models \Omega_2$. Notice that using this strategy $\vec{k}$ will
always be empty, as the rule used for samplings does not introduce any
coupling between the two samples.  Hence, we don't need to quantify
over those symbols, and we can just write $C$.  The set $C$ though
might very well be non empty, because it will potentially include
relational assumptions, e.g. adjacency of the inputs.
\subsection{Strategy B}
This strategy symbolically executes the program in order to find a
 specular trace for which no matter how we relate, within the budget, the various pairs of
samples $X^{i}_1, X^{i}_2$ in the two runs - using the relational
schema $X^i_1+K_i=X^i_2$ - the postcondition is always false. That is
\csre looks for specular relational traces $\srconf{m_1}{m_2}{\skipp}{p_1}{p_2}{\Omega}$ 
such that:
\[ \forall \vec{k}.\bigg[ (\Omega_1\wedge C(\vec{k}) \implies \Omega_2 )
\wedge \tocstr{\pair{m_1}{m_2}}{\epsilon_c \leq \epsilon )}\bigg ] \implies \tocstr{\pair{m_1}{m_2}}{o_1\neq o_2}
\]
\subsection{Strategy C}
This strategy looks for relational traces for which the output
variable takes the same value on the two runs but too much of the
budget was spent. That is \csre looks for traces  $\srconf{m_1}{m_2}{\skipp}{p_1}{p_2}{\Omega}$ 
such that:
\[
  \forall \vec{k}. \bigg[ \Omega_1 \wedge C(\vec{k}) \wedge \Omega_2 \implies \tocstr{\pair{m_1}{m_2}}{o_1=o_2} \bigg] \implies \tocstr{\pair{m_1}{m_2}}{\epsilon_c >\epsilon}
\]

\else
\section{Strategies for counterexample finding}
Lemma \ref{lem:refutations} is hard to use to find counterexamples
in practice. For this reasons we will now
describe three strategies that can help in reducing the effort in
counterexample finding. This strategies help in isolating traces that could
potentially lead to violations. For this we need first some notation.
Given a set of constraints $s$ we define the triple
$\Omega=\langle\Omega_1, \Omega_2, C(\vec{k})\rangle\equiv\langle
\proj{1}{s}, \proj{2}{s},s\setminus(\proj{1}{s}\cup \proj{2}{s})
\rangle$. We sometimes abuse notation
and consider $\Omega$ also as a set of constraints given by the union
of its first, second and third projection, and we will also
consider a set of constraints as a single proposition given by the conjunction of its elements.
The set $C(\vec{k})$ contains relational constraints coming from either
preconditions or invariants or, from the rule
$\rulestyle{Proof-Step-Lap-Gen}$. The, potentially empty, vector $\vec{k}=K_1,\dots K_n$
is the set of fresh symbols $K$ generated by that rule.
In the rest of the paper  we will assume the following simplifying assumption. 
\begin{assumption}
\label{ass:assumption}
  Consider $c\in\prcmd$ with output variable $o$, then $c$ is such that $ \{\{\{\srconf{m_1}{m_2}{c}{[]}{[]}{s}\}\}\}\proofstep^{*}\setproof$ and 
$ \forall \setsymsubdistr{\langle \Omega_1, C(\vec{k}), \Omega_2\rangle}\in\setdistr\in\setproof.\final{\setdistr}\wedge o_1=o_2 \implies \Omega_1\Leftrightarrow \Omega_2$.
\end{assumption}
This assumption allow us to consider only programs for which
it is necessary, for the output variable on both runs to assume the same value,
that the two runs follow the same branches. That is,  if the two output differ then the two executions must have, at some point, taken different branches.

The following definition will be used to distinguish
relational traces which are reachable on one run but not on the
other. We call these traces \emph{orthogonal}.
\begin{definition} A final relational symbolic trace is orthogonal
  when  its set of constraints is such that $\exists\sigma.\sigma \not\models \Omega_2$
  and $\sigma \models \Omega_1\wedge C(\vec{k})$. That is a trace for which $\neg(\Omega_1\wedge C(\vec{k})\implies \Omega_2)$ is satisiable. 
\end{definition}
The next definition, instead, will be used to isolate relational
traces for which it is not possible that the left one is executed but the right one
is not. We call these traces \emph{specular}.
\begin{definition} A final rel. symbolic trace
  is specular if 
  $\exists \vec{k}.\Omega_1\wedge C(\vec{k}) \implies \Omega_2$.
\end{definition}
The constraint $\Omega_1\wedge C(\vec{k})$ includes all the
constraints coming from the left projection's branching of the
symbolic execution and all the relational assumptions such as the
adjacency condition, and all constraints added by the potentially
fired $\rulestyle{Proof-Step-Lap-Gen}$ rule.  A specular trace
is such that its left projection constraints plus the relational
assumptions imply the right projection constraints. 
We will now describe our three strategies.

\noindent\emph{Strategy A}
In this strategy \csre uses only the rule
$\rulestyle{Proof-Step-Avoc}$ for sampling instructions, also
this strategy searches for orthogonal relational traces.  Under
assumption \ref{ass:assumption}, if this happens for a program then it
must be the case that the progam can output one value on one run with
some probability but the same value has 0 probability of being output
on the second run.  This implies that for some input the
program has an unbounded privacy loss. To implement this strategy
\csre looks for orthogonal relational traces
$\srconf{m_1}{m_2}{\skipp}{p_1}{p_2}{\Omega}$ such that: $ \exists
\sigma. \sigma \models \Omega_1\wedge C(\vec{k})$ but $\sigma
\not\models \Omega_2$. Notice that using this strategy $\vec{k}$ will
always be empty, as the rule used for samplings does not introduce any
coupling between the two samples.

\noindent\emph{Strategy B}
This strategy symbolically executes the program in order to find a
 specular trace for which no matter how we relate, within the budget, the various pairs of
samples $X^{i}_1, X^{i}_2$ in the two runs - using the relational
schema $X^i_1+K_i=X^i_2$ - the postcondition is always false. That is
\csre looks for specular relational traces $\srconf{m_1}{m_2}{\skipp}{p_1}{p_2}{\Omega}$ 
such that:
$ \forall \vec{k}.[ (\Omega_1\wedge C(\vec{k}) \implies \Omega_2 )
\wedge \tocstr{\pair{m_1}{m_2}}{\epsilon_c \leq \epsilon )} ] \implies \tocstr{\pair{m_1}{m_2}}{o_1\neq o_2}$.

\noindent\emph{Strategy C}
This strategy looks for relational traces for which the output
variable takes the same value on the two runs but too much of the
budget was spent. That is \csre looks for traces  $\srconf{m_1}{m_2}{\skipp}{p_1}{p_2}{\Omega}$ 
such that:
$\forall \vec{k}. [ \Omega_1 \wedge C(\vec{k}) \wedge \Omega_2 \implies \tocstr{\pair{m_1}{m_2}}{o_1=o_2} ] \implies \tocstr{\pair{m_1}{m_2}}{\epsilon_c >\epsilon}
$.

Of the
presented strategies only strategy A is sound with respect to counterexample
finding, while the other two apply when the algorithm cannot be proven
differentially private by any combination of the rules. In this second
case though, \crse provides counterexamples which agree with other
refutation oriented results in literature. This strategies are hence termed
\emph{useful} because they amount to heuristics that can be applied in
some situations.
\fi
\ifnum\full=1
\section{Examples}
In this section we will review the examples presented in Section \ref{sec:highlevel}
and variations thereof to show how \csre works. 
\subsection{Unsafe sparse vector implementation: Algorithm \ref{alg:wrongsvt-1}}
\label{subsec:unsafe-svt-1}
In this section we will describe in more detail how
\csre deals with Algorithm \ref{alg:wrongsvt-1}.
The algorithm is not $\epsilon$-differentially private.
An easy fix would be to add noise the output too, that is, substitute line 7 with
$\ass{o[i]}{\lapp{q[i](D)}{\frac{\epsilon}{2}}}$, giving us an $2\epsilon$-differentially private algorithm.
\begin{wrapfigure}[15]{L}{0.45\textwidth}
\begin{minipage}{0.45\textwidth}
  \includegraphics[width=\textwidth]{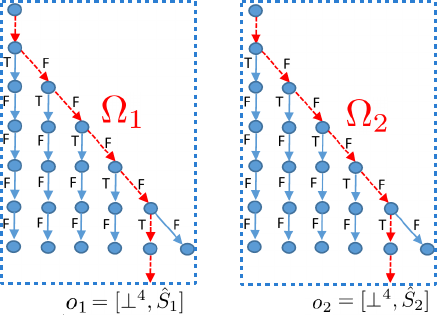}
  \caption{Two runs of Algorithm \ref{alg:wrongsvt-1} for 5 iterations}
  \label{img:tracessvt}
\end{minipage}
\end{wrapfigure}
Algorithm \ref{alg:wrongsvt-1} satisfies assumption
\ref{ass:assumption} because it outputs the whole array $o$ which takes
values of the form $\bot^{i},t$ or $\bot^{n}$ for $1\leq i\leq n$ and
$t\in\mathbb{R}$. The array, hence, encodes the whole trace.
So if two runs of the algorithm output the same
value it must be the case that they followed they same branching
instructions. 
Let's first notice that the algorithm is trivially $\epsilon$
differentially private, for any $\epsilon$, when the number of iterations $n$ is less than
or equal to 4.  Indeed it's enough to apply the sequential
composition theorem and get the obvious bound $\frac{\epsilon}{4}\cdot
n$. \csre can  prove this by applying the rule 
$\rulestyle{Proof-Step-Lap-Gen}$ $n$ times, and then 
choosing $K_1,\dots, K_n$ all equal to 0. This would imply
the statement of equality of the output variables spending less than
$\epsilon$.  Hence, it is obvious that if there is a potential
counterexample it can only be found after 4 iterations. In fact, a
potential counterexample can be found in 5 iterations. 
If we apply strategy B to this algorithm and follow the relational
symbolic trace that applies the rule
$\rulestyle{Proof-Step-Lap-Gen}$ for all the samplings we can
isolate the relational specular trace showed in Figure
\ref{alg:wrongsvt-1}, which corresponds to the left execution
following the false branch for the first four iterations and then
following the true branch and setting the fifth element of the array
to the sampled value.
Let's denote the respective final relational configuration by $\srconf{m_1}{m_2}{\skipp}{p_1}{p_2}{  s}$.
The set of constraints is as follows: $s=\langle \Omega_1, C(\vec{k}), \Omega_2\rangle=$
\begin{align*}
 \langle \{T_1>S^1_1, T_1>S^2_1, T_1>S^3_1, T_1>S^4_1, T_1&\leq S^5_1\},\\
  \{T_1+k_0=T_2, S^1_1+k_1=S^1_2, S^2_1+k_2=&S^2_2, \\
  S^3_1+k_3=S^3_2, S^4_1+k_4=S^4_2, S^5_1+k_5&=S^5_2, \\
  E_6=k_0\frac{\epsilon}{2}+\frac{\epsilon}{4}\displaystyle\sum_{i=1}^4k_i\dots\},\\
  \{T_2>S^1_2, T_2>S^2_2, T_2>S^3_2, T_2>S^4_2, T_2&\leq S^5_2\}\rangle
\end{align*}
with $m_1(\epsilon_c)=m_2(\epsilon_c)=E_6, m_1(o)=[S^1_1,\dots, S^5_1], m_2(o)=[S^1_2,\dots, S^5_2], m_1(t)=T_1, m_2(t)=T_2$.
We can see that strategy B applies, because
\begin{equation*}
  \models \forall \vec{k}.\bigg[ (\Omega_1\wedge C(\vec{k}) \implies \Omega_2 )
  \wedge \tocstr{\pair{m_1}{m_2}}{\epsilon_c \leq \epsilon )}\bigg ] \implies \tocstr{\pair{m_1}{m_2}}{o_1\neq o_2}
\end{equation*}
holds. The probability associated with these two traces can be
expressed as:
\begin{equation*}
\Gamma_{j}(\vec{q}(D_j), \epsilon, T, o)\equiv
  \int_{-\infty}^{+\infty}\textup{pdflap}^{T}_{\frac{\epsilon}{2}}(\rho)\bigg(\displaystyle\prod_{i=1}^{4}\textup{cdflap}^{q[i](D_{j})}_{\frac{\epsilon}{4}}(\rho)\Pr[\hat{s}^{5}_{j}=o \wedge \hat{s}^{5}_{j}\geq \rho]\bigg) d\rho
\end{equation*}
where $j\in\{1,2\}$ denotes which run (left or right) we are
considering, $\vec{q}$ is the vector of queries such that $\mid
q[i](D_1)-q[i](D_2)\mid \leq 1$ for adjacent databases $D_1, D_2$ and
$1\leq i\leq 5$. Also, $\textup{pdflap}^{T}_{\frac{\epsilon}{2}}(\rho)$
denotes the probability density at the point $\rho$ of a random
variable with Laplace distribution
with mean $T$ and scale $\frac{2}{\epsilon}$, and, finally, $\textup{cdflap}^{q[i](D_{j})}_{\frac{\epsilon}{4}}(\rho)$ denotes the
cumulative distributive function at the point $\rho$, of a random variable with Laplace distribution with mean $q[i](D_{j})$ and
scale $\frac{4}{\epsilon}$. Let's define $\Gamma(\vec{q}(D_1), \vec{q}(D_2), \epsilon, T, o)\equiv \frac{\Gamma_1(\vec{q}(D_1), \epsilon, T, o)}{\Gamma_2(\vec{q}(D_1), \epsilon, T, o)}$, then we can see that $\Gamma([00001],[11110], 1, 0, 0)>e^{1}$.
Where
\[ [q[1](D_1),q[2](D_1),q[3](D_1),q[4](D_1),q[5](D_1)]=\]$[00001]$
and, \[ [q[1](D_2),q[2](D_2),q[3](D_2),q[4](D_2),q[5](D_2)]=\]$[11110]$.
This pair of traces is, in fact, the same that has been found in
\cite{Lyu-2017} for a sligthly more general version of Algorithm
(\ref{alg:wrongsvt-1}).  Strategy B selects this relational trace
because, as already noticed in \cite{Barthe:2016} for a different
version of the algorithm, in order to make sure that the traces follow
the same branches, the coupling rules enforce necessarily that the two
samples released are different, preventing the \csre to prove equality
of the output variables in the two runs.

\subsection{Unsafe sparse vector implementation: Algorithm \ref{alg:wrongsvt-2}}
Algorithm \ref{alg:wrongsvt-2} also satisfies assumption
\ref{ass:assumption}: that is, the output encodes univocally the whole
history of the trace, and hence every trace corresponds injectively to
an event. The algorithm is trivially $\epsilon$ differentially private
for one iteration.  This because, intuitively, adding noise to the
threshold protects the result of the query as well at the branching
instruction, but only for one iteration, after that there is no
resampling. Indeed, the algorithm is not
$\epsilon$-differentially private, for any finite $\epsilon$ already
at the second iteration, and a witness for this can be found using \csre.
We can see this using strategy B.  Thanks to
this strategy we will isolate a relational orthogonal trace, similarly
to what has been found in \cite{Lyu-2017} for the same algorithm.
\csre will unfold the loop twice, and it will scan all relational traces
to see if there is an orthogonal trace. In particular, the relational trace that corresponds
to the output $o_1=o_2=[\bot,\top]$,
that is the the trace with set of constraints $\langle \Omega_1, C(\vec{k}), \Omega_2 \rangle=$ 
\begin{align*}
 \langle \{T_1>q_{1d1}, T_1\leq q_{2d1}\},\\
  \{|q_{1d1}- q_{1d2}|\leq 1, |q_{2d1}- q_{2d2}|\leq 1\}\\
  \{T_2>q_{1d2}, T_2\leq q_{2d2}\}\rangle 
\end{align*}

Since the vector $\vec{k}$ is empty we can omit it and just write
$C$. It is easy to see now that the following sigma:
$\sigma\equiv[q_{1d1}\mapsto 0, q_{2d1}\mapsto 1, q_{1d2}\mapsto 1,
q_{2d2}\mapsto 0]$, proves that this relational trace is orthogonal:
that is $\sigma \models \Omega_1\wedge C$, but $\sigma \not
\models\Omega_2$.  Indeed if we consider two inputs $D_1,D_2$ and two
queries $q_1, q_2$ such that: $q_1(D_1)=q_2(D_2)=0,
q_2(D_1)=q_1(D_2)=1$ we get that the probability of outputting the
value $o=[\bot,\top]$ is positive in the first run, but it is 0 on the second.
This implies that the algorithm can merely be proven to be $\infty$-differentially private.

\else
\paragraph{\bf Examples}
\emph{Unsafe sparse vector implementation: Algorithm \ref{alg:wrongsvt-1}}
\label{subsec:unsafe-svt-1}
We already discussed why this algorithm is not $\epsilon$-differentially private.
% An easy fix would be to add noise the output too, that is, substitute line 7 with
% $\ass{o[i]}{\lapp{q[i](D)}{\frac{\epsilon}{2}}}$, giving us an $2\epsilon$-differentially private algorithm.
Algorithm \ref{alg:wrongsvt-1} satisfies Assumption
\ref{ass:assumption} because it outputs the whole array $o$ which takes
values of the form $\bot^{i},t$ or $\bot^{n}$ for $1\leq i\leq n$ and
$t\in\mathbb{R}$. The array, hence, encodes the whole trace.
So if two runs of the algorithm output the same
value it must be the case that they followed they same branching
instructions. 
Let's first notice that the algorithm is trivially $\epsilon$
differentially private, for any $\epsilon$, when the number of iterations $n$ is less than
or equal to 4.
\begin{wrapfigure}[12]{L}{0.37\textwidth}
\vspace{-8mm}
  \begin{minipage}[t]{0.45\textwidth}
  \includegraphics[width=0.8\textwidth]{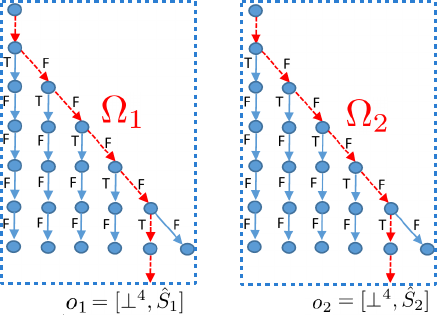}
  \caption{Two runs of Alg. \ref{alg:wrongsvt-1}.}
  \label{img:tracessvt}
\end{minipage}
\end{wrapfigure}

Indeed it's enough to apply the sequential
composition theorem and get the obvious bound $\frac{\epsilon}{4}\cdot
n$. \csre can  prove this by applying the rule 
$\rulestyle{Proof-Step-Lap-Gen}$ $n$ times, and then 
choosing $K_1,\dots, K_n$ all equal to 0. This would imply
the statement of equality of the output variables spending less than
$\epsilon$. A
potential counterexample can be found in 5 iterations. 
If we apply strategy B to this algorithm and follow the relational
symbolic trace that applies the rule
$\rulestyle{Proof-Step-Lap-Gen}$ for all the samplings we can
isolate the relational specular trace showed in Figure
\ref{alg:wrongsvt-1}, which corresponds to the left execution
following the false branch for the first four iterations and then
following the true branch and setting the fifth element of the array
to the sampled value.
Let's denote the respective final relational configuration by $\srconf{m_1}{m_2}{\skipp}{p_1}{p_2}{  s}$.
The set of constraints is as follows: $s=\langle \Omega_1, C(\vec{k}), \Omega_2\rangle=
 \langle \{T_1>S^1_1, T_1>S^2_1, T_1>S^3_1, T_1>S^4_1, T_1\leq S^5_1\},
  \{T_1+k_0=T_2, S^1_1+k_1=S^1_2, S^2_1+k_2=S^2_2, 
  S^3_1+k_3=S^3_2, S^4_1+k_4=S^4_2, S^5_1+k_5=S^5_2, 
  E_6=k_0\frac{\epsilon}{2}+\frac{\epsilon}{4}\displaystyle\sum_{i=1}^4k_i\dots\},
  \{T_2>S^1_2, T_2>S^2_2, T_2>S^3_2, T_2>S^4_2, T_2\leq S^5_2\}\rangle$
with $m_1(\epsilon_c)=m_2(\epsilon_c)=E_6, m_1(o)=[S^1_1,\dots, S^5_1], m_2(o)=[S^1_2,\dots, S^5_2], m_1(t)=T_1, m_2(t)=T_2$.
We can see that strategy B applies, because we have
$
\models \forall \vec{k}.[ (\Omega_1\wedge C(\vec{k}) \implies \Omega_2 )
  \wedge \tocstr{\pair{m_1}{m_2}}{\epsilon_c \leq \epsilon )} ] \implies \tocstr{\pair{m_1}{m_2}}{o_1\neq o_2}$. 
Computing the probability associated with these two traces we can verify that we have a counterexample.
% expressed as:
% $
% \Gamma_{j}(\vec{q}(D_j), \epsilon, T, o)\equiv
%   \int_{-\infty}^{+\infty}\textup{pdflap}^{T}_{\frac{\epsilon}{2}}(\rho)\bigg(\displaystyle\prod_{i=1}^{4}\textup{cdflap}^{q[i](D_{j})}_{\frac{\epsilon}{4}}(\rho)\Pr[\hat{s}^{5}_{j}=o \wedge \hat{s}^{5}_{j}\geq \rho]\bigg) d\rho
% $
% where $j\in\{1,2\}$ denotes which run (left or right) we are
% considering, $\vec{q}$ is the vector of queries such that $\mid
% q[i](D_1)-q[i](D_2)\mid \leq 1$ for adjacent databases $D_1, D_2$ and
% $1\leq i\leq 5$. Also, $\textup{pdflap}^{T}_{\frac{\epsilon}{2}}(\rho)$
% denotes the probability density at the point $\rho$ of a random
% variable with Laplace distribution
% with mean $T$ and scale $\frac{2}{\epsilon}$, and, finally, $\textup{cdflap}^{q[i](D_{j})}_{\frac{\epsilon}{4}}(\rho)$ denotes the
% cumulative distributive function at the point $\rho$, of a random variable with Laplace distribution with mean $q[i](D_{j})$ and
% scale $\frac{4}{\epsilon}$. Let's define $\Gamma(\vec{q}(D_1), \vec{q}(D_2), \epsilon, T, o)\equiv \frac{\Gamma_1(\vec{q}(D_1), \epsilon, T, o)}{\Gamma_2(\vec{q}(D_1), \epsilon, T, o)}$, then we can see that $\Gamma([00001],[11110], 1, 0, 0)>e^{1}$.
% Where
% $ [q[1](D_1),q[2](D_1),q[3](D_1),q[4](D_1),q[5](D_1)]=[00001]$
% and, $ [q[1](D_2),q[2](D_2),q[3](D_2),q[4](D_2),q[5](D_2)]=[11110]$.
This pair of traces is, in fact, the same that has been found in
\cite{Lyu-2017} for a sligthly more general version of Algorithm
(\ref{alg:wrongsvt-1}).  Strategy B selects this relational trace
since in order to make sure that the traces follow
the same branches, the coupling rules enforce necessarily that the two
samples released are different, preventing \csre to prove equality
of the output variables in the two runs.

\paragraph{Unsafe sparse vector implementation: Algorithm \ref{alg:wrongsvt-2}}
Also this algorithm satisfies Assumption \ref{ass:assumption}. The
algorithm is  $\epsilon$ differentially private for one
iteration. This because, intuitively, adding noise to the threshold
protects the result of the query as well at the branching instruction,
but only for one iteration. The algorithm is not $\epsilon$-differentially private, for any finite
$\epsilon$ already at the second iteration, and a witness for this can
be found using \csre.  We can see this using strategy B.  Thanks to
this strategy we will isolate a relational orthogonal trace, similarly
to what has been found in \cite{Lyu-2017} for the same algorithm.
\csre will unfold the loop twice, and it will scan all relational
traces to see if there is an orthogonal trace. In particular, the
relational trace that corresponds to the output $o_1=o_2=[\bot,\top]$,
that is the the trace with set of constraints
$\langle \Omega_1, C(\vec{k}), \Omega_2 \rangle= \langle
\{T_1>q_{1d1}, T_1\leq q_{2d1}\}, \{|q_{1d1}- q_{1d2}|\leq 1,
|q_{2d1}- q_{2d2}|\leq 1\} \{T_2>q_{1d2}, T_2\leq q_{2d2}\}\rangle$.
Since the vector $\vec{k}$ is empty we can omit it and just write
$C$. It is easy to see now that the following sigma:
$\sigma\equiv[q_{1d1}\mapsto 0, q_{2d1}\mapsto 1, q_{1d2}\mapsto 1,
q_{2d2}\mapsto 0]$, proves that this relational trace is orthogonal:
that is $\sigma \models \Omega_1\wedge C$, but $\sigma \not
\models\Omega_2$.  Indeed if we consider two inputs $D_1,D_2$ and two
queries $q_1, q_2$ such that: $q_1(D_1)=q_2(D_2)=0,
q_2(D_1)=q_1(D_2)=1$ we get that the probability of outputting the
value $o=[\bot,\top]$ is positive in the first run, but it is 0 on the second.
Hence, the algorithm can only be proven to be $\infty$-differentially private.

\fi
\ifnum\full=1
\paragraph{A safe sparse vector implementation}
As already mentioned two variations of Algorithm \ref{alg:wrongsvt-1} can be proven secure.
The first one substitutes $\rass{o[i]}{\lapp{q[i](D)}{\frac{\epsilon}{2}}}$ in place of line 7,
while the second one substitutes  $\ass{o[i]}{\top}$.
The former version can be proven $2\epsilon$-dp, while the latter: $\epsilon$-dp.
We will explain a proof of this last statement for a constant $n$, for example 5.
The proof presented is based on that in \cite{Barthe:2016}, but will
have a relational symbolic execution style instead of an apRHL$^{+}$ one.
\csre will try to prove the following postconditions:
\begin{enumerate}
\item $o_1=[\top,\bot,\dots,\bot] \implies o_2=[\top,\bot,\dots,\bot] \land\epsilon_c\leq \epsilon$
\item $o_1=[\bot,\top,\dots,\bot] \implies o_2=[\bot,\top,\dots,\bot] \land\epsilon_c\leq \epsilon$
\item $o_1=[\bot,\dots,\top,\bot] \implies o_2=[\bot,\dots,\top,\bot] \land\epsilon_c\leq \epsilon$
\item $\dots$
\item $o_1=[\bot,\dots,\bot,\top] \implies o_2=[\bot,\dots,\bot,\top] \land\epsilon_c\leq \epsilon$
\end{enumerate}
When trying to prove the i-th one the only interesting iteration will be the i-th one. This because
all the others the postcondition will be vacuously true, and also the budget spent will be $k_0\frac{\epsilon}{2}$,
the one spent for the threshold, and for all the other sampling instruction we can spend 0 by just setting $k_j=q[j](D_2)-q[j](D_1)$ for all
$j\neq i$, that is by couplin in this way the samples: $\hat{s}_1+k_j=\hat{s}_2$, with $k_j=q[j](D_2)-q[j](D_1)$, spending $\lvert k_j+ q[j](D_2)-q[j](D_1)\rvert=0$.
So, at the i-th iteration the samples are coupled $\hat{s}_1+k_i=\hat{s}_2$, with $k_i=1$.
So if $\hat{s}_1\geq \hat{t}_1$ then also $\hat{s}_2\geq \hat{t}_2$, and also, if
$\hat{s}_1< \hat{t}_1$ then also $\hat{s}_2< \hat{t}_2$. This implies that at th i-th iteration
we enter on the right run the true branch iff we enter the true branch on the left one.
This by spending $\lvert k_i + q[i](D_2)-q[i](D_1)\rvert\frac{\epsilon}{4}\leq 2\frac{\epsilon}{4}$.
For a total of $\epsilon$.
\paragraph{Unsafe Laplace mechanism: Algorithm \ref{alg:wrongnoise} }
Algorithm \ref{alg:wrongnoise} is not $\epsilon$ differentially
private for any finite $\epsilon$.  The intuition is that not enough
noise is added to hide the difference of the result of a query applied
to two adjacent databases. This translates in any possible potential proof based
on the coupling rules in using too much of the budget.
The program of Algorithm \ref{alg:wrongnoise} has only one possible
final relational trace: $\srconf{m_1}{m_2}{\skipp}{p_1}{p_2}{\langle
\Omega_1, C(\vec{k}, \Omega_2)\rangle}$. Since there are no branching
instructions $\Omega_1=\{ \proj{1}{2E} >0\}$ and $\Omega_2=\emptyset$,
where $m_1(\epsilon)=m_2(\epsilon)=E$. Since there is one sampling
instruction $C(\vec{k})$ will include the following set of constraints
$\{|Q_{d1}- Q_{d2}|\leq 1, R_1+K=R_2, E_{c}=\mid K\mid\cdot 2\cdot
K'\cdot E,O_1=R_1+ Q_{d1}, O_2= R_2+ Q_{d2}, E_c=K'\cdot E\}$, with
$m_1(o)=O_1, m_2(o)=O_2, m_1(\epsilon_c)=m_2(\epsilon_c)=E_c$.
Intuitively we can see that, given this set of constraints, if it has
to be the case that $O_1=O_2$ then, $Q_{d1}-Q_{d_2}=K$. But
$Q_{d1}-Q_{d_2}$ can be 1 and hence, $E_{c}$ is at least 2. This tells
us that if we want to equate the two output variables we need to spen
at least twice the budget. Any relational input satisfying the precondition will give us
a counterexample, provided the two projections are different.
\paragraph{A safe Laplace mechanism}
By substituting line 2 in Algorithm  \ref{alg:wrongnoise} with  $\rass{\rho}{\lapp{0}{\epsilon}}$
we get an $\epsilon$-dp algorithm. Indeed when executing that line \csre would generate the following
constraint $p_1+k_0=p_2 \land\mid k_0 + 0 - 0\mid\leq k_1\land o_1=v_1+p_1\land o_2=v_2+p_2$.
Which by instantiating $k=0,k_1=v_2-v_1$ implies $o_1=o_2\land \epsilon_c\leq \epsilon$. 

\else
\paragraph{A safe sparse vector implementation}
Algorithm \ref{alg:wrongsvt-1} can be proven $\epsilon$-differentially private if we replace  $\ass{o[i]}{\top}$ to  line 7.
Let us consider a proof of this statement for $n=5$.
\csre will try to prove the following 5 sentences:
$o_1=[\top,\bot,\dots,\bot] \implies o_2=[\top,\bot,\dots,\bot] \land\epsilon_c\leq \epsilon$, $\dots$,
$o_1=[\bot,\dots,\bot,\top] \implies o_2=[\bot,\dots,\bot,\top] \land\epsilon_c\leq \epsilon$.
The only interesting iteration will be the i-th one. This because
all the others the postcondition will be vacuously true, and also the budget spent will be $k_0\frac{\epsilon}{2}$,
the one spent for the threshold, and for all the other sampling instruction we can spend 0 by just setting $k_j=q[j](D_2)-q[j](D_1)$ for all
$j\neq i$, that is by couplin in this way the samples: $\hat{s}_1+k_j=\hat{s}_2$, with $k_j=q[j](D_2)-q[j](D_1)$, spending $\lvert k_j+ q[j](D_2)-q[j](D_1)\rvert=0$.
So, at the i-th iteration the samples are coupled $\hat{s}_1+k_i=\hat{s}_2$, with $k_i=1$.
So if $\hat{s}_1\geq \hat{t}_1$ then also $\hat{s}_2\geq \hat{t}_2$, and also, if
$\hat{s}_1< \hat{t}_1$ then also $\hat{s}_2< \hat{t}_2$. This implies that at th i-th iteration
we enter on the right run the true branch iff we enter the true branch on the left one.
This by spending $\lvert k_i + q[i](D_2)-q[i](D_1)\rvert\frac{\epsilon}{4}\leq 2\frac{\epsilon}{4}$.
For a total of $\epsilon$.
\fi
%

% \section{Experimental evaluation}

%
%
%
% 
\ifnum\full=1
\section{Related Works}
The closest work to ours is \cite{Barthe2019AutomatedMF} where authors
devise a decision logic for differential privacy. The logic allows to soundly prove
or disprove $\epsilon$ and $(\epsilon,\delta)$ differential privacy
by encoding the semantics of the progam into a decidable fragment of the first-order
theory of the reals with exponentiation. The programs considered
don't allow assignemnts to real and integer variables inside the body
of while loops.
Other two works very related to ours are \cite{DingWWZK18} and
\cite{Bichsel:2018}.  Their approach to finding counterexamples to
differential privacy differs from ours in two main ways. First of all
they use a statistical approach by approximating the output
distributions of the program on two related inputs and then smartly
checking whether for some events these output distributions provide an
out of bound ratio. Secondly they are by nature numerical methods
providing results stating that an algorithm is not $\epsilon$
differential private for some actual concrete $\epsilon$. These kind
of results obviously imply that the algorithm is not differentially
private for all other concrete $\epsilon'$ such that
$\epsilon'>\epsilon$, but they can only suggest that the algorithm is not private
for also the $\epsilon'$, such that $\epsilon'< \epsilon$, if that
is indeed the case. Our work instead is a purely symbolic
technique which provides results stating that an algorithm is not
$\epsilon$-differentially private for any finite $\epsilon$.  An
advantage of our approach is obviously the speed of the analysis which
does not require any sampling. 
In \cite{LiuWZ18} authors add model checking to the tools for counterexample finding
to differential privacy. The main difference with our work is that, as usual, model checking analyzes a
model of the code and not directly the code. Also, in the specific case
of the sparse vector algorithm family, their work seem to be able to handle only
a finite number of iterations.

This work can be seen as a non trivial probabilistic extension of the
framework presented in \cite{Farina2019}, where sampling instructions
in the relational symbolic semantics are handled through an
adaptation, in a symbolic execution framework, of the apRHL$^{+}$
rules first presented in \cite{Barthe:2016}. Their logic 
proves judgments implying differential privacy but does not help in
finding counterexamples when the program is not private.
This work is also close to \cite{Albarghouthi:2017} where authors
devised a framework to automatically discover proofs of privacy using
coupling rules, but again thier work does not help in refuting privacy of
buggy programs.

\else
\section{Related Works}
The closest works to ours are
\cite{Albarghouthi:2017,Barthe2019AutomatedMF}.  In
\cite{Albarghouthi:2017} the authors devised a synthesis framework to
automatically discover proofs of privacy using coupling rules similar
to ours. Their approach is not based on symbolic execution but on
synthesis technique. Moreover, their framework cannot be directly used
to refuting privacy of buggy programs.
In \cite{Barthe2019AutomatedMF} the authors
devise a decision logic for differential privacy which can to soundly prove
or disprove differential privacy. The programs considered
don't allow assignemnts to real and integer variables inside the body
of while loops. While their technique is different from our, their logic could be potentially integrated in our framework as a decision procedure.
Another related work is \cite{LiuWZ18}, where the authors study  model checking as a tool for counterexample finding
to differential privacy. The main difference with our work is in the basic technique and in the fact that model checking reason about a
model of the code, rather than the code itself. They also consider the above threshold example and they are able to handle only
a finite number of iterations.
Other works have studied how to find violations to differential privacy~\cite{DingWWZK18,Bichsel:2018}.  Their approach differs from ours in two ways: first,
they use a statistical approach; second, they  look at concrete values of
the data and the privacy parameters. By using an approach based on symbolic
execution we are able to reason about symbolic values, and so consider
$\epsilon$-differential privacy for any finite $\epsilon$. Moreover, our
technique does not need sampling - although we still need to compute
distributions to confirm a violation. 
Our work can be seen as a probabilistic extension of the
framework presented in \cite{Farina2019}, where sampling instructions
in the relational symbolic semantics are handled through rules inspired by
the logic apRHL$^{+}$~\cite{Barthe:2016}. This logic can be used to prove
differential privacy but does not directly help in
finding counterexamples when the program is not private.

\fi
\ifnum\full=1
\section{Conclusion and Future Work}
In this work we presented \csre: a symbolic execution engine framework
which integrates relational reasoning and probabilistic couplings.
The framework allows both proving differential privacy of the most
known differentially private algorithms, and refutation of buggy
versions thereof which are particularly trick to distinguish from the
correct ones.  While refuting it is also able isolate traces and
events which leads to counterexamples to differnetial privacy.  When
proving \csre uses a similar approach to apRHL$^+$ but follows a
strong postcondion approach instead of the more standard weak
precondition style of proof, as it is common in symbolic execution
style of proofs. \csre uses refuting principles, or
strategies, to isolate potentially \emph{dangerous} traces. Of the
presented strategies only one is sound with respect to counterexample
finding, while the other two apply when the algorithm cannot be proven
differentially private by any combination of the rules. In this second
case though \csre provides counterexamples which agree with other
refutation oriented results in literature.
Future work includes interfacing more efficiently \csre with numeric
solvers to find maximums of ratios of probabilities of traces.

\else
\section{Conclusion}
We presented \csre: a symbolic execution engine framework
integrating relational reasoning and probabilistic couplings.
The framework allows both proving and refuting differential privacy.
When
proving \csre can be seen as strong postcondion calculus. When refuting \csre
uses several strategies to isolate potentially \emph{dangerous} traces. 
Future work includes interfacing more efficiently \csre with numeric
solvers to find maximums of ratios of probabilities of traces.

\fi

\bibliographystyle{splncs04}
\ifnum\full=1
\bibliography{IEEEabrv,biblio}
\else
\bibliography{IEEEabrv,biblio-short}

\begin{thebibliography}{10}
\providecommand{\url}[1]{\texttt{#1}}
\providecommand{\urlprefix}{URL }
\providecommand{\doi}[1]{https://doi.org/#1}

\bibitem{Albarghouthi:2017}
Albarghouthi, A., Hsu, J.: Synthesizing coupling proofs of differential
  privacy. Proc. ACM Program. Lang.  \textbf{2}(POPL),  58:1--58:30 (Dec 2017).
  \doi{10.1145/3158146}, \url{http://doi.acm.org/10.1145/3158146}

\bibitem{AndryscoKMJLS15}
Andrysco, M., Kohlbrenner, D., Mowery, K., Jhala, R., Lerner, S., Shacham, H.:
  On subnormal floating point and abnormal timing. In: 2015 {IEEE} Symposium on
  Security and Privacy, {SP} 2015, San Jose, CA, USA, May 17-21, 2015. pp.
  623--639 (2015). \doi{10.1109/SP.2015.44},
  \url{https://doi.org/10.1109/SP.2015.44}

\bibitem{Barthe2019AutomatedMF}
Barthe, G., Chadha, R., Jagannath, V., Sistla, A.P., Viswanathan, M.: Deciding
  differential privacy for programs with finite inputs and outputs. In:
  Hermanns, H., Zhang, L., Kobayashi, N., Miller, D. (eds.) {LICS} '20: 35th
  Annual {ACM/IEEE} Symposium on Logic in Computer Science, Saarbr{\"{u}}cken,
  Germany, July 8-11, 2020. pp. 141--154. {ACM} (2020).
  \doi{10.1145/3373718.3394796}, \url{https://doi.org/10.1145/3373718.3394796}

\bibitem{BartheFGAGHS16}
Barthe, G., Farina, G.P., Gaboardi, M., Arias, E.J.G., Gordon, A., Hsu, J.,
  Strub, P.: Differentially private bayesian programming. In: Proceedings of
  the 2016 {ACM} {SIGSAC} Conference on Computer and Communications Security,
  Vienna, Austria, October 24-28, 2016. pp. 68--79 (2016).
  \doi{10.1145/2976749.2978371}, \url{https://doi.org/10.1145/2976749.2978371}

\bibitem{BartheFGGHS16}
Barthe, G., Fong, N., Gaboardi, M., Gr{\'{e}}goire, B., Hsu, J., Strub, P.:
  Advanced probabilistic couplings for differential privacy. In: Weippl, E.R.,
  Katzenbeisser, S., Kruegel, C., Myers, A.C., Halevi, S. (eds.) Proceedings of
  the 2016 {ACM} {SIGSAC} Conference on Computer and Communications Security,
  Vienna, Austria, October 24-28, 2016. pp. 55--67. {ACM} (2016).
  \doi{10.1145/2976749.2978391}, \url{https://doi.org/10.1145/2976749.2978391}

\bibitem{BartheGAHKS14}
Barthe, G., Gaboardi, M., Arias, E.J.G., Hsu, J., Kunz, C., Strub, P.: Proving
  differential privacy in hoare logic. In: {IEEE} 27th Computer Security
  Foundations Symposium, {CSF} 2014, Vienna, Austria, 19-22 July, 2014. pp.
  411--424. {IEEE} Computer Society (2014). \doi{10.1109/CSF.2014.36},
  \url{https://doi.org/10.1109/CSF.2014.36}

\bibitem{BartheGAHRS15}
Barthe, G., Gaboardi, M., Arias, E.J.G., Hsu, J., Roth, A., Strub, P.:
  Higher-order approximate relational refinement types for mechanism design and
  differential privacy. In: Proceedings of the 42nd Annual {ACM}
  {SIGPLAN-SIGACT} Symposium on Principles of Programming Languages, {POPL}
  2015, Mumbai, India, January 15-17, 2015. pp. 55--68 (2015).
  \doi{10.1145/2676726.2677000}, \url{https://doi.org/10.1145/2676726.2677000}

\bibitem{Barthe:2016}
Barthe, G., Gaboardi, M., Gr{\'e}goire, B., Hsu, J., Strub, P.Y.: Proving
  differential privacy via probabilistic couplings. In: Proceedings of the 31st
  Annual ACM/IEEE Symposium on Logic in Computer Science. pp. 749--758. LICS
  '16, ACM, New York, NY, USA (2016). \doi{10.1145/2933575.2934554},
  \url{http://doi.acm.org/10.1145/2933575.2934554}

\bibitem{barthe2012probabilistic}
Barthe, G., K{\"o}pf, B., Olmedo, F., Zanella~Beguelin, S.: Probabilistic
  relational reasoning for differential privacy. ACM SIGPLAN Notices
  \textbf{47}(1),  97--110 (2012)

\bibitem{Bichsel:2018}
Bichsel, B., Gehr, T., Drachsler-Cohen, D., Tsankov, P., Vechev, M.: Dp-finder:
  Finding differential privacy violations by sampling and optimization. In:
  Proceedings of the 2018 ACM SIGSAC Conference on Computer and Communications
  Security. pp. 508--524. CCS '18, ACM, New York, NY, USA (2018).
  \doi{10.1145/3243734.3243863},
  \url{http://doi.acm.org/10.1145/3243734.3243863}

\bibitem{Burke17}
Burke, M.: Vulnerability in floating point implementation of exponential
  mechanism (2017), {Poster at Theory and Practice of Differential Privacy
  (TPDP 17)}

\bibitem{ChatzikokolakisGPX14}
Chatzikokolakis, K., Gebler, D., Palamidessi, C., Xu, L.: Generalized
  bisimulation metrics. In: Baldan, P., Gorla, D. (eds.) {CONCUR} 2014 -
  Concurrency Theory - 25th International Conference, {CONCUR} 2014, Rome,
  Italy, September 2-5, 2014. Proceedings. Lecture Notes in Computer Science,
  vol.~8704, pp. 32--46. Springer (2014). \doi{10.1007/978-3-662-44584-6\_4},
  \url{https://doi.org/10.1007/978-3-662-44584-6\_4}

\bibitem{ChistikovMP18}
Chistikov, D., Murawski, A.S., Purser, D.: Bisimilarity distances for
  approximate differential privacy. In: Lahiri, S.K., Wang, C. (eds.) Automated
  Technology for Verification and Analysis - 16th International Symposium,
  {ATVA} 2018, Los Angeles, CA, USA, October 7-10, 2018, Proceedings. Lecture
  Notes in Computer Science, vol. 11138, pp. 194--210. Springer (2018).
  \doi{10.1007/978-3-030-01090-4\_12},
  \url{https://doi.org/10.1007/978-3-030-01090-4\_12}

\bibitem{ChistikovMP19}
Chistikov, D., Murawski, A.S., Purser, D.: Asymmetric distances for approximate
  differential privacy. In: Fokkink, W.J., van Glabbeek, R. (eds.) 30th
  International Conference on Concurrency Theory, {CONCUR} 2019, August 27-30,
  2019, Amsterdam, the Netherlands. LIPIcs, vol.~140, pp. 10:1--10:17. Schloss
  Dagstuhl - Leibniz-Zentrum f{\"{u}}r Informatik (2019).
  \doi{10.4230/LIPIcs.CONCUR.2019.10},
  \url{https://doi.org/10.4230/LIPIcs.CONCUR.2019.10}

\bibitem{DeMoura:2008}
De~Moura, L., Bj{\o}rner, N.: Z3: An efficient smt solver. In: Proceedings of
  the Theory and Practice of Software, 14th International Conference on Tools
  and Algorithms for the Construction and Analysis of Systems. pp. 337--340.
  TACAS'08/ETAPS'08, Springer-Verlag, Berlin, Heidelberg (2008),
  \url{http://dl.acm.org/citation.cfm?id=1792734.1792766}

\bibitem{DingWWZK18}
Ding, Z., Wang, Y., Wang, G., Zhang, D., Kifer, D.: Detecting violations of
  differential privacy. In: Proceedings of the 2018 {ACM} {SIGSAC} Conference
  on Computer and Communications Security, {CCS} 2018, Toronto, ON, Canada,
  October 15-19, 2018. pp. 475--489 (2018). \doi{10.1145/3243734.3243818},
  \url{https://doi.org/10.1145/3243734.3243818}

\bibitem{DworkMNS16}
Dwork, C., McSherry, F., Nissim, K., Smith, A.D.: Calibrating noise to
  sensitivity in private data analysis. J. Priv. Confidentiality
  \textbf{7}(3),  17--51 (2016). \doi{10.29012/jpc.v7i3.405},
  \url{https://doi.org/10.29012/jpc.v7i3.405}

\bibitem{EbadiAS16}
Ebadi, H., Antignac, T., Sands, D.: Sampling and partitioning for differential
  privacy. In: 14th Annual Conference on Privacy, Security and Trust, {PST}
  2016, Auckland, New Zealand, December 12-14, 2016. pp. 664--673 (2016).
  \doi{10.1109/PST.2016.7906954},
  \url{https://doi.org/10.1109/PST.2016.7906954}

\bibitem{Farina2019}
Farina, G.P., Chong, S., Gaboardi, M.: Relational symbolic execution. In:
  Proceedings of the 21st International Symposium on Principles and Practice of
  Programming Languages 2019. pp. 10:1--10:14. PPDP '19, ACM, New York, NY, USA
  (2019). \doi{10.1145/3354166.3354175},
  \url{http://doi.acm.org/10.1145/3354166.3354175}

\bibitem{Gaboardi2013}
Gaboardi, M., Haeberlen, A., Hsu, J., Narayan, A., Pierce, B.C.: Linear
  dependent types for differential privacy. SIGPLAN Not.  \textbf{48}(1),
  357--370 (Jan 2013). \doi{10.1145/2480359.2429113},
  \url{http://doi.acm.org/10.1145/2480359.2429113}

\bibitem{GaboardiNP20}
Gaboardi, M., Nissim, K., Purser, D.: The complexity of verifying loop-free
  programs as differentially private. In: Czumaj, A., Dawar, A., Merelli, E.
  (eds.) 47th International Colloquium on Automata, Languages, and Programming,
  {ICALP} 2020, July 8-11, 2020, Saarbr{\"{u}}cken, Germany (Virtual
  Conference). LIPIcs, vol.~168, pp. 129:1--129:17. Schloss Dagstuhl -
  Leibniz-Zentrum f{\"{u}}r Informatik (2020).
  \doi{10.4230/LIPIcs.ICALP.2020.129},
  \url{https://doi.org/10.4230/LIPIcs.ICALP.2020.129}

\bibitem{Haeberlen11}
Haeberlen, A., Pierce, B.C., Narayan, A.: Differential privacy under fire. In:
  Proceedings of the 20th USENIX Security Symposium (Aug 2011)

\bibitem{Mathematica}
Inc., W.R.: Mathematica, {V}ersion 12.0,
  \url{https://www.wolfram.com/mathematica}, champaign, IL, 2019

\bibitem{Jonsson0L01}
Jonsson, B., Yi, W., Larsen, K.G.: Probabilistic extensions of process
  algebras. In: Bergstra, J.A., Ponse, A., Smolka, S.A. (eds.) Handbook of
  Process Algebra, pp. 685--710. North-Holland / Elsevier (2001).
  \doi{10.1016/b978-044482830-9/50029-1},
  \url{https://doi.org/10.1016/b978-044482830-9/50029-1}

\bibitem{King:1976}
King, J.C.: Symbolic execution and program testing. Commun. ACM
  \textbf{19}(7),  385--394 (Jul 1976). \doi{10.1145/360248.360252},
  \url{http://doi.acm.org/10.1145/360248.360252}

\bibitem{Lindvall1992LecturesOT}
Lindvall, T.: Lindvall1992lecturesot. In: Lectures on the Coupling Method
  (1992)

\bibitem{LiuWZ18}
Liu, D., Wang, B., Zhang, L.: Model checking differentially private properties.
  In: Programming Languages and Systems - 16th Asian Symposium, {APLAS} 2018,
  Wellington, New Zealand, December 2-6, 2018, Proceedings. pp. 394--414
  (2018). \doi{10.1007/978-3-030-02768-1\_21},
  \url{https://doi.org/10.1007/978-3-030-02768-1\_21}

\bibitem{Lyu-2017}
Lyu, M., Su, D., Li, N.: Understanding the sparse vector technique for
  differential privacy. Proc. VLDB Endow.  \textbf{10}(6),  637--648 (Feb
  2017). \doi{10.14778/3055330.3055331},
  \url{https://doi.org/10.14778/3055330.3055331}

\bibitem{MatlabOTB}
Matlab optimization toolbox, the MathWorks, Natick, MA, USA

\bibitem{Mironov12}
Mironov, I.: On significance of the least significant bits for differential
  privacy. In: the {ACM} Conference on Computer and Communications Security,
  CCS'12, Raleigh, NC, USA, October 16-18, 2012. pp. 650--661 (2012).
  \doi{10.1145/2382196.2382264},
  \url{http://doi.acm.org/10.1145/2382196.2382264}

\bibitem{NearDASGWSZSSS19}
Near, J.P., Darais, D., Abuah, C., Stevens, T., Gaddamadugu, P., Wang, L.,
  Somani, N., Zhang, M., Sharma, N., Shan, A., Song, D.: Duet: an expressive
  higher-order language and linear type system for statically enforcing
  differential privacy. Proc. {ACM} Program. Lang.  \textbf{3}({OOPSLA}),
  172:1--172:30 (2019). \doi{10.1145/3360598},
  \url{https://doi.org/10.1145/3360598}

\bibitem{pottier2002information}
Pottier, F., Simonet, V.: Information flow inference for ml. In: ACM SIGPLAN
  Notices. vol.~37, pp. 319--330. ACM (2002)

\bibitem{ReedP10}
Reed, J., Pierce, B.C.: Distance makes the types grow stronger: a calculus for
  differential privacy. In: Hudak, P., Weirich, S. (eds.) Proceeding of the
  15th {ACM} {SIGPLAN} international conference on Functional programming,
  {ICFP} 2010, Baltimore, Maryland, USA, September 27-29, 2010. pp. 157--168.
  {ACM} (2010). \doi{10.1145/1863543.1863568},
  \url{https://doi.org/10.1145/1863543.1863568}

\bibitem{SatoBGHK19}
Sato, T., Barthe, G., Gaboardi, M., Hsu, J., Katsumata, S.: Approximate span
  liftings: Compositional semantics for relaxations of differential privacy.
  In: 34th Annual {ACM/IEEE} Symposium on Logic in Computer Science, {LICS}
  2019, Vancouver, BC, Canada, June 24-27, 2019. pp. 1--14. {IEEE} (2019).
  \doi{10.1109/LICS.2019.8785668},
  \url{https://doi.org/10.1109/LICS.2019.8785668}

\bibitem{TschantzKD11}
Tschantz, M.C., Kaynar, D.K., Datta, A.: Formal verification of differential
  privacy for interactive systems (extended abstract). In: Mislove, M.W.,
  Ouaknine, J. (eds.) Twenty-seventh Conference on the Mathematical Foundations
  of Programming Semantics, {MFPS} 2011, Pittsburgh, PA, USA, May 25-28, 2011.
  Electronic Notes in Theoretical Computer Science, vol.~276, pp. 61--79.
  Elsevier (2011). \doi{10.1016/j.entcs.2011.09.015},
  \url{https://doi.org/10.1016/j.entcs.2011.09.015}

\bibitem{Vadhan827361}
Vadhan, S.: The Complexity of Differential Privacy (2016)

\bibitem{Wang19}
Wang, Y., Ding, Z., Wang, G., Kifer, D., Zhang, D.: Proving differential
  privacy with shadow execution. In: Proceedings of the 40th ACM SIGPLAN
  Conference on Programming Language Design and Implementation. pp. 655--669.
  PLDI 2019, ACM, New York, NY, USA (2019). \doi{10.1145/3314221.3314619},
  \url{http://doi.acm.org/10.1145/3314221.3314619}

\bibitem{ZhangK17}
Zhang, D., Kifer, D.: Lightdp: towards automating differential privacy proofs.
  In: Castagna, G., Gordon, A.D. (eds.) Proceedings of the 44th {ACM} {SIGPLAN}
  Symposium on Principles of Programming Languages, {POPL} 2017, Paris, France,
  January 18-20, 2017. pp. 888--901. {ACM} (2017),
  \url{http://dl.acm.org/citation.cfm?id=3009884}

\end{thebibliography}
\fi
%

%\import{appendix.tex}
\end{document}